\documentclass{aa}
\usepackage{graphicx}
\usepackage{txfonts}
\usepackage{hyperref}
\usepackage{url}
\usepackage{xcolor}
\hypersetup{
colorlinks,
linkcolor={red!80!black},
citecolor={blue!80!black},
urlcolor={blue!80!black}
}

\def\cm3{cm$^{-3}$}
\def\kms{km~s$^{-1}$}

\def\msun{M$_{\odot}$}

\def\one{\ts {\,\sc i}}
\def\two{\ts {\,\sc ii}}
\def\three{\ts {\,\sc iii}}
\def\four{\ts {\,\sc iv}}
\def\five{\ts {\sc v}}
\def\six{\ts {\sc vi}}
\def\beq{\begin{equation}}
\def\eeq{\end{equation}}

\def\lesssim{\mathrel{\hbox{\rlap{\hbox{\lower4pt\hbox{$\sim$}}}\hbox{$<$}}}}
\def\gtrsim{\mathrel{\hbox{\rlap{\hbox{\lower4pt\hbox{$\sim$}}}\hbox{$>$}}}}

\def\one{{\,\sc i}}
\def\two{{\,\sc ii}}
\def\three{{\,\sc iii}}
\def\four{{\,\sc iv}}
\def\five{{\sc v}}
\def\six{{\sc vi}}

\def\v1d{{\sc v1d}}

\def\cmfgen{{\sc cmfgen}}
\def\heracles{{\sc heracles}}

\def\ergs{erg\,s$^{-1}$}

\newcommand{\iso}[2]{\ensuremath{^{#1}\rm{#2}}}

\def\pasp{PASP}

\def\apj{ApJ}
\def\apjs{ApJS}
\def\apjl{ApJL}
\def\aap{A\&A}

\def\mnras{MNRAS}
\def\nat{Nature}

\def\nifs{\iso{56}Ni}

\begin{document}

   \title{
   Simulations of light curves and spectra for superluminous Type Ic supernovae
   powered by magnetars
   }

   \titlerunning{
   Type Ic supernovae powered by magnetars
   }

\author{
   Luc Dessart
}

\institute{Unidad Mixta Internacional Franco-Chilena de Astronom\'ia (CNRS UMI 3386),
    Departamento de Astronom\'ia, Universidad de Chile,
    Camino El Observatorio 1515, Las Condes, Santiago, Chile.
  }

   \date{Received; accepted}

  \abstract{
Numerous superluminous supernovae (SLSNe) of Type Ic have been discovered and monitored in the last decade. The favored mechanism at their origin is a sustained power injection from a magnetar. This study presents non-local thermodynamic equilibrium (nLTE) time-dependent radiative transfer simulations of various single carbon-rich Wolf-Rayet star explosions influenced by magnetars of diverse properties and covering from a few days to one or two years after explosion. Nonthermal processes are treated; the magnetar-power deposition profile is prescribed;  dynamical effects are ignored. In this context, the main influence of the magnetar power is to boost the internal energy of the ejecta on week-long time scales, enhancing the ejecta temperature and ionization, shifting the spectral energy distribution to the near-UV (even for the adopted solar metallicity), creating blue optical colors. Varying the ejecta and magnetar properties introduces various stretches and shifts to the light curve (rise time, peak or nebular luminosity, light curve width). At maximum, all models show the presence of O\two\ and C\two\ lines in the optical, and more rarely O\three\ and C\three\ lines. Non-thermal effects are found to be negligible during the high-brightness phase. After maximum, higher energy explosions are hotter and more ionized, and produce spectra that are optically bluer. Clumping is a source of spectral diversity after maximum. Clumping is essential to trigger ejecta recombination and yield the presence of O\one, Ca\two, and Fe\two\ lines from a few weeks after maximum until nebular times. The UV and optical spectrum of Gaia16apd at maximum or the nebular spectrum of LSQ14an at +410\,d are compatible with some models that assume no clumping. However, most observed SLSNe Ic seem to require clumping from early post-maximum to nebular times (e.g., SN\,2007bi at +46 and +367\,d; Gaia16apd at +43\,d).
  }

\keywords{
  radiative transfer --
  supernovae: general --
  magnetars
}

   \maketitle
%

\section{Introduction}

Over the last decade, numerous superluminous Type Ic supernovae (SNe) have been discovered and monitored extensively (see e.g., \citealt{quimby_slsnic_11}, \citealt{inserra_slsn_13}, \citealt{nicholl_slsn_13}, \citealt{de_cia_slsn_ic_17}, \citealt{lunnan_slsn_ic_18}). These superluminous SNe (SLSNe) are generally found in low-mass low-metallicity host galaxies \citep{lunnan_slsn_14,leloudas_slsn_15,perley_slsn_ic_16,schulze_slsn_18}. They reach peak luminosities of up to a few 10$^{44}$\,\ergs\ and visual magnitudes of about $-21$\,mag \citep{quimby_slsnic_11,de_cia_slsn_ic_17,nicholl_mosfit_mag_17,lunnan_slsn_ic_18}. Their rise times to bolometric maximum ranges from 20 to 125\,d, while their post-maximum brightness decline rates cover from 0.01 to 0.06\,mag\,d$^{-1}$. At maximum, their optical spectra have a blue color. Apart from a few weak features associated with O\two\ \citep{pasto_10gx_10,quimby_slsnic_11}, these maximum-light optical spectra are nearly featureless. After maximum, the optical spectra show lines attributed to O\one, Ca\two, and Fe\two\ (e.g., \citealt{galyam_07bi_09}; \citealt{nicholl_slsn_13}), with similar properties to broad-line SNe Ic \citep{liu_slsnic_17}. At late times, their optical spectra exhibit forbidden lines of O\one\ and Ca\two, as well as Fe\two\ lines, and are similar to the contemporaneous spectra of GRB SNe such as 1998bw \citep{nicholl_15bn_16,jerkstrand_slsnic_17}. However, SLSNe Ic also show O\two\ and O\three\ lines, which are strong in LSQ14an \citep{inserra_slsn_ic_17} or PS1-14bj \citep{lunnan_slsnic_16}, and weak in SN\,2007bi or SN\,2015bn \citep{jerkstrand_slsnic_17}).  In rare instances, SLSNe Ic exhibit a strong and broad H$\alpha$ line at $100-200$\,d after maximum \citep{yan_slsnic_ha_15,yan_slsn_ic_17}.

\vspace{1cm}

\begin{figure}
   \includegraphics[width=\hsize]{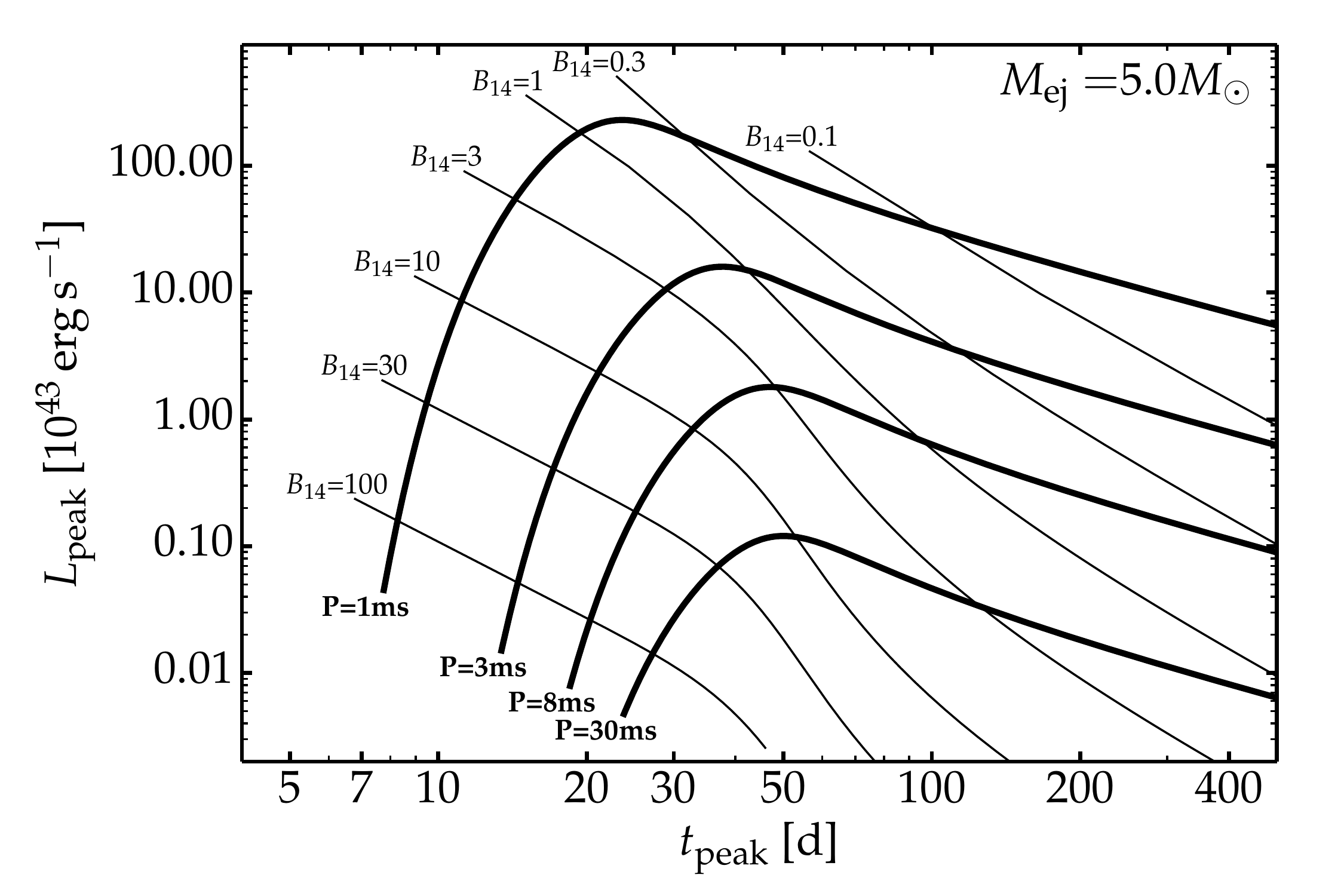}
\caption{Distribution of the rise time to maximum ($t_{\rm peak}$) and bolometric luminosity at maximum ($L_{\rm peak}$) for a range of magnetar field strengths and initial spin periods (the ejecta mass is 5\,\msun, the ejecta kinetic energy is 10$^{51}$\,erg, and the adopted opacity is 0.2\,cm$^2$\,g$^{-1}$). The figure is adapted from \citet[using their Equations 15 and 16]{KB10}.
\label{fig_kb10}
}
\end{figure}

The current consensus is that SLSNe Ic owe their exceptional luminosities, color evolution, and spectral properties to the power contribution from a compact remnant \citep{bodenheimer_ostriker_74,KB10,woosley_pm_10}. In standard core-collapse SNe, the explosion is thought to result from complex circumstances involving neutrino absorption and multi-dimensional fluid instabilities (for recent results from numerical simulations, see e.g., \citealt{lentz_ccsn_3d_15,mueller_ccsn_3d_17,glas_ccsn_3d_18,oconnor_couch_ccsn_3d_18,vartanyan_ccsn_3d_19}). The resulting SN luminosity then corresponds to the release of previously stored energy (left by shock passage) and the continuous release of energy by the decay of unstable isotopes, in particular \nifs. Although expected to be rare, the iron core of a massive star may be fast spinning at the time of collapse \citep{hirschi_rot_04,yoon_grb_05,WH06,georgy_snibc_09}, establishing suitable conditions for the enhancement of the magnetic field during the collapse and the early post-bounce phase. Depending on the initial spin period and the magnetic field amplification, the proto-neutron star may lead to the formation of a proto-magnetar with a range of magnetic field strengths. For the strongest fields, the magneto-rotational effects may lead to a highly energetic explosion, perhaps associated with the formation of a baryon-free relativistic jet and a $\gamma$-ray burst \citep{usov_pm_92,wheeler+00,akiyama_mri_03,thompson_pm_04}. For weaker fields, the magnetar spin-down timescale is longer, the dynamical impact of the magnetar is weaker, but its influence on the SN luminosity can then be large \citep{KB10,woosley_pm_10}, and completely swamp the contribution from \nifs. Figure~\ref{fig_kb10} illustrates the predicted outcome for SN peak luminosity and rise time to peak as a function of magnetar spin period and field strength, and for an ejecta of 5\,\msun\ and a kinetic energy of 10$^{51}$\,erg (and no \nifs). The maximum luminosity is reached for millisecond spin periods (i.e., large rotational energy budget) and moderate field strengths of 10$^{14}$\,G (corresponding to week-long spin-down timescales). In the context of the magnetar model, the current sample of SLSNe Ic yields inferred magnetar properties spanning the range $1.2-4$\,ms initial spin periods, 0.2 to $1.8 \times 10^{14}$\,G field strengths, while the associated ejecta  spans masses from 2.2 to 12.9\,\msun, and kinetic energies from 1.9 to $9.8 \times 10^{51}$\,erg \citep{nicholl_mosfit_mag_17}. Sophisticated numerical simulations of magnetar-powered SNe will probably not change these numbers much. Slower rotators and less magnetized magnetars can also appreciably alter SN radiation because the time-integrated luminosity of a SN is only 10$^{49}$\,erg, which is comparable to the rotational energy of a $\sim$\,40\,ms period magnetar. Magnetars may also influence the properties of some He-rich explosions (e.g., SN\,2005bf; \citealt{maeda_05bf_07}) as well as some H-rich explosions (e.g., iPTF14hls; \citealt{d18_iptf14hls}). Future observations will probably extend the distribution of events lying between superluminous SNe and standard-luminous SNe (there may be a continuum of values for initial magnetar spins and magnetic field).

One-dimensional radiation-hydrodynamics modeling has been done to characterize the influence of the magnetar on the SN radiation and the ejecta structure \citep{KB10, woosley_pm_10,bersten_15lh_16,moriya_pm_bh_16,tolstov_slsnic_17}. These simulations yield results in agreement with analytical predictions \citep{KB10}. When extended to 2-D, hydrodynamical simulations exhibit strong differences with key features obtained in 1-D. In 2-D, the magnetar power leads to strong turbulence in the inner ejecta, which alters the shock properties and prevents the formation in the inner ejecta of a narrow dense and fast moving shell. The inner ejecta thus contains material down to low velocities rather than a low-density high temperature cocoon bounded by a fast-moving dense shell. In 3-D, the effect should persist but will probably be quantitatively different.  The power released from the magnetar is likely aspherical and may not follow the smooth evolution of a dipole, which is generally assumed throughout the early magnetar life. Another concern is the efficiency with which the radiation from the magnetar is thermalized by the ejecta (see., e.g., \citealt{kasen_pm_16}). This power may eventually leak out from the ejecta, and indeed, leakage has been invoked to explain the late-time light curve (for the case of SN\,2015bn, see \citealt{nicholl_15bn_1000d_18}).

Spectral modeling at the time of maximum provided strong evidence for energy injection from a compact remnant compared to a scenario in which the ejecta is influenced by an exceptionally large mass of \nifs\ \citep{d12_magnetar}, later confirmed by similar modeling at the time of maximum \citep{howell_slsnic_13,mazzali_slsn_16} and at nebular times \citep{d13_pisn,jerkstrand_slsnic_17}.

This work is a follow-up to \citet{d12_magnetar}, but this time using a time-dependent rather than a steady-state approach; the magnetar-powered SN can thus be followed from about a day until one or two years after explosion in a continuous fashion. This evolution is followed with \cmfgen\ \citep{HD12}, using the approach of \citet{d18_iptf14hls} for the treatment of the magnetar power.  The next section presents the carbon-rich Wolf-Rayet progenitors and Type Ic SNe models used, as well as the grid of magnetar properties considered. Section~\ref{sect_r0e2} discusses the results for the magnetar powered SN model r0e2. This model is used as a reference for additional comparisons and investigations. The subsequent sections investigate how the SN properties vary when various ejecta or magnetar characteristics are modified. Section~\ref{sect_edep} considers the influence of the adopted deposition profile for the magnetar power. Section~\ref{sect_ekin} presents how a magnetar-powered ejecta is influenced by a change in ejecta kinetic energy. Sections~\ref{sect_epm} and \ref{sect_bpm} present the impact of varying the magnetar initial spin and field strength on the SN radiation and ejecta properties. Section~\ref{sect_nonth} discusses the impact of the SN radiation when non-thermal processes are ignored (here, all simulations treat non-thermal processes unless otherwise stated). Section~\ref{sect_cl} quantifies the influence of clumping on the SN radiation, both at photospheric and nebular epochs. Section~\ref{sect_obs} presents a comparison between these models and a few observed SLSNe Ic. Section~\ref{sect_conc} presents the conclusions.

\begin{table*}
\caption{
Summary of properties for the progenitor, the ejecta, the magnetar, as well as results at bolometric maximum obtained with the \cmfgen\ simulations. Numbers in parenthesis correspond to powers of ten. Each group of models (separated by a horizontal line) has a specific purpose. Models 5p11Bx2 and 5p11Bx2th address the influence of non-thermal processes. The reference magnetar model r0e2 is confronted to model r0e2n (no magnetar), to model r0e2s (narrow magnetar-power deposition profile), to models r0e1 and r0e4 (different explosion energies), to model r0e2ba (different magnetar field strength), and to models r0e2ea, r0e2eb, and r0e2ec (different magnetar spins). Also discussed is the influence of ejecta clumping for model r0e2 (Section~\ref{sect_cl}).
 \label{tab_sum}}
\begin{center}
\begin{tabular}{l@{\hspace{4mm}}c@{\hspace{4mm}}c@{\hspace{4mm}}
c@{\hspace{4mm}}c@{\hspace{4mm}}c@{\hspace{4mm}}c@{\hspace{4mm}}
c@{\hspace{4mm}}c@{\hspace{4mm}}c@{\hspace{4mm}}
c@{\hspace{4mm}}c@{\hspace{4mm}}
}
\hline
Model                & \nifs    & $M_{\rm ej}$ &   $E_{\rm kin}$    &   $E_{\rm pm}$      & $P_{\rm pm}$      & $B_{\rm pm}$  & $V_0$  & $dV$  & $t_{\rm pm}$
                          &    $t_{\rm peak}$ &  $L_{\rm peak}$   \\
                          & [\msun]            &     [\msun]    &  [erg] &  [erg]  &  [ms] &   [G]  &    [\kms] & [\kms] & [d]    &       [d]               &     [erg\,s$^{-1}$] \\
\hline
5p11Bx2th       &  0.19                &        3.63        &        2.49(51)      &    0.4(51) & 7.0 & 3.5(14) &   6000  & 3000  & 19.1 &    26.0      & 4.83(43)      \\
5p11Bx2   &  0.19                &        3.63        &        2.49(51)      &    0.4(51)  & 7.0 & 3.5(14) &   6000  & 3000  & 19.1 &    26.7      & 4.86(43)      \\
\hline
r0e2               &    0.13              &   9.69               &       4.12(51)     &   0.4(51)  & 7.0 & 3.5(14) &    5700  & 2850  & 19.1  &    31.7  &  3.75(43)      \\
r0e2n           &    0.13              &   9.69               &       4.12(51)     &   \dots  & \dots & \dots &    \dots & \dots & \dots  &    44.4  & 2.25(42)      \\
\hline
r0e2               &    0.13              &   9.69               &       4.12(51)     &   0.4(51)  & 7.0 & 3.5(14) &    5700  & 2850  & 19.1  &    31.7  &  3.75(43)      \\
r0e2s              &    0.13              &   9.69               &       4.12(51)     &   0.4(51)  & 7.0 & 3.5(14) &    1425  & 713  & 19.1  &  51.7   &  2.81(43)      \\
\hline
r0e1               &   0.09              &    9.52              &       1.14(51)      &   0.4(51)  & 7.0 & 3.5(14) &   2700  & 1350  & 19.1 &     43.0   & 2.53(43)      \\
r0e2               &    0.13              &   9.69               &       4.12(51)     &   0.4(51)  & 7.0 & 3.5(14) &    5700  & 2850  & 19.1  &    31.7  &  3.75(43)      \\
r0e4                &   0.17              &   9.86                & 1.23(52)         &    0.4(51)  & 7.0 & 3.5(14) &  10000  & 5000  &   19.1  &    25.8   &  4.86(43)      \\
\hline
r0e2               &    0.13              &   9.69               &       4.12(51)     &   0.4(51)  & 7.0 & 3.5(14) &    5700  & 2850  & 19.1  &    31.7  &  3.75(43)      \\
r0e2ba          &   0.13              &   9.69               &       4.12(51)     &   0.4(51) & 7.0 & 1.0(14) &    5700  & 2850  &  234.4 &    48.9  & 1.61(43)      \\
\hline
r0e2               &    0.13              &   9.69               &       4.12(51)     &   0.4(51)  & 7.0 & 3.5(14) &    5700  & 2850  & 19.1  &    31.7   &  3.75(43)      \\
r0e2ea         &    0.13              &   9.69               &       4.12(51)     &   0.8(51)  & 5.0 & 3.5(14) &    6800  & 3400  & 9.6    &    25.0     &   7.20(43)      \\
r0e2eb            &   0.13              &   9.69               &       4.12(51)     &   1.2(51)  & 4.1 & 3.5(14) &    8000  & 4000   & 6.4   &    21.6   &   1.03(44)     \\
r0e2ec            &   0.13              &   9.69               &       4.12(51)     &   5.0(51)  & 2.0 & 2.0(14) &    8000  & 4000   &   4.7 &    24.6    &   3.21(44)   \\
\hline
\end{tabular}
\end{center}
\end{table*}


\section{Numerical setup}
\label{sect_setup}

\subsection{Progenitor and explosion models}

	The simulations presented here are based on Type Ic SN ejecta that have been used in previous studies on SNe Ibc and GRB SNe \citep{D15_SNIbc_I,D16_SNIbc_II,dessart_98bw_17}. Two progenitor models and a variety of ejecta properties were used. One progenitor is model 5p11 from \citet[model 31 in Table~1 of that paper]{yoon_ibc_10} and the other progenitor is model r0 \citep{dessart_98bw_17}. These two models correspond to solar metallicity carbon-rich Wolf-Rayet stars that die with a final mass of 5.11\,\msun\ (rather than 4.95\,\msun\ quoted in \citet{yoon_ibc_10}; see \citealt{D15_SNIbc_I} for discussion) and 11.4\,\msun. The influence of metallicity, and the use of a low metallicity progenitor more suitable for SLSNe Ic will be explored later (this is not essential given the large metal content of such Wolf-Rayet stars). These were exploded with \v1d\ \citep{livne_93,dlw10a,dlw10b} by means of a piston to produce ejecta with different kinetic energies and \nifs\ masses.

	Model 5p11Bx2 (see \citealt{D16_SNIbc_II} for details, including the mixing procedure, which is applied to all species) corresponds to an ejecta with a mass of 3.63\,\msun, a \nifs\ mass of 0.19\,\msun, and a kinetic energy of $2.49 \times 10^{51}$\,erg. The ejecta contains 0.34\,\msun\ of helium, 0.9\,\msun\ of carbon,  1.38\,\msun\ of oxygen, 0.43\,\msun\ of neon, and 0.14\,\msun\ of magnesium.

\begin{figure}[t!]
   \includegraphics[width=\hsize]{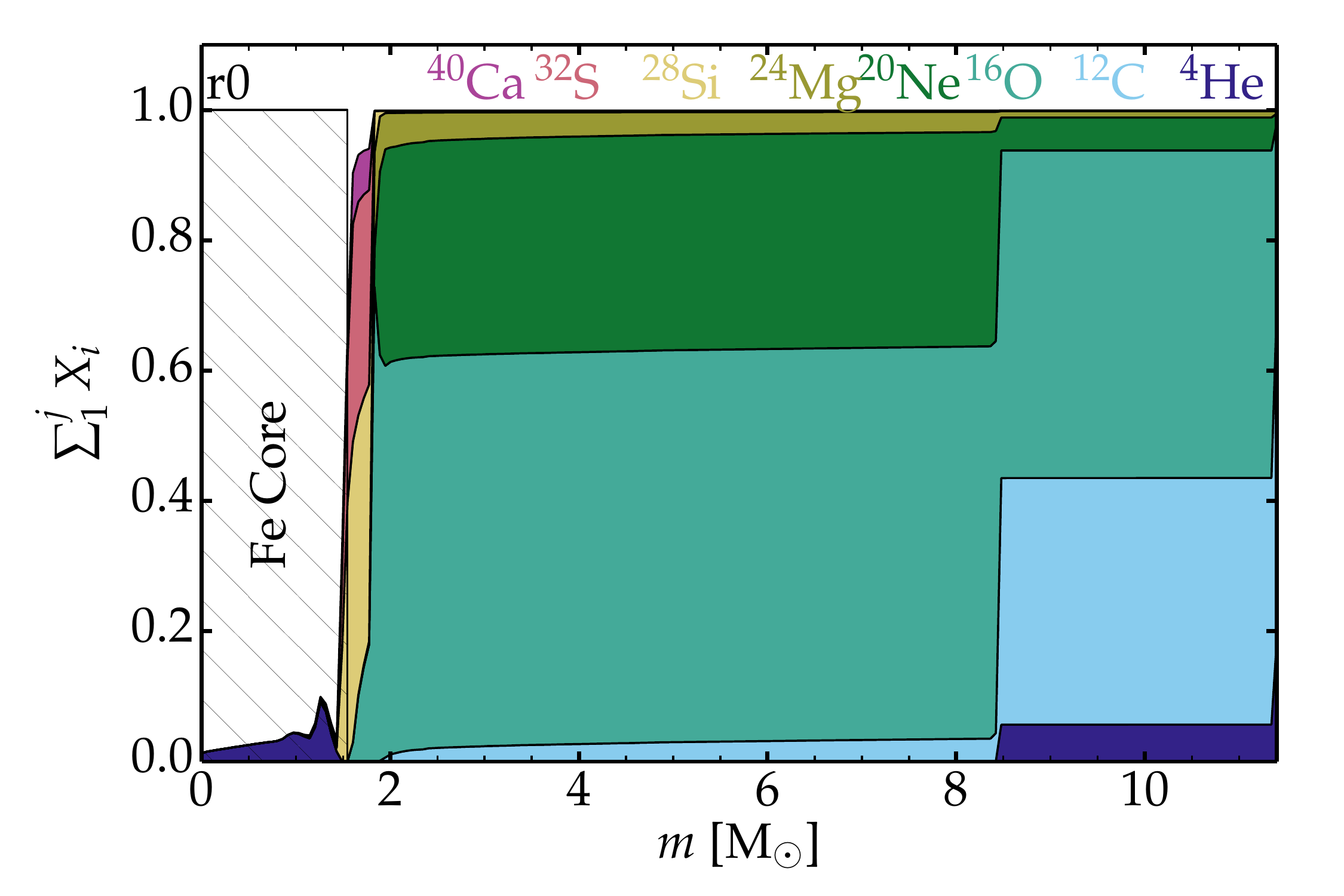}
   \includegraphics[width=\hsize]{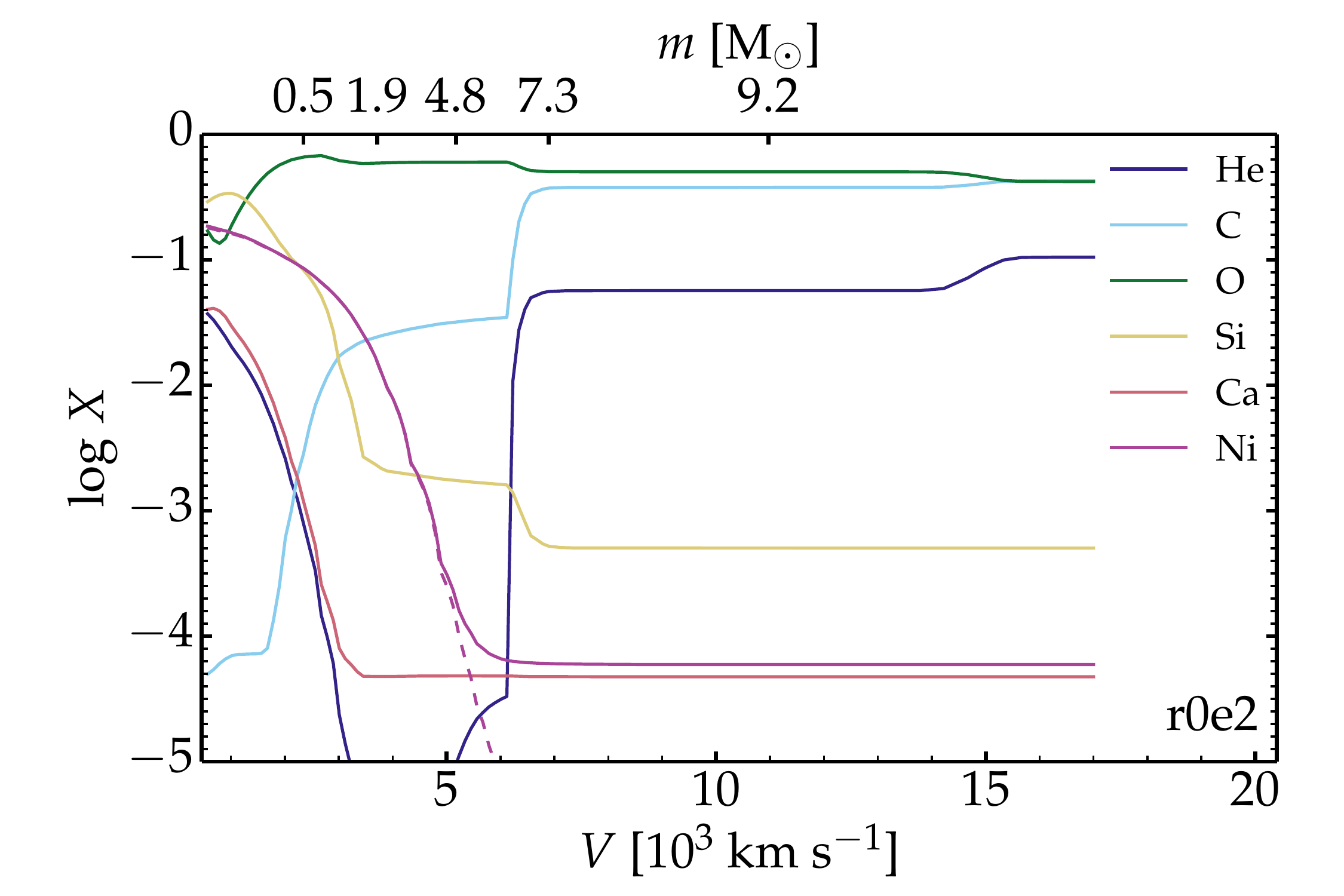}
   \includegraphics[width=\hsize]{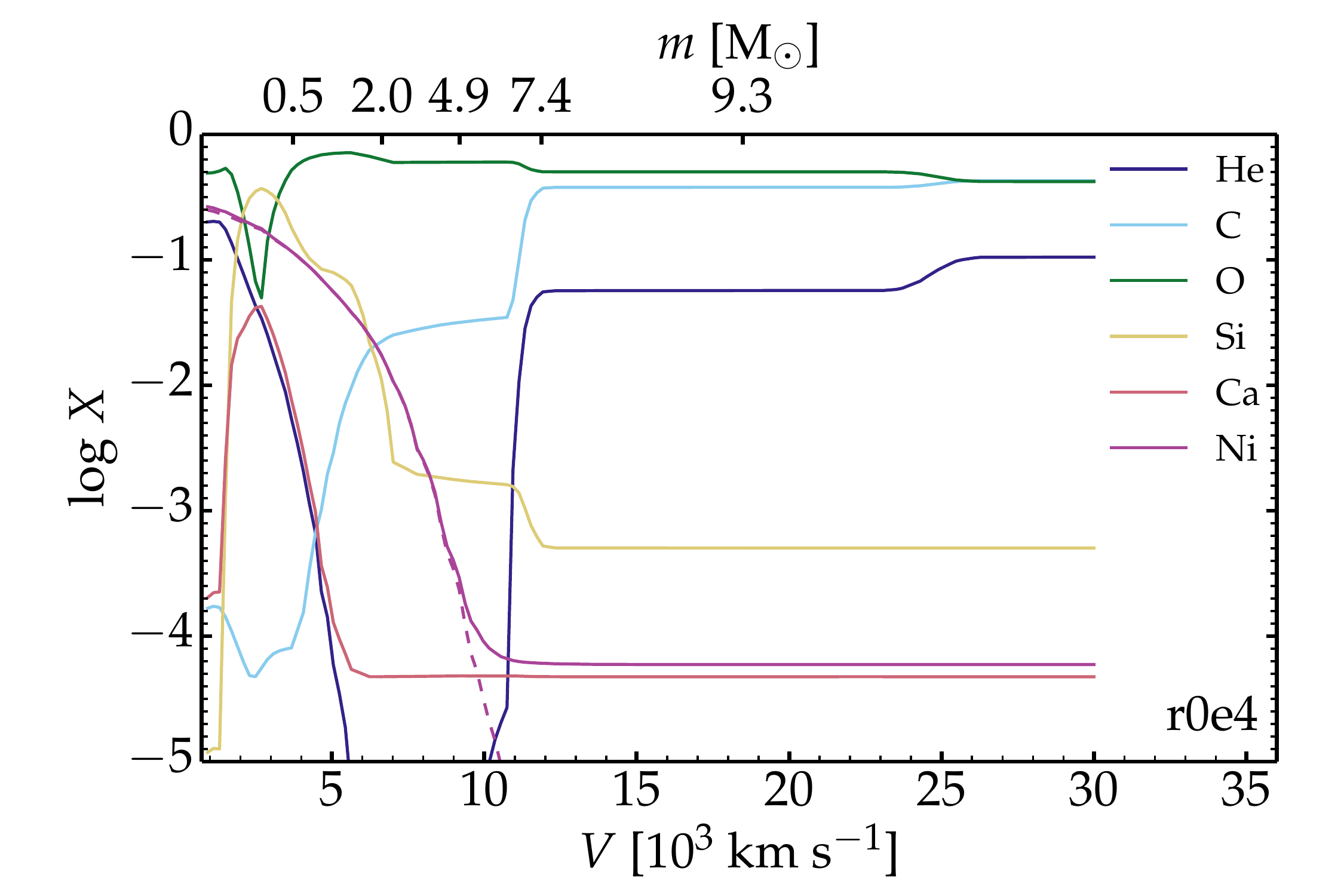}
\caption{Top: Envelope composition for model r0 at the onset of core collapse as a function of Lagrangian mass $m$. The cumulative sum of mass fractions is shown versus depth, starting from light elements. The iron core is the hatched area and its outer edge at 1.55\,\msun\ corresponds closely to the adopted mass cut at 1.57\,\msun\ for the piston trigger. Middle: Ejecta composition for model r0e2 versus velocity and Lagrangian mass (the innermost ejecta shell is at  $m=0$ and the dashed line corresponds to \nifs).  Bottom: Same as middle but now for model r0e4. The profiles are similar apart from a shift in velocity and a slightly different composition resulting from the explosive nucleosynthesis in the innermost layers (a greater mass of $\alpha$-rich freeze-out products, including helium, is present in higher-energy explosions).
\label{fig_init_comp}
}
\end{figure}

	Models r0e1/r0e2/r0e4 derive from progenitor model r0 and correspond to ejecta with a total mass $M_{\rm ej}$ of about 9.52/9.69/9.86\,\msun, a \nifs\ mass of 0.09/0.13/0.17\,\msun, and a kinetic energy $E_{\rm kin}$ of $1.14/4.12/12.3 \times 10^{51}$\,erg. Ejecta model r0 and its variants have a composition dominated by oxygen (a cumulative amount of about 5.4\,\msun), neon (about 1.9\,\msun), carbon (about 1.3\,\msun), and magnesium (about 0.35\,\msun). The helium content rises from 0.18 (r0e1) to 0.22\,\msun\ (r0e4) because of the extra helium produced for higher energies in the inner envelope during explosive nucleosynthesis -- most of the helium is however present in the outermost layers (see Fig.~\ref{fig_init_comp} for an illustration of the composition in the progenitor model r0 and of the explosion models r0e2 and r0e4; the left part of Table~\ref{tab_sum} summarizes the model properties). In these ejecta models based on the progenitor model r0, the mixing was applied only to \nifs\ -- the composition profile for other species was left as in the progenitor model (see Section~2 of \citealt{dessart_98bw_17} for discussion).

\subsection{Magnetar power in \cmfgen\ simulations}
\label{sect_pm_setup}

   At 1\,d after explosion,  models 5p11Bx2, r0e1, r0e2, and r0e4 are remapped into \cmfgen. These simulations use the same model atoms as in \citet{D15_SNIbc_I,D16_SNIbc_II,dessart_98bw_17}. The grid typically consists of 100 points to cover evenly in optical-depth scale from 500 to $30000-50000$\,\kms. The simulations treat radioactive decay from the \nifs\ chain. The energy deposition from radioactive decay is computed using a gray-absorption (opacity $\kappa_\gamma=0.06 Y_{\rm e}$\,cm$^2$\,g$^{-1}$, where $Y_{\rm e}$ is the local electron fraction) solution to the radiative transfer of $\gamma$-rays \citep{swartz_gray_95}. Positrons are absorbed locally.  For the magnetar power, the approach is similar to that presented in \citet{d18_iptf14hls} for the study of the Type II SN iPTF14hls.

	Starting at day one (the magnetar power released prior to day one is ignored), a magnetar power  $\dot{e}_{\rm pm}$ is introduced within the ejecta in homologous expansion, and is given by  the expression
 \begin{equation}
\dot{e}_{\rm pm} = (E_{\rm pm} / t_{\rm pm}) \,  / \left( 1 + t / t_{\rm pm} \right)^2    \, ,
        t_{\rm pm} = \frac{6 I_{\rm pm} c^3}{B_{\rm pm}^2  R_{\rm pm}^6  \omega_{\rm pm}^2} \,\, ,
\end{equation}
where $E_{\rm pm}$, $B_{\rm pm}$, $R_{\rm pm}$, $I_{\rm pm}$ and $\omega_{\rm pm}$ are the initial rotational energy, magnetic field, radius, moment of inertia, and angular velocity of the magnetar; $c$ is the speed of light. In all cases, $I_{\rm pm}=10^{45}$\,g\,cm$^2$ and $R_{\rm pm}=10^6$\,cm (see \citealt{KB10} for details).

To determine how the magnetar power is deposited within the ejecta is difficult because the process is dynamical and multi-dimensional. In contrast, \cmfgen\ does not treat dynamics (the ejecta structure is frozen) and is 1-D.  In 1-D, a hydrodynamical simulation of a magnetar powered SN naturally leads to the formation of a dense shell if the energy is deposited in the innermost layers of the SN ejecta \citep{KB10,woosley_pm_10,bersten_15lh_16,dessart_audit_18,orellana_pm_18}. In \cmfgen, adopting this deposition procedure would be  unpractical since a huge temperature spike would form in the \cmfgen\ simulation, producing large gradients that would be hard to resolve and would likely impede the convergence. It would also be physically inconsistent since the swept-up mass in the dense shell is accelerated as a result of magnetar power, while \cmfgen\ cannot treat this dynamical effect (neither the snow-plow effect nor the acceleration).

   In 2-D, hydrodynamical simulations indicate that this swept-up dense shell does not form \citep{chen_pm_2d_16,suzuki_pm_2d_17}. Hence, the dense shell is an artifact of the 1-D treatment. Without an imposed spherical symmetry, the magnetar power triggers convection in the inner ejecta. This turbulent medium also causes advection of the magnetar energy faster than achieved by diffusion. An angle-dependent and radial-dependent (i.e., over a broad range of radii rather than in the innermost shells) energy deposition might also prevent the formation of a shock. The impact of the magnetar energy injection on the density structure is much weaker in 2-D (and probably also in 3-D) than in 1-D.

    In the 1-D radiation hydrodynamical simulations with \heracles, \citet{dessart_audit_18} used this result to adopt a broad profile for the deposition of magnetar energy. A consequence is that then, even in 1-D, no dense shell forms. Furthermore, the broad deposition causes an earlier brightening of the SN luminosity, capturing in a simplistic way the energy advection seen in 2-D hydrodynamical simulations. This approach is therefore qualitatively sound  but quantitatively uncertain. For example, in the process of advection of magnetar energy deposited at depth, the heated material would cool by stretching out in radius. By adopting a broad deposition profile, the cooling from expansion is not properly accounted for since the energy is deposited directly at large velocities.

\begin{figure}[t!]
 \includegraphics[width=\hsize]{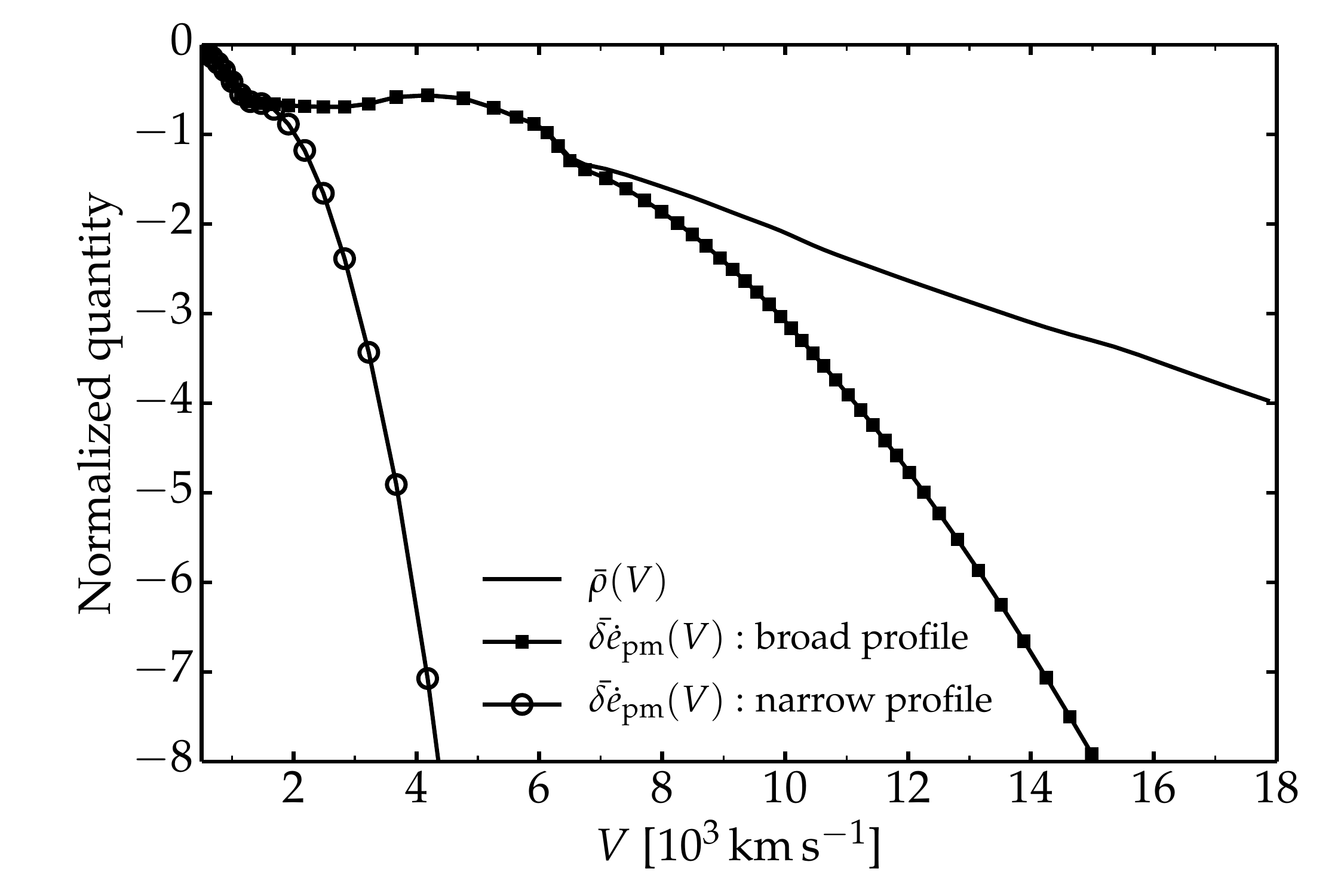}
  \caption{Illustration of the normalized ejecta density ${\bar \rho}(V)$ for model r0e2, together with the normalized magnetar energy deposition ${\bar{\delta \dot{e}}}_{\rm pm}(V)$ for a broad ($V_0 = 5700$\,\kms\ and $dV=2850$\,\kms) and a narrow ($V_0 = 1425$\,\kms\ and $dV=713$\,\kms) distribution. In general, the models adopt a broad deposition profile, but one model (r0e2s) was done with a narrower energy deposition profile to check the impact on the simulation (see Section~\ref{sect_edep} for discussion).
\label{fig_edep_profile}
  }
\end{figure}

    The present simulations with \cmfgen\ use  a relatively extended energy deposition as parametrized in \citet{dessart_audit_18} and \citet{d18_iptf14hls}. Numerically, the deposition profile follows the density $\rho$ for $V <V_0$, and $\rho \exp \left(- [(V-V_0)/dV]^2 \right)$ for $V >V_0$. This profile is frozen in time. Figure.~\ref{fig_edep_profile} shows two deposition profiles, one broad and one narrow. A normalization is applied so that the volume integral of this deposition profile yields the instantaneous magnetar power at that time.

   The impact of the choice of $V_0$ and $dV$ is explored in one model (r0e2s; Section~\ref{sect_edep} and Table~\ref{tab_sum}). In other simulations, $V_0$ and $dV$ scale with the magnetar energy or power since this seems plausible from the point of view of hydrodynamics. With this choice, the energy deposition profile influences the model luminosity mostly before maximum \citep{dessart_audit_18}. After maximum, the conditions are close to nebular so that it is the total magnetar power that primarily matters rather than its precise distribution within the ejecta.

   While the $\gamma$-rays from radioactive decay can escape at late times, the entire magnetar power is designed to remain trapped at all times and is deposited according to the prescription given above. This approach is not optimal but a proper handing of the process would require transport modeling for the deposition of the high-energy radiation and particles released by the magnetar together with the multi-dimensional hydrodynamical modeling of its impact on the 3D structure and dynamics of the ejecta. At late times, one may adjust the trapping efficiency of the magnetar power so that the power absorbed equals the observed luminosity. This could be considered when comparing models to specific observations.

    The approach in \cmfgen\ currently requires that the composition is homogeneous at a given ejecta depth (or velocity). There is no chemical segregation. In the context of core-collapse SNe exploded through the neutrino-driven mechanism, this choice is not optimal (see e.g., \citealt{wongwathanarat_15_3d}). However, in the context of magnetar-powered SNe, the inner ejecta becomes very turbulent \citep{chen_pm_2d_16,suzuki_pm_2d_17} and chemical mixing may take place down to small scales. So, the assumption of complete microscopic mixing in \cmfgen\ may in fact be more suitable than assuming no microscopic mixing at all. The issue is most relevant in Type II SNe because of the strong composition stratification of the progenitor massive stars. In Type Ic SN progenitor like model r0, the progenitor envelope is essentially a large ONeMg rich envelope on top of an iron core.

\begin{figure}[t!]
   \includegraphics[width=\hsize]{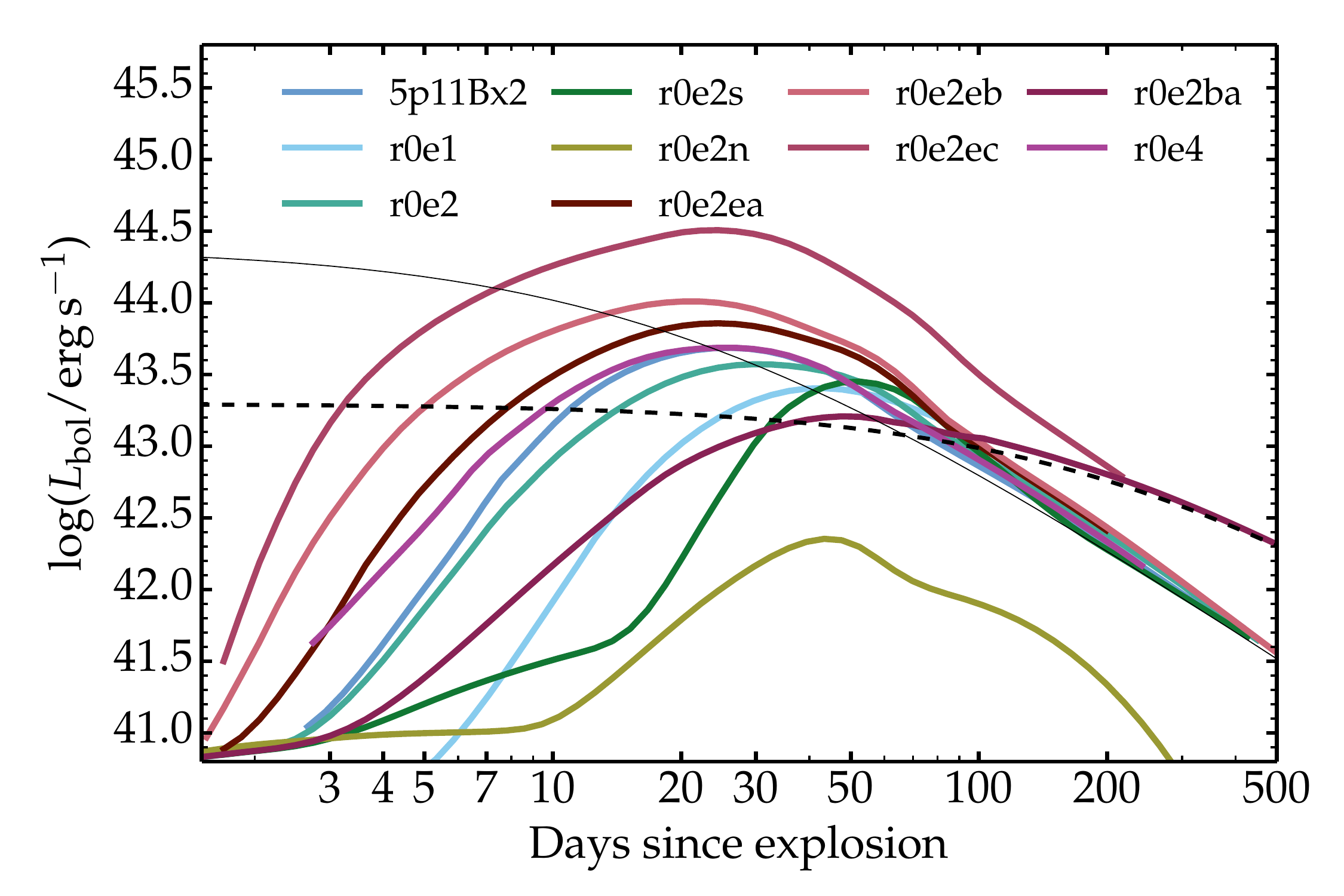}
\caption{Bolometric light curves for the full grid of models computed in this study (with the exception of models 5p11Bx2th and r0e2cl, which have the same bolometric light curve as models 5p11Bx2 and r0e2, respectively). Magnetar-power is applied in all models apart from model r0e2n in which the only power source after explosion is radioactive decay. The thin solid (thick dashed) black line corresponds to the magnetar power for $B_{\rm pm}=3.5\times 10^{14}$\,G ($B_{\rm pm}=10^{14}$\,G), and $E_{\rm pm} = 4 \times 10^{50}$\,erg. A logarithmic scale is used for the time axis.
\label{fig_lbol_all}
}
\end{figure}

\subsection{Grid of models}

In this work, magnetars with a range of initial spin and magnetar field strengths are considered. The initial magnetar energy $E_{\rm pm}$ covers from 0.4 to $5.0 \times 10^{51}$\,erg. The magnetar field strength $B_{\rm pm}$ covers a small range from 1 to $3.5 \times 10^{14}$\,G. Larger values are a problem because the main effect from magnetar power would be dynamical, which is not treated in \cmfgen. Lower values would produce brighter light curves at nebular times, but with fainter maxima so the SN would no longer be superluminous (this is still of interest but out of the scope of the present study).  These choices of $E_{\rm pm}$ and $B_{\rm pm}$ correspond to magnetar spin-down timescales in the range $4.7-234.4$\,d. The energy released during the first day, which is neglected, is  $0.4-17.6$\,\% of the total magnetar energy, which is quite substantial for the highly magnetized fast-spinning models. Because the dynamical influence of magnetar power is ignored, all the energy liberated by the magnetar goes into ejecta internal energy (and eventually radiation). Throughout the time sequence, our magnetar-powered model will retain the same ejecta structure (mass, density, velocity, and radius) and thus the same kinetic energy.

In practice, the code treats the magnetar power the same way as radioactive decay. Energy is injected as 1\,keV electrons for which the degradation spectrum is computed. The contribution to the heating of thermal electrons, and to non-thermal excitation and ionization is computed explicitly in all but one model -- one simulation is run with non-thermal processes turned off (i.e. all the energy is in this case entirely channeled into heat).

The model nomenclature is the following. The same magnetar properties ($B_{\rm pm}=3.5\times 10^{14}$\,G and $E_{\rm pm} = 4 \times 10^{50}$\,erg) are used in the ``reference'' models 5p11Bx2, r0e1, r0e2, and r0e4. There is no specific suffix to characterize the magnetar in these 4 models. The model r0e2 is extensively discussed and used for various comparisons. The properties of its magnetar are the same as those used in the study of \citet{dessart_audit_18} --- this particular choice was in part arbitrary but also driven by the need to use a magnetar with a modest dynamical influence (i.e., with $E_{\rm pm}$ significantly smaller than $E_{\rm kin}$, which is $\sim$\,10\% in the case of model r0e2). Variants of these models are then explored. Model 5p11Bx2th is the same model as 5p11Bx2 except that non-thermal processes are switched off (all magnetar and decay power goes into heat). Model r0e2n is equivalent to model r0e2 except that the magnetar power is turned off.  Model r0e2s is equivalent to model r0e2 except that the magnetar energy deposition profile is narrower. Model r0e2ba is equivalent to model r0e2 except that the magnetar field strength is lower. Models r0e2ea, r0e2eb, and r0e2ec are equivalent to model r0e2 except that the initial magnetar rotational energy is enhanced (model r0e2ec also uses a weaker field). Model r0e2cl is  equivalent to model r0e2 except that the material is clumped. This simulation is started around the time of maximum. In addition, in models r0e1, r0e2, and r0e4, various levels of clumping are enforced to explore the effects at one nebular epoch (details and results are given in Section~\ref{sect_cl}).

	This set of models includes only one magnetar with a large rotational energy because in that regime, the neglect of dynamical effects in \cmfgen\  becomes a severe limitation. Consequently, most of the present simulations are limited in maximum luminosity and optical brightness. And the quantitative results are less robust than the trends.

   Figure~\ref{fig_lbol_all} shows the bolometric light curve for the full set of time-dependent
 \cmfgen\  simulations. The peak luminosity covers from about $10^{43}$ up to few times $10^{44}$\,\ergs, with rise times from about 25 up to about 50\,d after explosion. The slight offset at $100-200$\,d between the model curves and the magnetar power is related to the small contribution from radioactive decay and residual optical-depth effects (i.e., there is still stored energy released in addition to the instantaneous power absorbed by the ejecta). Eventually, the bolometric luminosity equals the instantaneous magnetar power. The contribution from decay power depends on the \nifs\ mass of each model (see Table~\ref{tab_sum}), as well as the amount of $\gamma$-ray escape (function of $E_{\rm kin}/M_{\rm ej}$).

  \begin{figure}[t!]
   \includegraphics[width=\hsize]{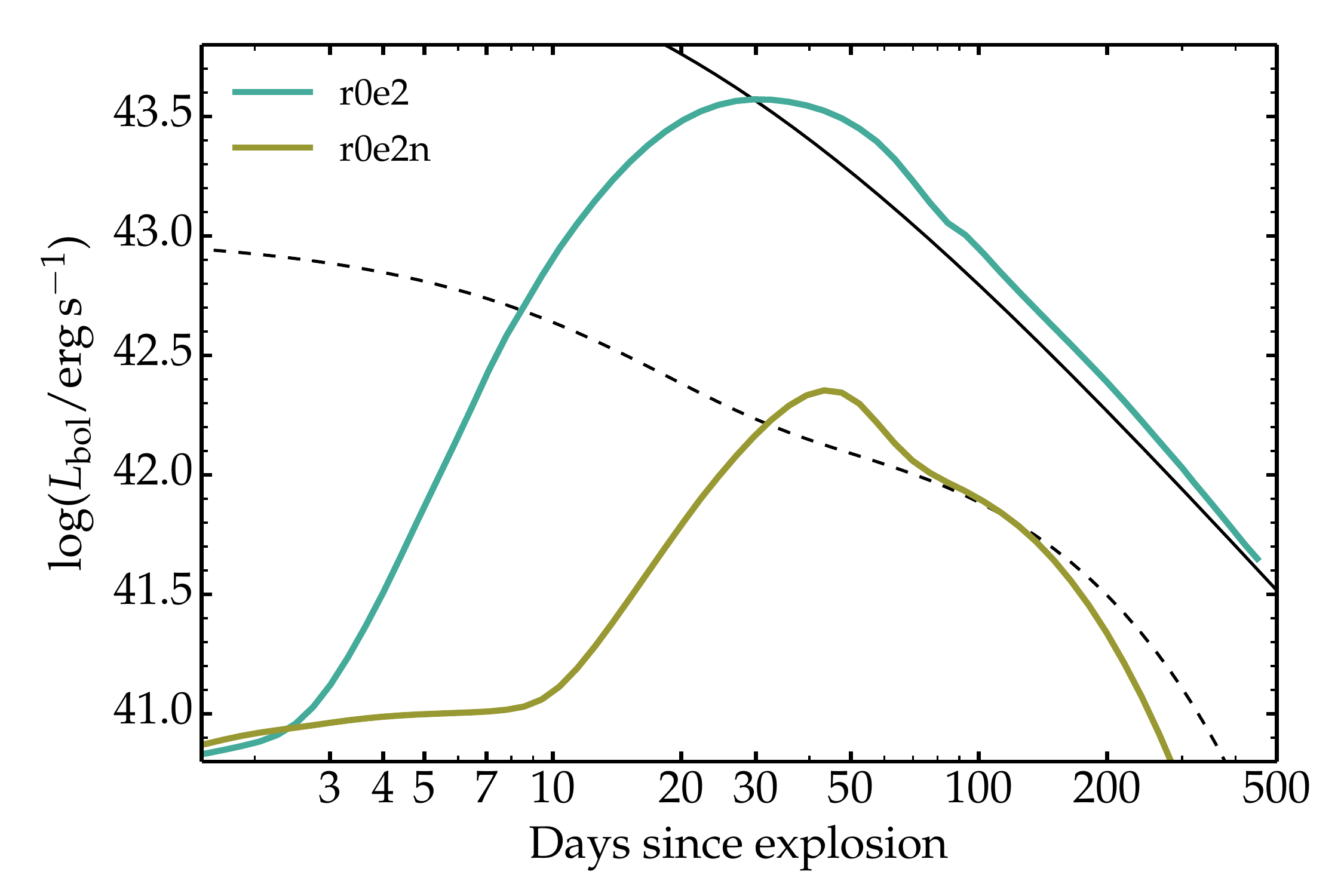}
   \includegraphics[width=\hsize]{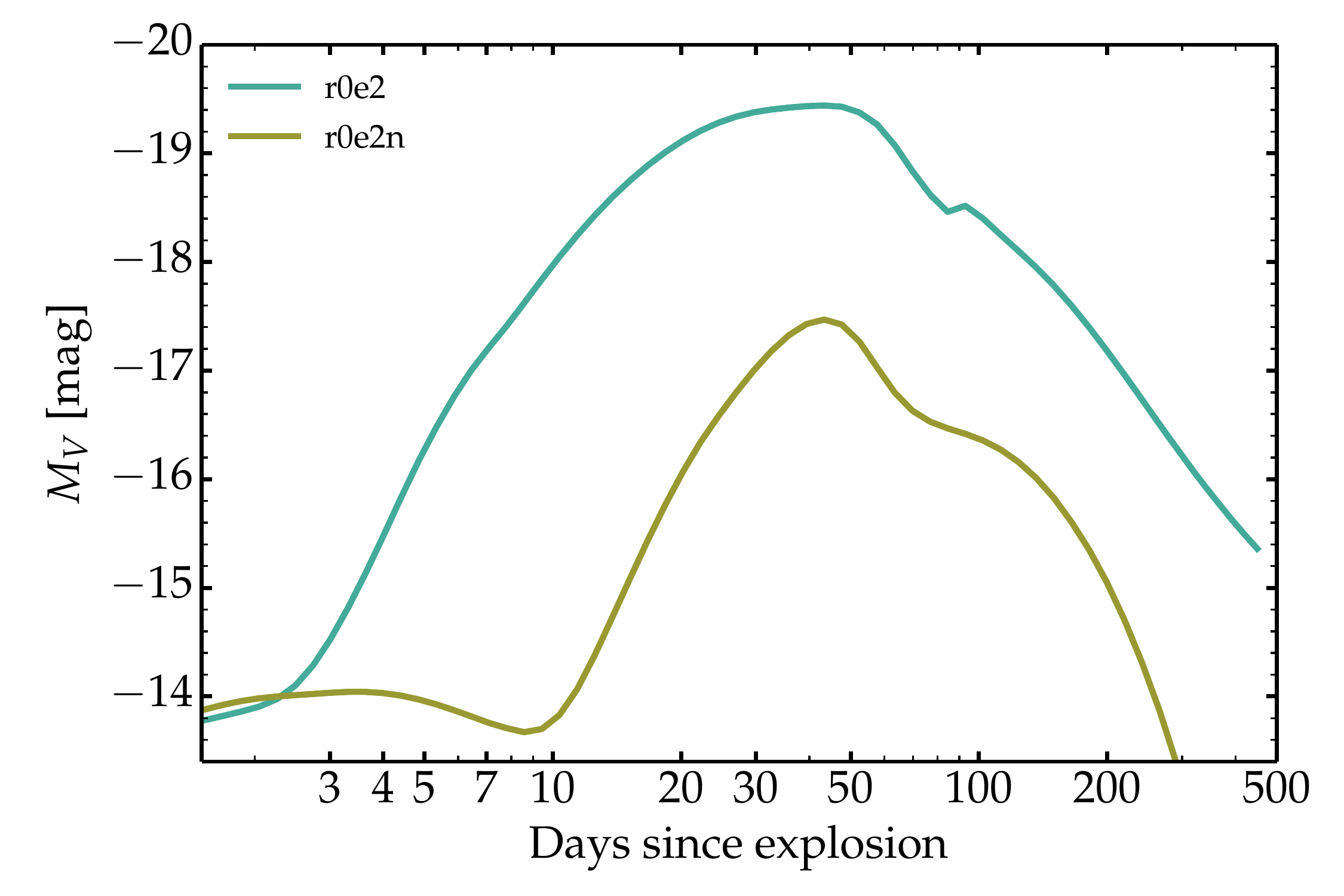}
   \includegraphics[width=\hsize]{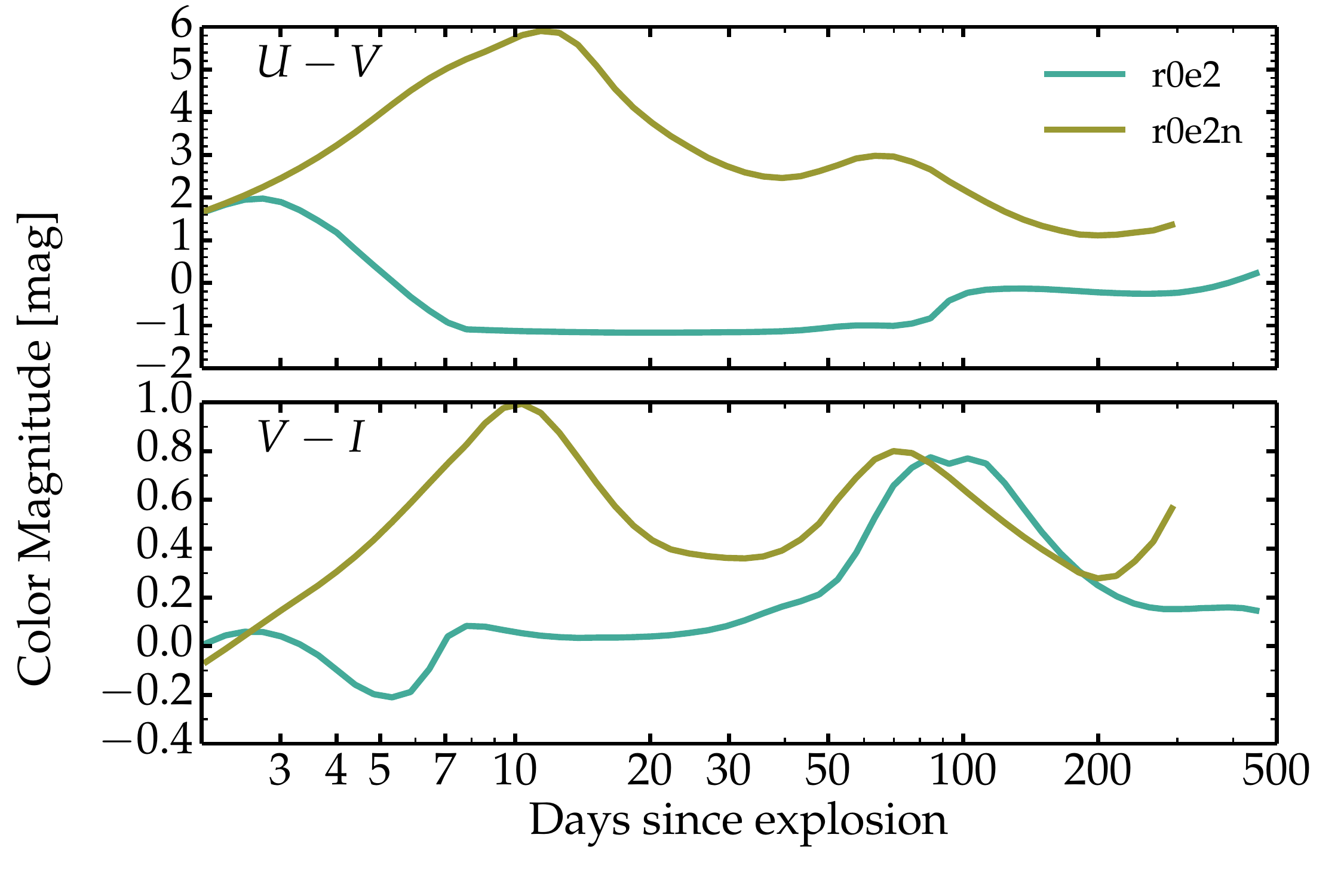}
\caption{Bolometric light curve (top; overplotted are the absorbed power from the magnetar [solid black line] and from \nifs\ decay [dashed black line]), $V$-band light curve (middle), and  $U-V$ and $V-I$ color evolution (bottom) for models r0e2 and its unmagnetized counterpart r0e2n. The glitch at 90\,d in the $V$-band light curve of model r0e2 corresponds to a rapid change in photospheric conditions (temperature and ionization), at a time when the ejecta optical depth is $5-10$. The offset between $L_{\rm bol}$ and $\dot{e}_{\rm pm}$ decreases with time in model r0e2 as the $\gamma$-rays from radioactive decay increasingly escape (the power from the magnetar is, by design, fully absorbed by the ejecta so the model eventually coasts to that value).
\label{fig_YN_mag}
}
\end{figure}

\begin{figure}
	\vspace{-0.32cm}
   \includegraphics[width=\hsize]{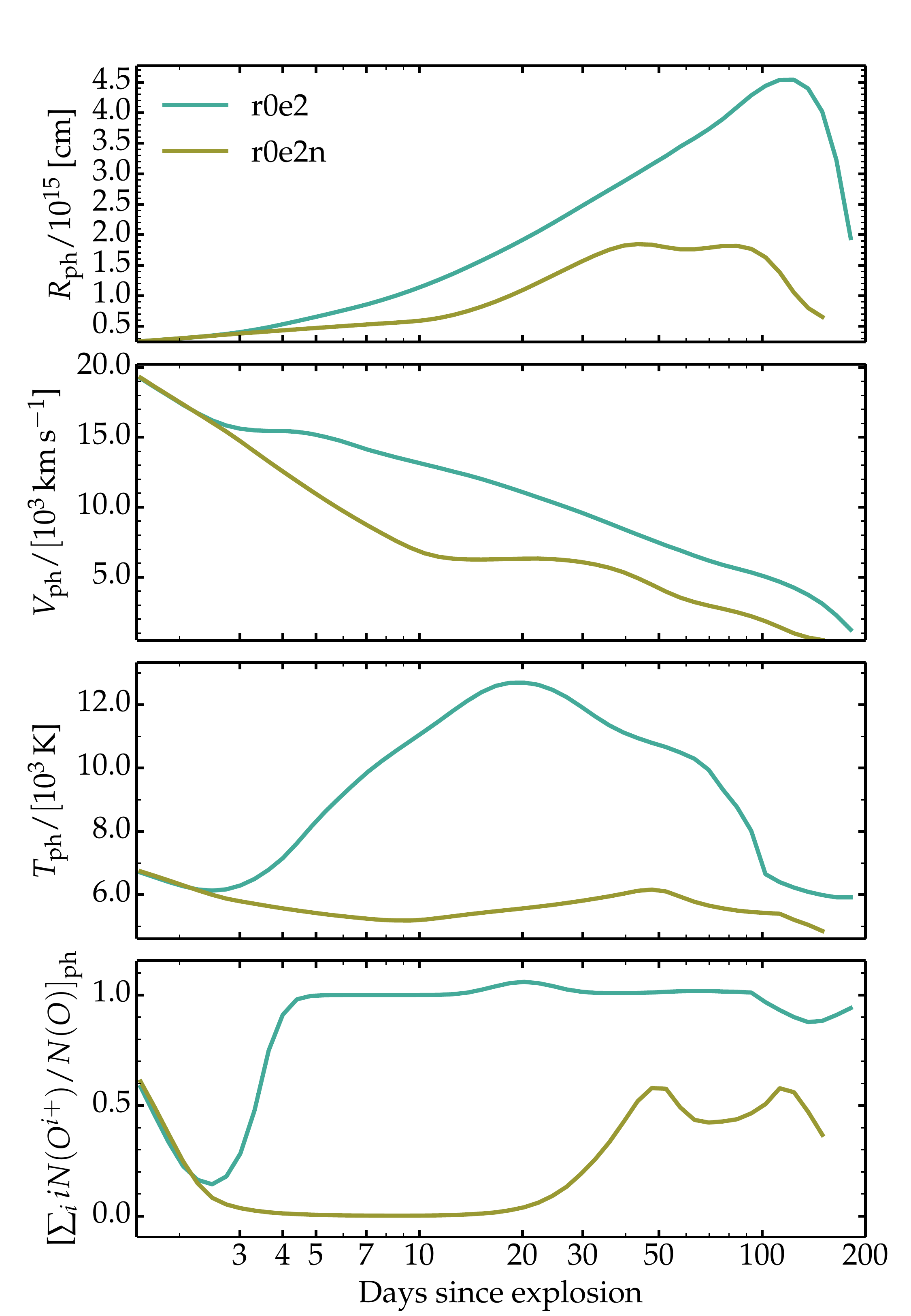}
   \caption{Evolution of photospheric properties for model r0e2 powered by a magnetar and model r0e2n without a magnetar (the sustained source of power comes from radioactive decay alone in model r0e2n). The bottom panel shows the ionization of oxygen at the photosphere (i.e., [$\sum_i i N(O^{i+})/N(O)]_{\rm ph}$) where $N(O^{i+})$ is the oxygen number density when $i$-times ionized and $N(O)$ is the  total oxygen number density. Oxygen is chosen since it is the main constituent of the ejecta and because it is associated with the emblematic spectroscopic signatures of SLSNe Ic. Only the electron-scattering opacity is used to locate the photosphere. Compared to its unmagnetized counterpart, model r0e2 shows photospheric properties that are hotter and more ionized over the period $3-200$\,d.
\label{fig_YN_phot}
}
\end{figure}

\begin{figure*}
\begin{center}
   \includegraphics[width=0.45\hsize]{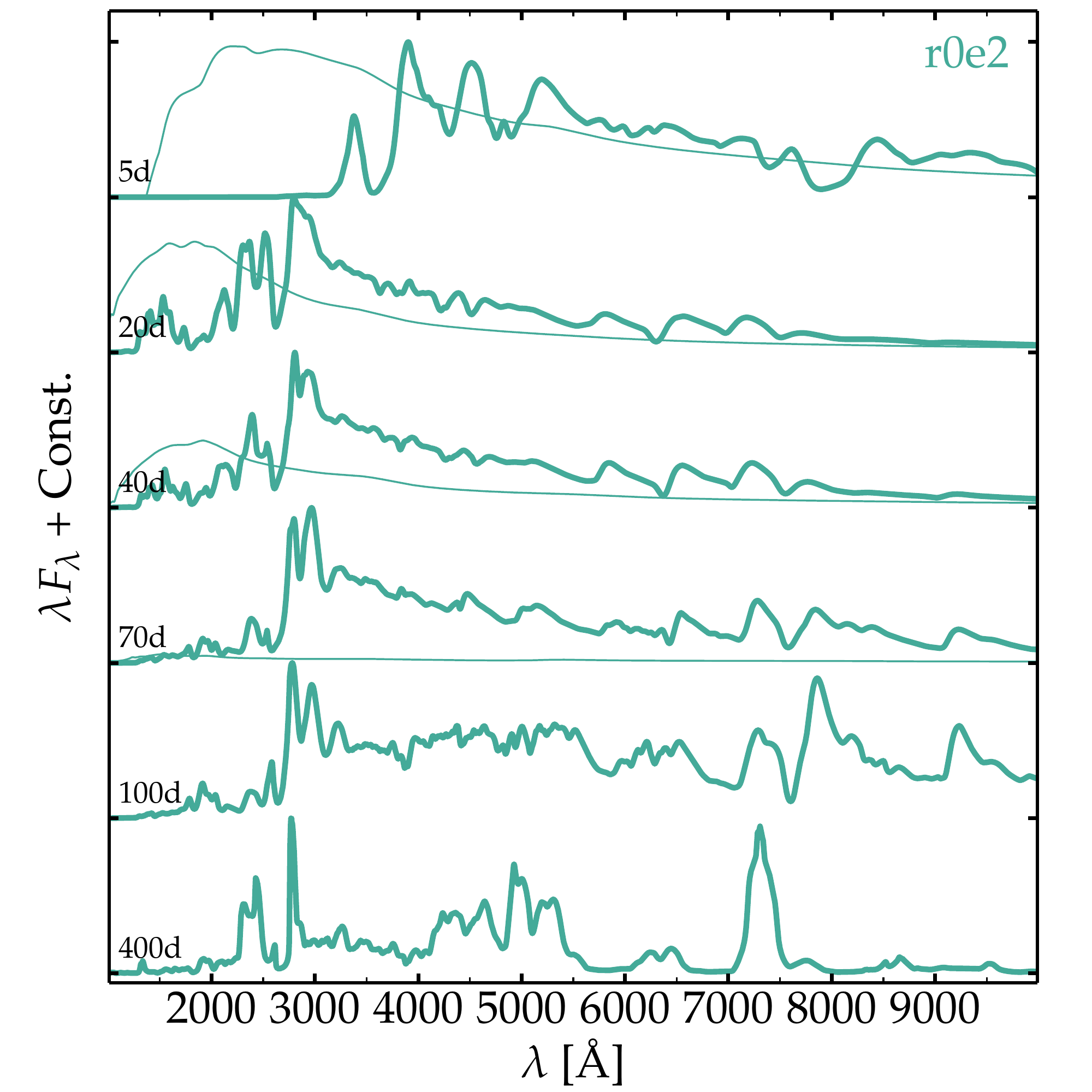}
   \includegraphics[width=0.45\hsize]{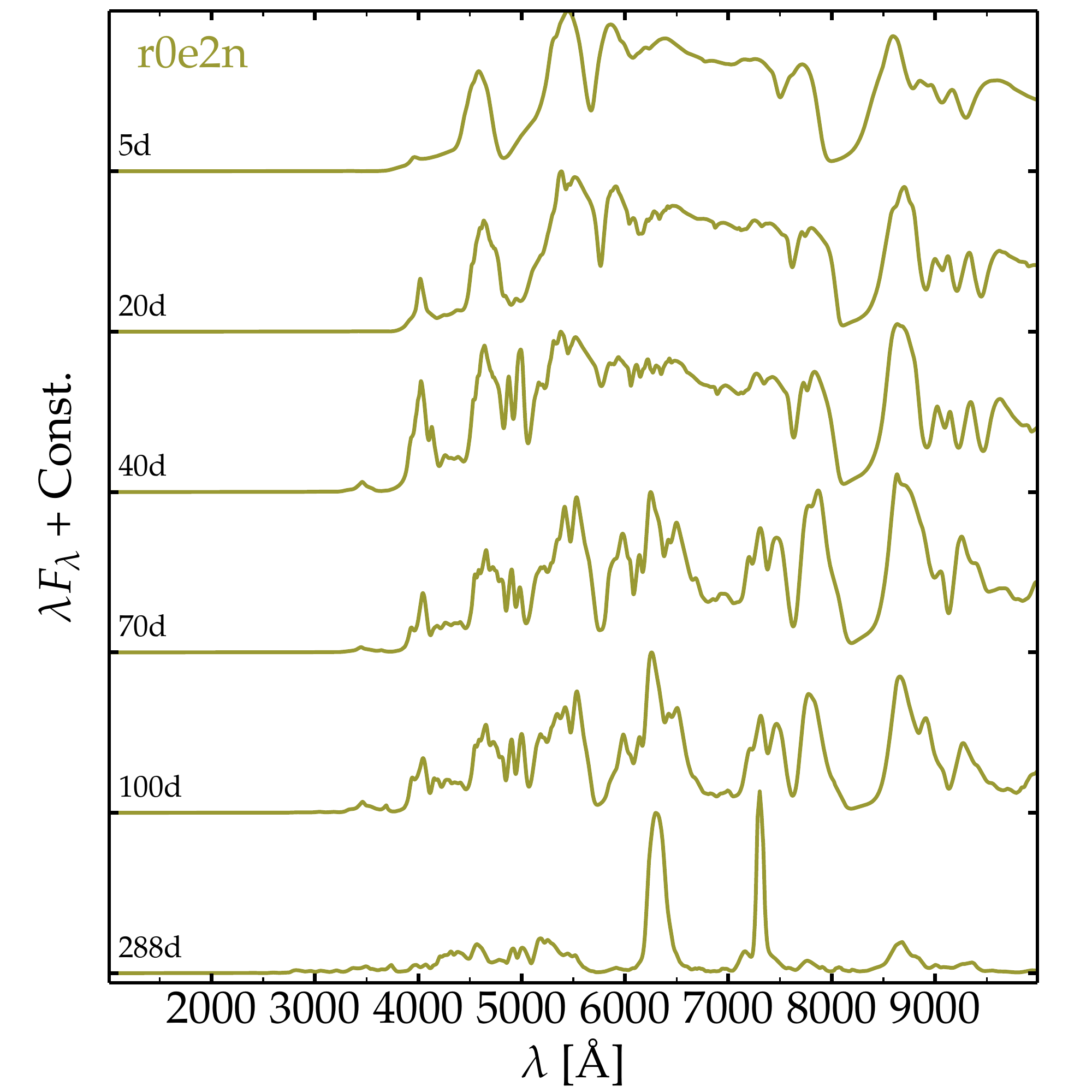}
\end{center}
\vspace{-0.3cm}
\caption{Multi-epoch spectra for model r0e2 (left) and model r0e2n (right). The quantity $\lambda F_\lambda$ is shown to better reveal the fainter flux at longer wavelength. In the left panel, the thin line corresponds to the continuum flux (the offset with the full spectrum arises from the contribution from lines). What drives the difference in spectral and color properties between the two models is magnetar heating since the same ejecta (composition, mass, kinetic energy) is used for both. [See discussion in Section~\ref{sect_r0e2}.]
\label{fig_YN_montage}
}
\end{figure*}

\begin{figure}
\begin{center}
   \includegraphics[width=0.9\hsize]{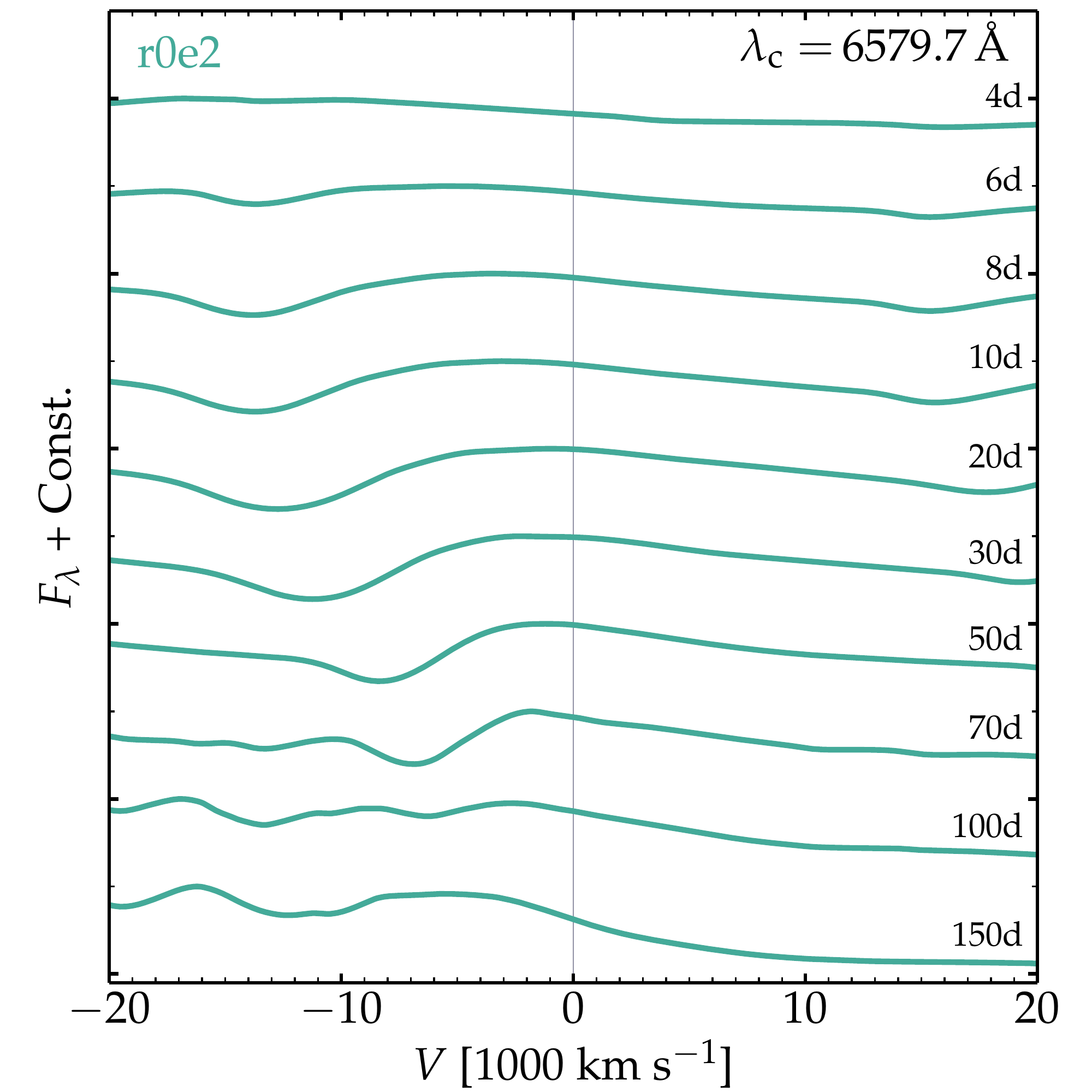}
\end{center}
\vspace{-0.5cm}
\caption{Evolution of the spectral region around $\lambda_{\rm c}=$\,6579.7\,\AA\ for model r0e2 to illustrate the evolution of the Doppler velocity at maximum absorption as well as the peak blueshift. The value of $\lambda_{\rm c}$ corresponds to the air wavelength of the $gf$-weighted mean of the corresponding C\,\two\ doublet. The Doppler velocity at the location of maximum absorption in this line follows closely the velocity of the photosphere shown in Fig.~\ref{fig_YN_phot}.
\label{fig_r0e2_c2_6581}
}
\end{figure}

\begin{figure}
\begin{center}
   \includegraphics[width=\hsize]{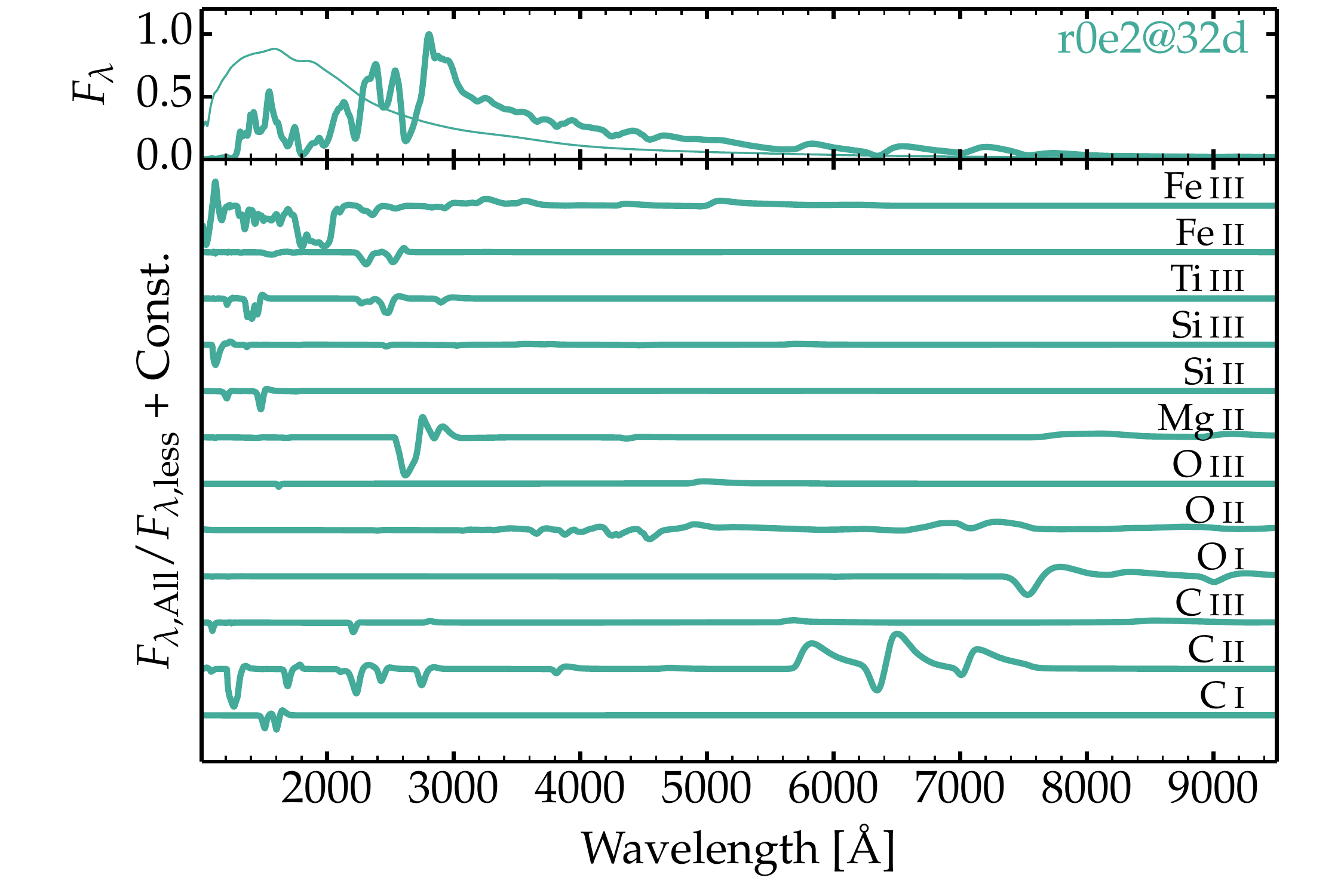}
   \includegraphics[width=\hsize]{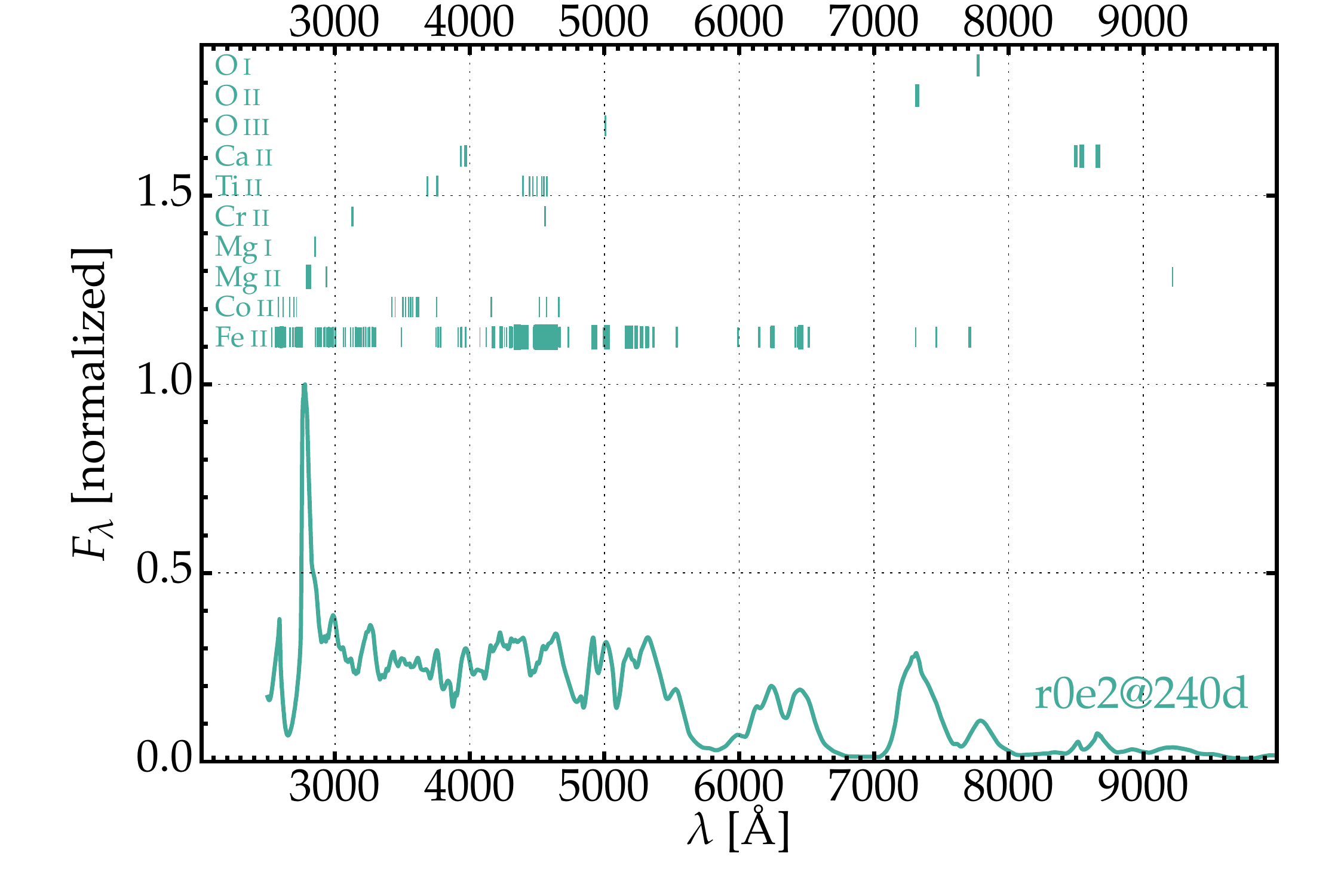}
\end{center}
\vspace{-0.4cm}
\caption{Top: Ladder plot for model r0e2 at 32\,d after explosion (which corresponds to the time of maximum), covering the UV and optical ranges (this model yields a good match to Gaia16apd at bolometric maximum; see Section~\ref{sect_comp_obs}). The upper part shows the full spectrum $F_{\lambda}$ together with the continuum (thin line). Line emission contributes throughout the optical and causes the excess flux above the continuum level. There are no true ``continuum'' windows -- the total model flux is 50\% greater everywhere except in two P-Cygni troughs. This excess line flux impacts the interpretation of a color temperature based on a blackbody fit. In the UV, lines cause a strong blanketing, with the total flux depleted below the continuum level. The lower part shows the ratio of the full spectrum ($F_{\lambda, {\rm All}}$) and  the spectrum computed by omitting the bound-bound transitions of a given ion ($F_{\lambda, {\rm less}}$). Bottom: Optical spectrum (normalized to the  maximum flux in the range $2500-10000$\,\AA) of model r0e2 at 240.0\,d after explosion. Vertical marks at the top correspond to the air wavelength of lines with a Sobolev equivalent width (EW) greater than 7000\,\AA\ (the ions shown at top are displayed in the order of appearance of transitions, starting at the shortest wavelength displayed). Here, 251 lines with an EW between 7000 and 49340\,\AA\ are shown. The thickness of each bar scales linearly with the line EW. The name of the corresponding ion lies at the left of each series of marks.
\label{fig_r0e2_spec_early_late}
}
\end{figure}

\section{Detailed discussion for magnetar-powered SN model r0e2 -- the reference case}
\label{sect_r0e2}

Figure~\ref{fig_YN_mag} illustrates the photometric properties of the model r0e2 with a magnetar ($B_{\rm pm}=3.5\times 10^{14}$\,G and $E_{\rm pm} = 4 \times 10^{50}$\,erg) relative to its counterpart r0e2n without a magnetar. Model r0e2 reaches a peak luminosity of $3.75 \times 10^{43}$\,\ergs\ at 31.7\,d after explosion while model r0e2n peaks later at 44.4\,d at a smaller peak luminosity of $2.26 \times 10^{42}$\,\ergs. The  $V$-band light curve is similar to the bolometric light curve (middle panel of Figure~\ref{fig_YN_mag}). The  $V$-band maximum occurs in both at 43.3\,d but it is $-19.4$\,mag in model r0e2 and $-17.5$\,mag in model r0e2n. The width of the light curve (bolometric or visual) is much broader in the magnetar-powered model, because the magnetar power in model r0e2 (black thin line Fig.~\ref{fig_YN_mag}) overwhelms the decay power from 0.13\,\msun\ of \nifs\ (dashed thin line in Figure~\ref{fig_YN_mag}; $\gamma$-ray escape is allowed for) at all times. The boost in luminosity or visual brightness in model r0e2 is accompanied by a shift to blue optical colors, which is maintained throughout the photospheric phase (bottom panel of Fig.~\ref{fig_YN_mag}).

Figure~\ref{fig_YN_phot} shows the evolution of the photospheric properties (the electron scattering opacity is used for the computation of the optical depth and the location of the photosphere). Although models r0e2 and r0e2n have the same ejecta mass, kinetic energy, density structure, the magnetar-powered model r0e2 has a photosphere at larger velocities $V_{\rm ph}$, (which extends to larger radii, with a maximum of about $4 \times 10^{15}$\,cm at 120\,d) and larger temperatures $T_{\rm ph}$. Around the time of maximum,  $T_{\rm ph}$ is $\sim$\,12,000\,K in model r0e2, compared to $\sim$\,6000\,K in model r0e2n. The higher temperature in model r0e2 causes a greater ionization of the photospheric layers, so that oxygen is once ionized from 4\,d onwards in model r0e2, while it is neutral up to 20\,d and only half ionized afterwards in model r0e2n (bottom panel of Fig.~\ref{fig_YN_phot}).

Since the spectrum formation region tracks approximately the photospheric layers\footnote{This is true up to bolometric maximum. After that, the spectrum forms over an increasingly extended region in the present simulations, in part because the ionization remains large above the photosphere so that no recombination front forms.} as long as the ejecta is optically thick (hence up until $\sim$\,200\,d after explosion in model r0e2, and $\sim$\,140\,d in model r0e2n), the contrast in photospheric properties shown in Fig.~\ref{fig_YN_phot} should lead to a modest change in line profile width and strength but a much bluer optical spectrum with lines from once ionized elements (if one considers that O is a representative element). Figure~\ref{fig_YN_montage} confirms this. The magnetar powered model shows a rapid shift from a red to a blue optical spectrum after a few days past explosion, with a spectral energy distribution (SED) that peaks around 2500\,\AA\ (Fig.~\ref{fig_YN_montage} shows the quantity $\lambda F_\lambda$, which gives more importance to the long-wavelength flux) from $\sim$\,10 to beyond 100\,d. The onset of that shift to the blue (and to greater luminosity and visual brightness is function, for example, of the magnetar power and deposition profiles --- see next sections for various dependencies). The contrast with the model without magnetar is striking (right panel of Fig.~\ref{fig_YN_montage}). In model r0e2n, the SED stays red at all times, with strong signs of blanketing in the UV and in the optical (model 5p11Bx2 and variants have been extensively discussed in \citealt{D15_SNIbc_I,D16_SNIbc_II}). Spectral differences between models r0e2 and r0e2n remain throughout the photospheric and nebular phases.

The left panel of Figure~\ref{fig_YN_montage} also shows the predicted continuum flux (thin line). At the earliest epoch (here, at 5\,d), the continuum flux follows the optical flux closely (outside strong-line regions), but it is strongly blanketed in the UV. As time proceeds, the continuum peaks more and more in the blue and the continuum flux weakens relative to the total flux. By 70\,d, the continuum flux is essentially zero. The difference between the total flux and the continuum flux  arises from the multitude of broad and weak overlapping lines, which cannot be clearly identified in the full spectrum. These lines contribute both in emission and absorption, although as time passes, the emission part dominates. A consequence is that the color temperature obtained by fitting the full spectrum with a blackbody overestimates the temperature of the thermalization layer. A lower gas temperature can accommodate the observed flux because of the line flux contribution (which a blackbody ignores). In these conditions, blackbody fits, which are often used, overestimate the temperature. The weakness of the continuum flux arises from the metal-dominated composition and the weakness of continuum processes relative to line processes. A similar effect holds in SNe Ia (see e.g., \citealt{hillier_hedp_13}).

Figure~\ref{fig_r0e2_c2_6581} shows the evolution of the spectral region centered on the C\,\two\,6579.7\,\AA\ (doublet) transition. This line is strong (to assess the strength of the line, see Figs.~\ref{fig_YN_montage} and \ref{fig_r0e2_spec_early_late}) and appears clearly as a P-Cygni profile as soon as the photospheric conditions are hot and ionized (this line is weak or absent in the model r0e2n at the corresponding epochs). The Doppler velocity at the maximum absorption in this feature tracks closely the velocity of the electron-scattering photosphere for up to 70\,d (for this illustration, this C\,\two\ doublet is a better line to use than those in the $4000-5000$\,\AA\ region, where O\,\two\ lines are numerous and overlap). The location of maximum absorption stays the same from about 60\,d until 100\,d because the photosphere then recedes to inner ejecta layers where C is under-abundant: the C-rich shell is located beyond 6000\,\kms\ in model r0e2 (analogous to the effect on the H$\alpha$ line profile caused by the H stratification in SNe IIb; see Fig.~\ref{fig_init_comp}). After $\sim$\,100\,d, the C\,\two\ lines disappear. In our WR star models, a C-rich shell is always present in the outermost layers of the ejecta and is the cause of the strong C\,\two\ lines in our models.

Figure~\ref{fig_r0e2_spec_early_late} uses two different ways to illustrate the main line contributions in the spectra of models r0e2 at 32.0 and 240.0\,d after explosion. The top panel shows the contribution of individual ions, obtained by comparing the full spectrum ($F_{\lambda, {\rm All}}$) with the spectrum computed by omitting the bound-bound transitions of a given ion ($F_{\lambda, {\rm less}}$). The bottom panel marks the strongest transitions with a Sobolev equivalent width (EW) greater than 7000\,\AA\ (at nebular times, the very weak continuum flux and the presence of strong forbidden emission lines explain these large EWs). The top panel of Fig.~\ref{fig_r0e2_spec_early_late} shows the contributions from a large number of lines associated with once and twice ionized states of C, O, Mg,  and Fe. At 32.0\,d, the spectrum forms around 10000\,\kms, in the C-rich part of the progenitor CO core (see middle panel of Fig.~\ref{fig_init_comp} and Fig.~\ref{fig_YN_phot}), hence the spectrum naturally shows signatures of these intermediate mass elements, although the number of line transitions is large and line overlap is strong, often preventing an association between a feature and a given transition. The photospheric temperature and ionization favor the presence of these elements in their once or twice ionized state. The region $4000-5000$\,\AA\ is primarily shaped by O\two\, which has about one hundred lines with an EW greater than 100\,\AA\ (3p--3s, 3d--3p, and 4f--3d transitions). At longer wavelength, the strongest features are due to C\two, mostly doublets or triplets at 5889.7,  6579.7,  7234.8\,\AA. He\one\,5875.6\,\AA\ contributes weakly (about 1/5 of the total Sobolev EW) to the feature at 5900\,\AA. O\one\,7773.4\,\AA\ is the main contributor of the feature at 7800\,\AA, with a weak contribution at 7896.4\,\AA\ by Mg\two\ -- the main Mg\two\ contribution is from multiple lines around 2800\,\AA, which also overlap with C\two\ and C\three\ transitions.  There are weaker Mg\two\ transitions around 2930 and 4480\,\AA. In the UV range, most lines are caused by similar ions, together with a few regions of strong blanketing by iron-group elements (Ti\three\ around 1500\,\AA; Fe\three\ in the range $1700-2200$\,\AA). The ups and downs in the UV range are not actual lines, not even blends of lines. These peaks and dips are regions of reduced and enhanced opacity, and the entire UV flux is subject to blanketing.

The bottom panel of Fig.~\ref{fig_r0e2_spec_early_late} shows the spectrum of model r0e2 at 240\,d. The total ejecta optical depth to electron scattering at that time is 0.48, and the conditions for the spectrum formation are hybrid, with the presence of  permitted transitions (similar to those seen at 24.5\,d, as well as a forest of Fe\two\ lines, including strong lines at 6149.3, 6247.6, and 6456.4\,\AA\ --- there are numerous weaker components overlapping with these transitions) and forbidden lines, most notably of O\,\two\,7324.3\,\AA\ (multiplet) and O\three\ at 5006.8\,\AA. At later times, the permitted transitions weaken and the spectrum becomes more and more dominated by forbidden lines (which ions contribute depend on how the ionization state evolves). Oxygen absorbs most of the magnetar power (the power absorbed scales roughly linearly with the mass of each element), although oxygen lines are not always the strongest lines (this depends on the composition mixture and on atomic  properties of the ions and atoms present).

\begin{figure}
  \includegraphics[width=\hsize]{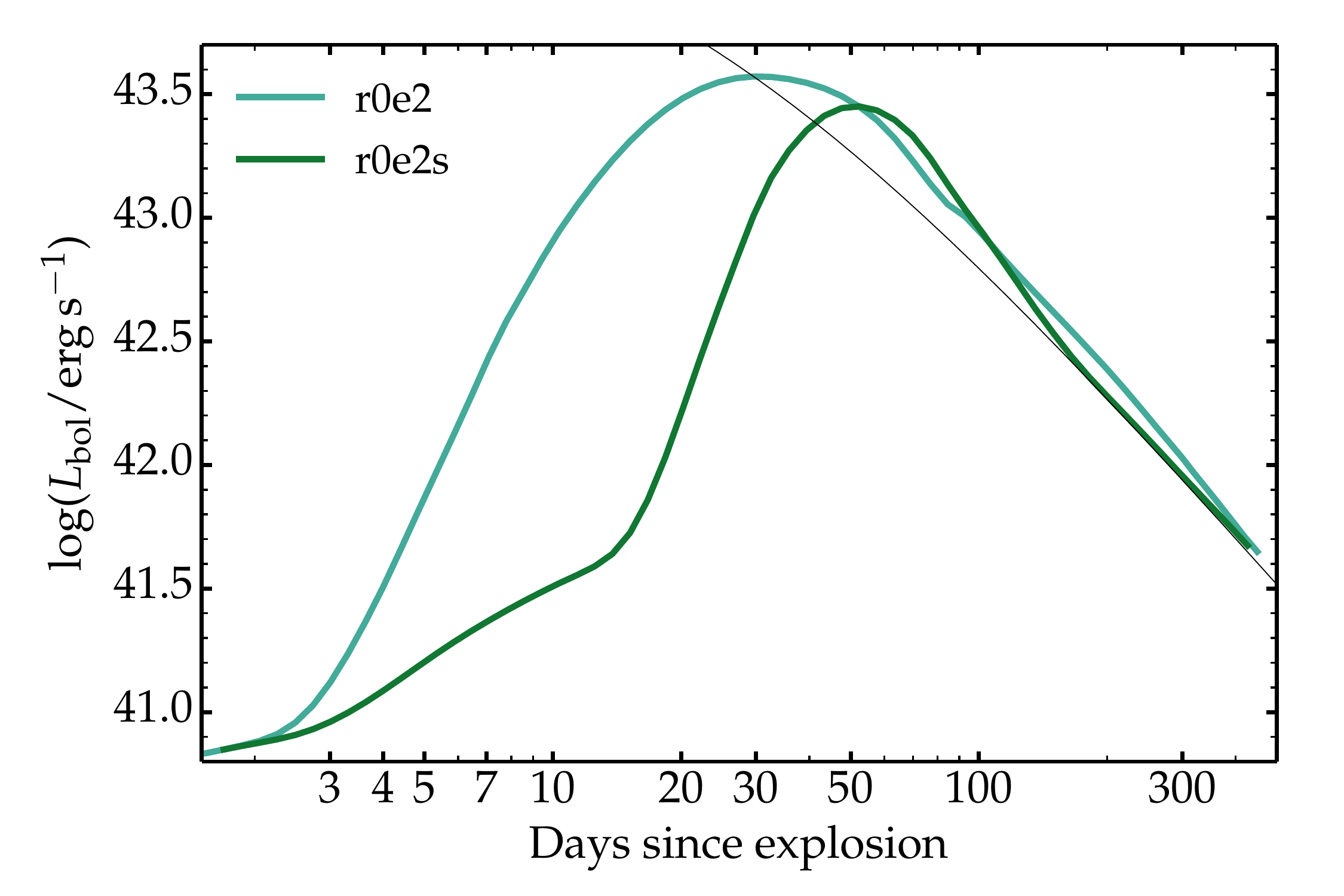}
   \includegraphics[width=\hsize]{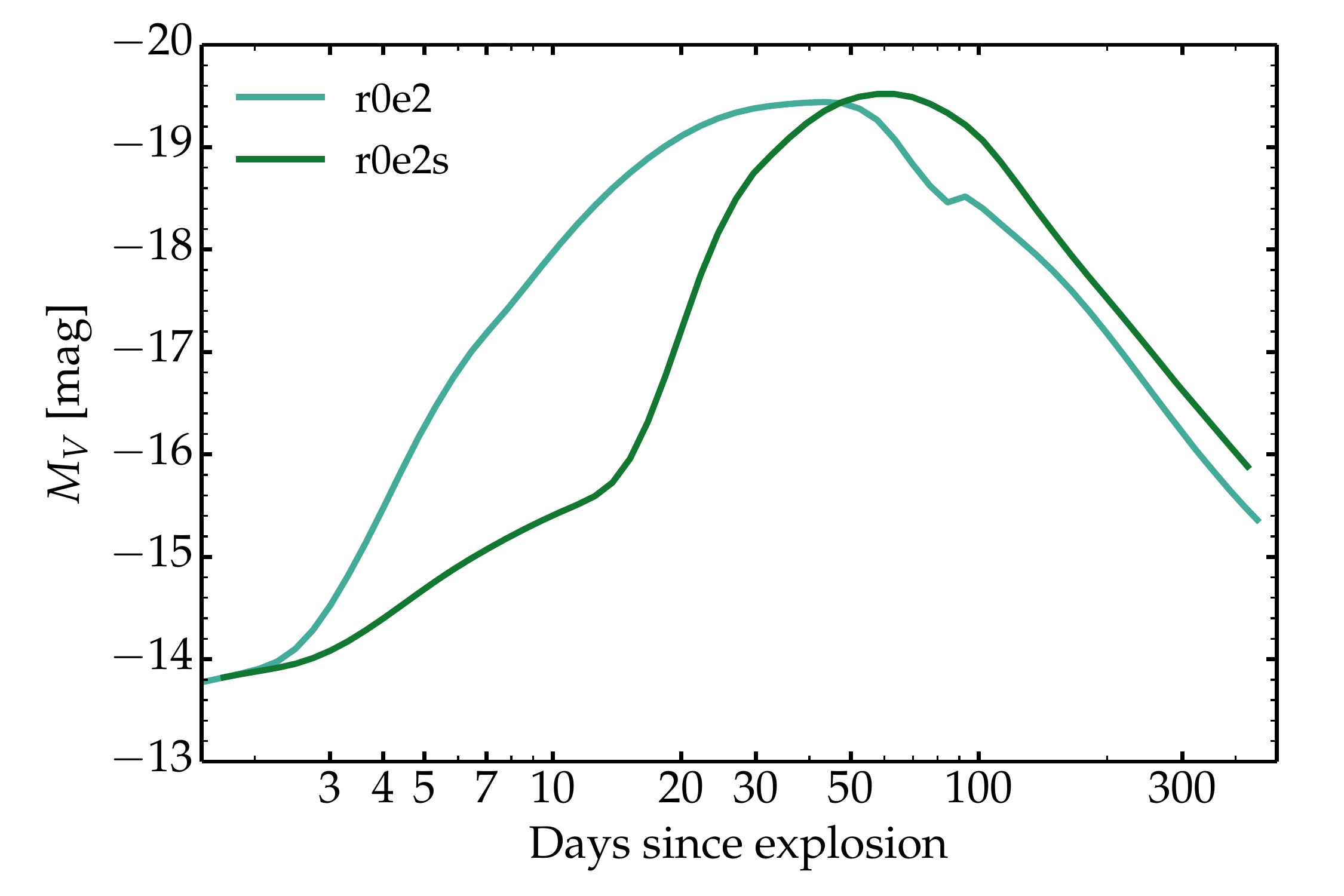}
   \includegraphics[width=\hsize]{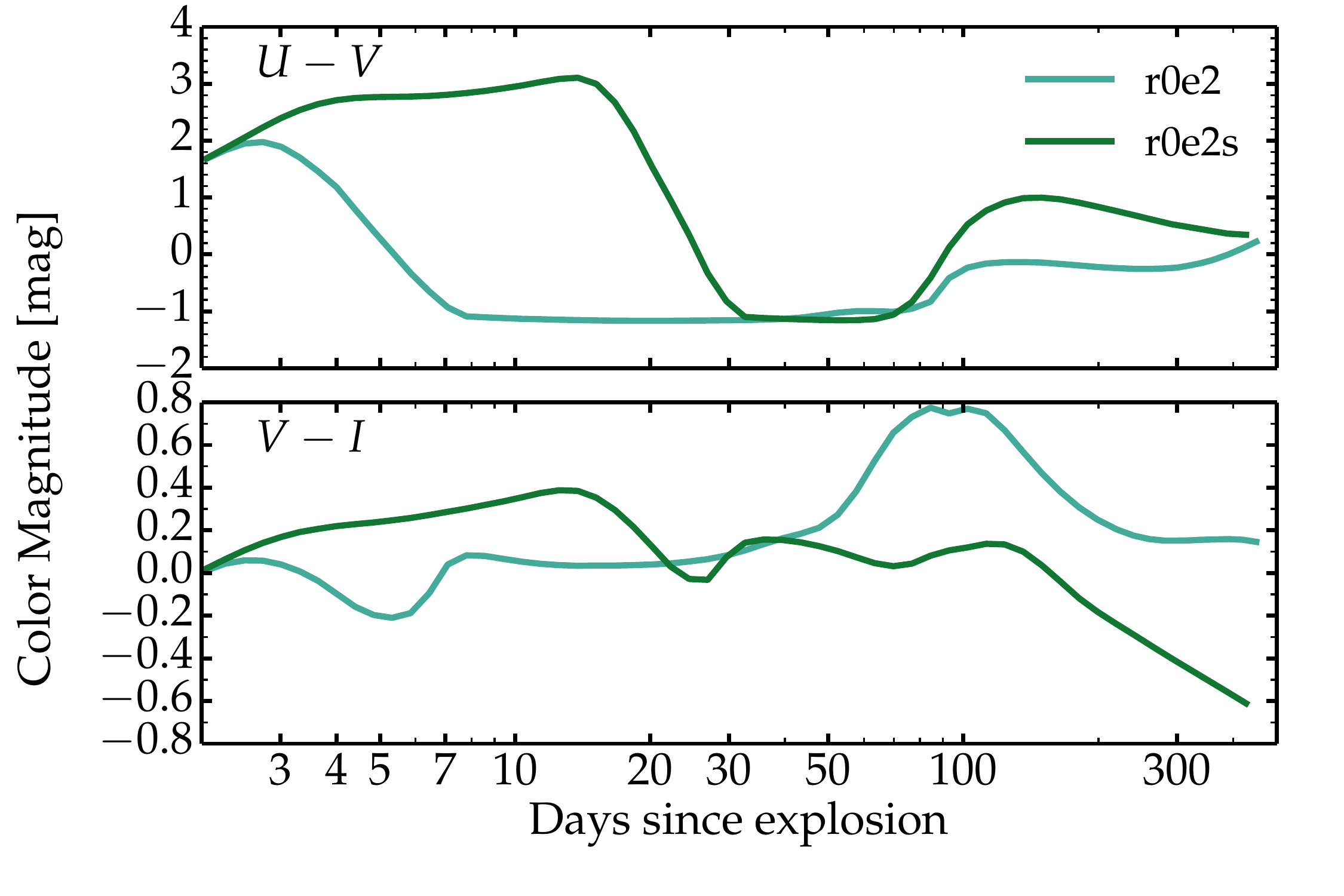}
  \caption{Same as Fig.~\ref{fig_YN_mag}, but now for model r0e2 and r0e2s, which differ in the adopted magnetar-power deposition profiles.
 \label{fig_edep_mag}
  }
\end{figure}

\begin{figure}
	\vspace{-0.32cm}
   \includegraphics[width=\hsize]{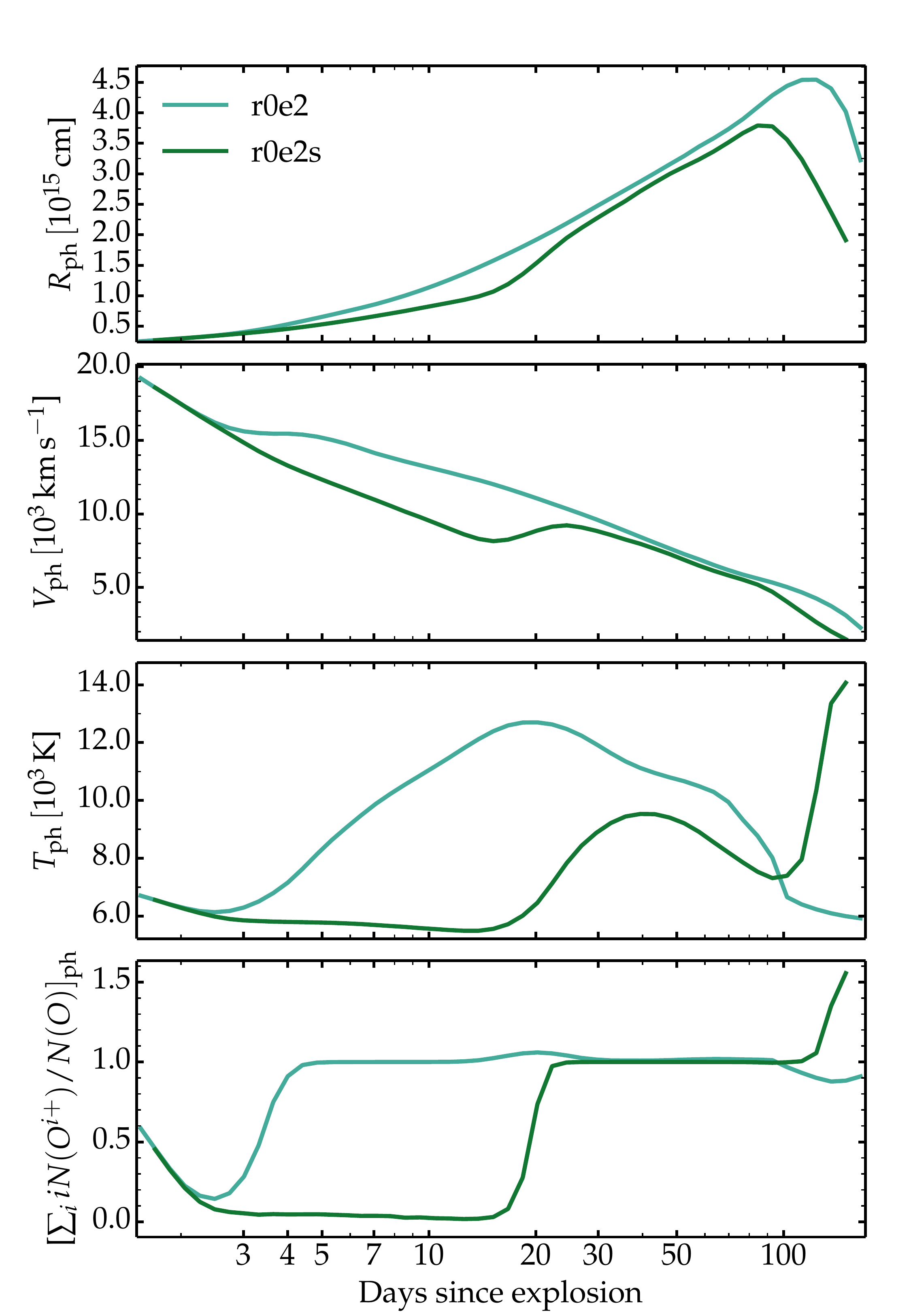}
   \caption{Same as Fig.~\ref{fig_YN_phot}, but now for models r0e2 and r0e2s.
\label{fig_edep_phot}
}
\end{figure}

\begin{figure}
  \includegraphics[width=\hsize]{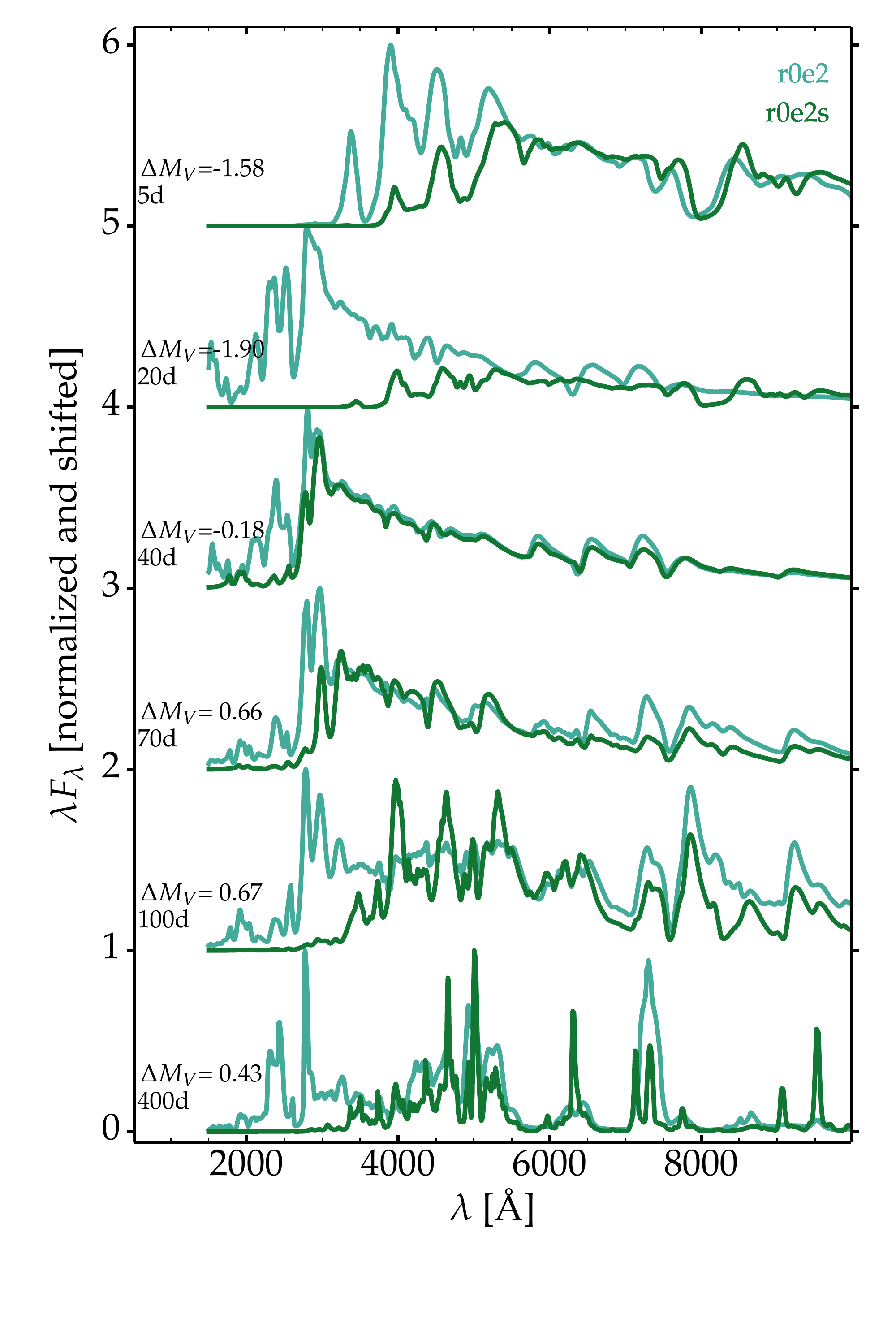}
     \vspace{-0.3cm}
  \includegraphics[width=\hsize]{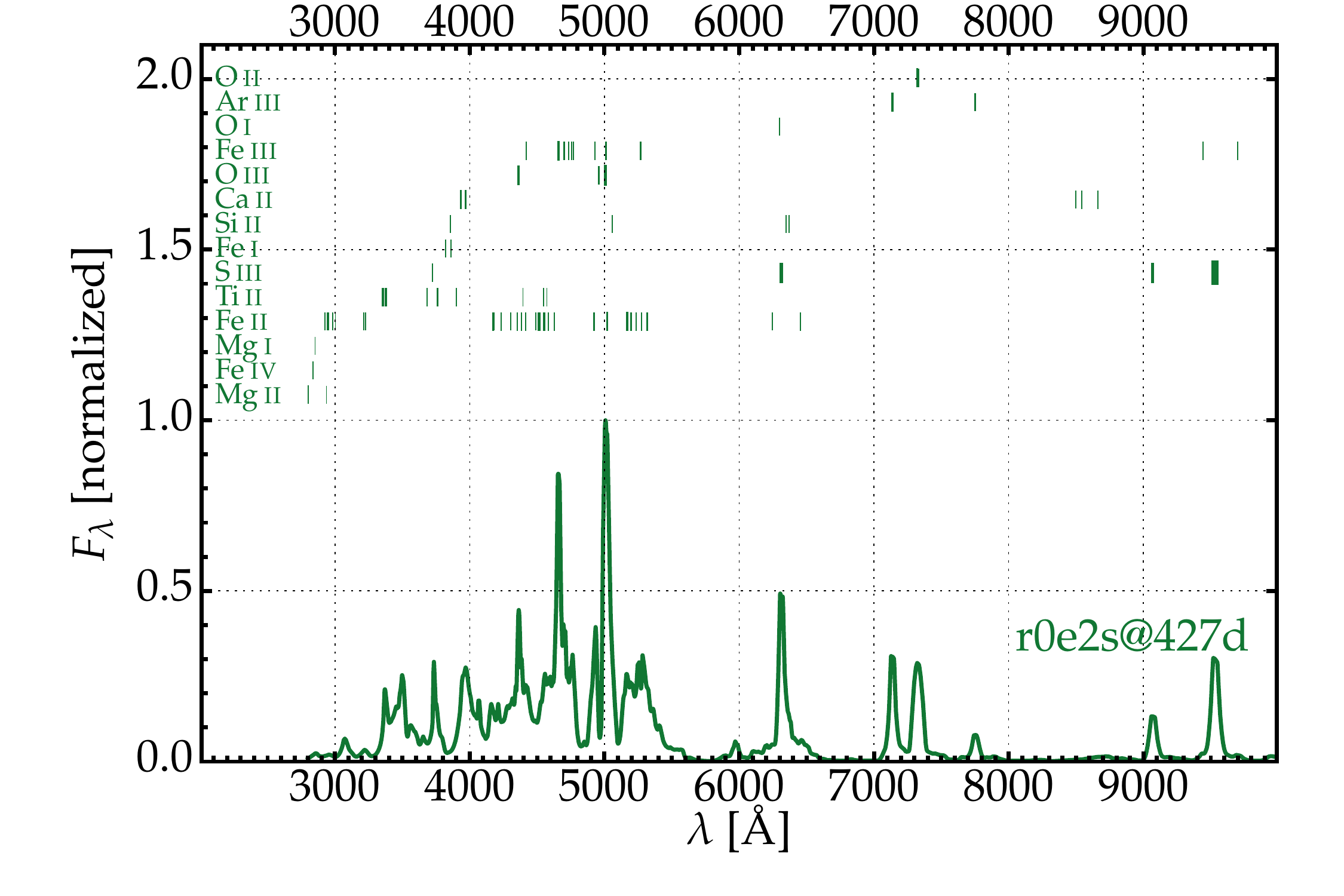}
\caption{Top: Comparison of multi-epoch spectra for models r0e2 and r0e2s. Bottom: Same as the bottom panel of Fig.~\ref{fig_r0e2_spec_early_late}, but now for model r0e2s at  427\,d after explosion.
  \label{fig_edep_spec}
  }
\end{figure}

\section{Influence of the magnetar energy deposition profile}
\label{sect_edep}

This section discusses the influence of the adopted deposition profile. Model r0e2s uses $V_0 = 1425$\,\kms\ and $dV=713$\,\kms, compared to $V_0 = 5700$\,\kms\ and $dV=2850$\,\kms\ in model r0e2 (see  Fig.~\ref{fig_edep_profile} and Section~\ref{sect_pm_setup}). The resulting SN ejecta and radiation properties computed with \cmfgen\ are qualitatively similar but there are quantitative differences (Fig.~\ref{fig_edep_mag}). At early times, the magnetar influence is delayed in model r0e2s, so that the increase in luminosity and visual brightness and the shift of the optical color to the blue are all delayed. The sudden rise in luminosity occurs at $\sim$\,3\,d in model r0e2 but is delayed until nearly 20\,d in model r0e2s. From model r0e2 to r0e2s, the rise time increases from 31.7 to 51.7\,d, the bolometric maximum drops by 25\%, the $V$-band magnitude rises by 0.2\,mag. During the rise to maximum, model r0e2s stays red longer but it has roughly the same $U-V$ and $V-I$ colors at maximum as model r0e2. In model r0e2, more energy is radiated (i.e., the time-integrated luminosity is greater) because the magnetar power is released over a broader range of velocities. By depositing more power at smaller optical depth, radiation is allowed to escape earlier with a reduced degradation by expansion. As discussed in Section~\ref{sect_pm_setup}, this approach is not fully consistent since magnetar energy advected out would be subject to expansion cooling. So, this neglect tends to overestimate the influence of magnetar power. However, our neglect of dynamics means that the inner ejecta layers are not accelerated, which would contribute to bringing them faster to a lower optical depth where they could radiate their stored energy (mass conservation also implies that this swept-up shell becomes very dense; a radial compression, however, does not change the radial column density).

  Figure~\ref{fig_edep_phot} illustrates the impact of the deposition profile on the photospheric properties. With a narrower profile, the photosphere is located deeper in (smaller radii and velocities), is cooler and less ionized on the rise to maximum. Around the time of maximum (which differs by 20\,d between the two models), the difference in photospheric properties is small, mostly limited to the temperature ($\sim$\,11000\,K in r0e2 compared to $\sim$\,9000\,K in model r0e2s). This offset likely results from the longer rise time in model r0e2s, which implies that the volume occupied by the ejecta in model r0e2s is greater at maximum (smaller equilibrium radiation temperature in the optically-thick volume) and the magnetar power at that later time is smaller. After maximum, the deposition profile has less impact on the SN properties because the ejecta is not very optically thick, the spectrum forms over a more extended volume, and the outer ejecta layers contribute less.

\begin{figure}[t!]
   \includegraphics[width=\hsize]{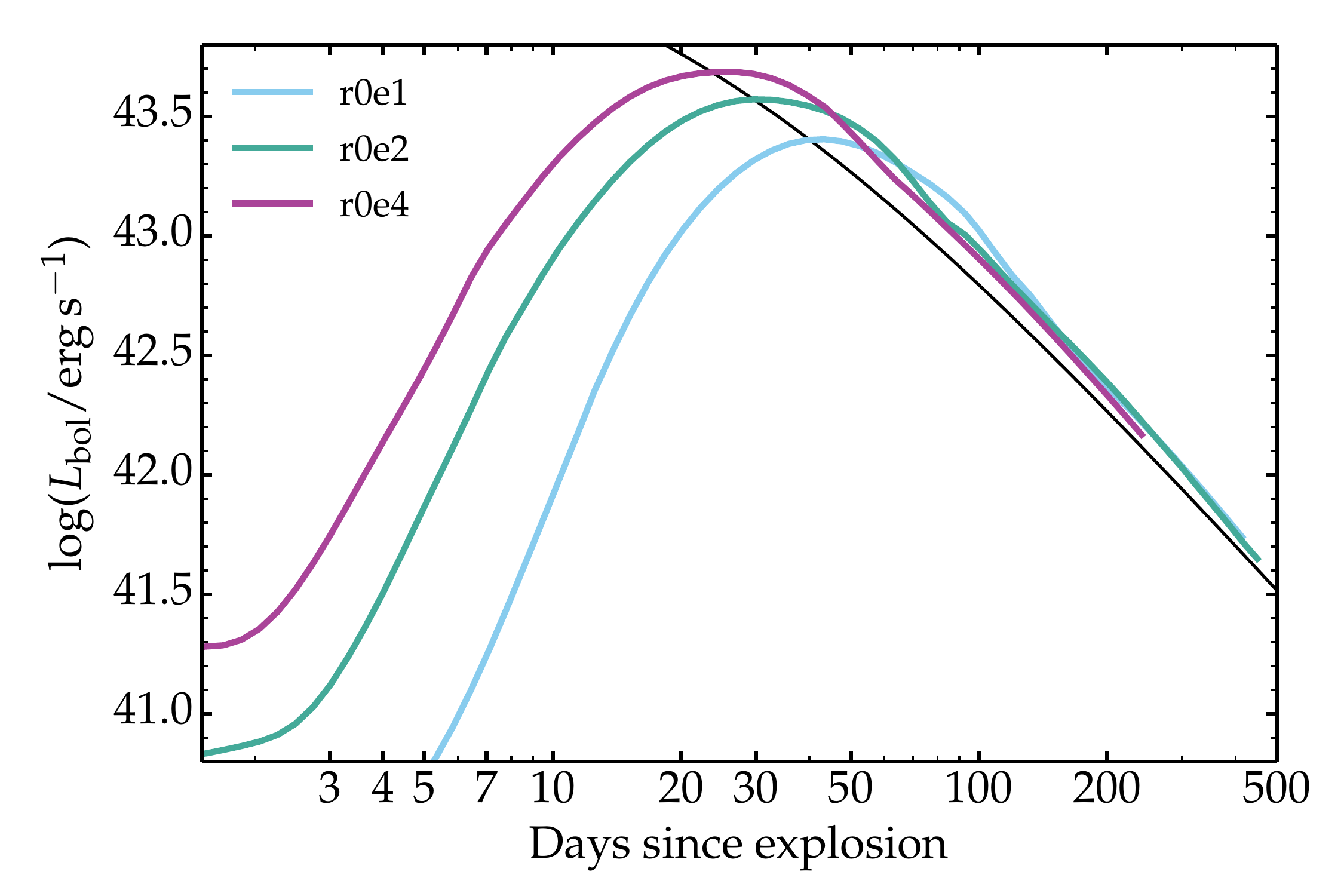}
   \includegraphics[width=\hsize]{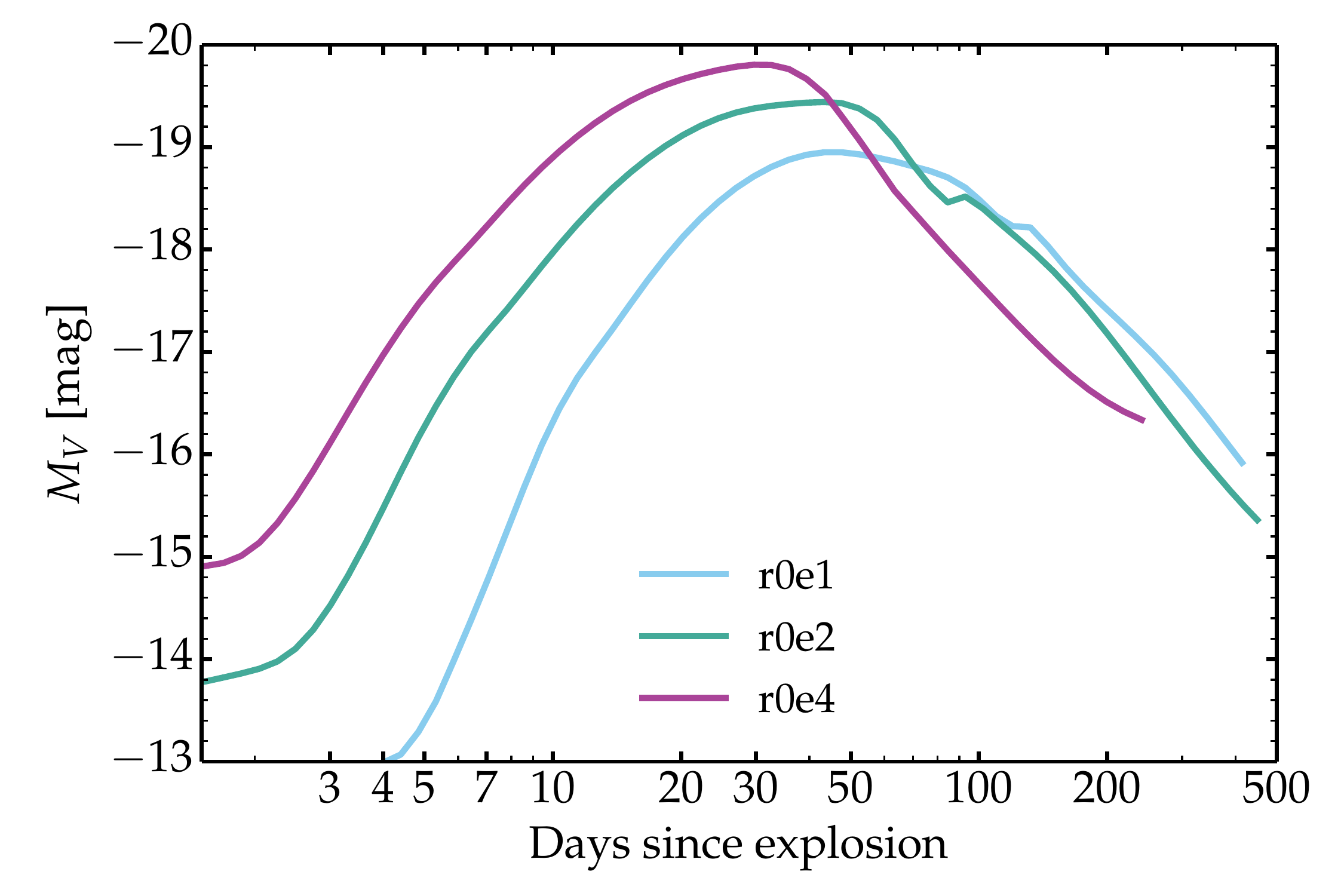}
   \includegraphics[width=\hsize]{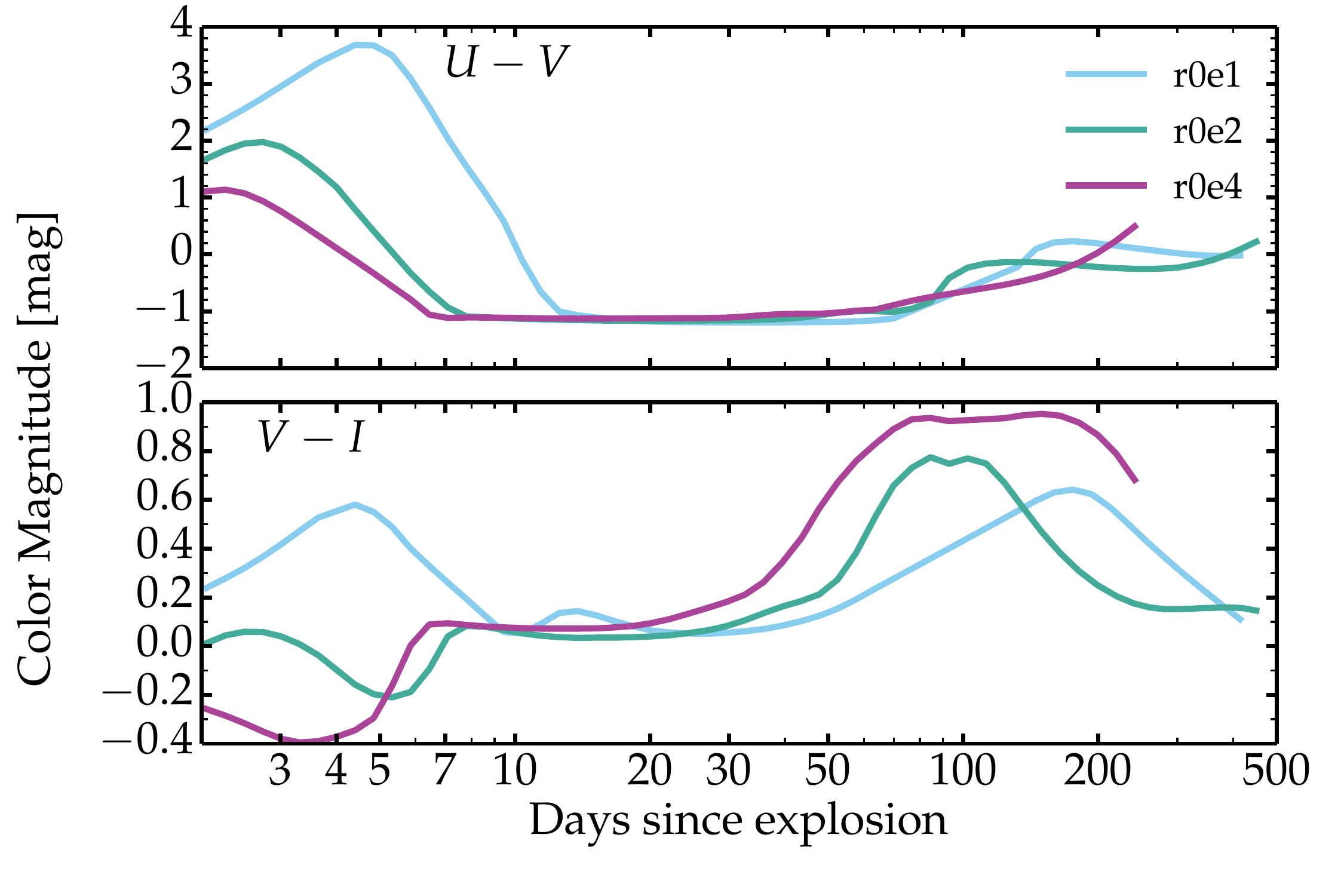}
\caption{Same as Fig.~\ref{fig_YN_mag}, but now for model r0e1, r0e2 and r0e4, which differ in ejecta kinetic energy primarily (as well as slightly in composition; see Table~\ref{tab_sum}).
\label{fig_ekin}
}
\end{figure}

The difference in spectral evolution reflects the delay in the magnetar influence, especially the shift to a hotter and more ionized photosphere radiating a bluer optical color (top panel of Fig.~\ref{fig_edep_spec}). Model r0e2s exhibits a similar spectrum to model r0e2n (without magnetar) for about 30\,d. By 40\,d, the spectra of models r0e2 and r0e2s show similar lines in the optical, while model r0e2 has a greater flux in the UV (its photosphere is hotter). Line widths are comparable around maximum (the photospheric velocity is similar in both models from about 30 to 100\,d; Fig.~\ref{fig_edep_phot}). At nebular times, model r0e2s shows narrower spectral lines, which therefore form deeper in the ejecta where the deposited power is relatively greater. A marked difference that persists at all times is the excess UV flux in model r0e2 compared to r0e2s. This arises from the excess power deposited in the outer lower-density ejecta layers, which tends to produce a greater temperature (and ionization).

\begin{figure}[t!]
	\vspace{-0.32cm}
   \includegraphics[width=\hsize]{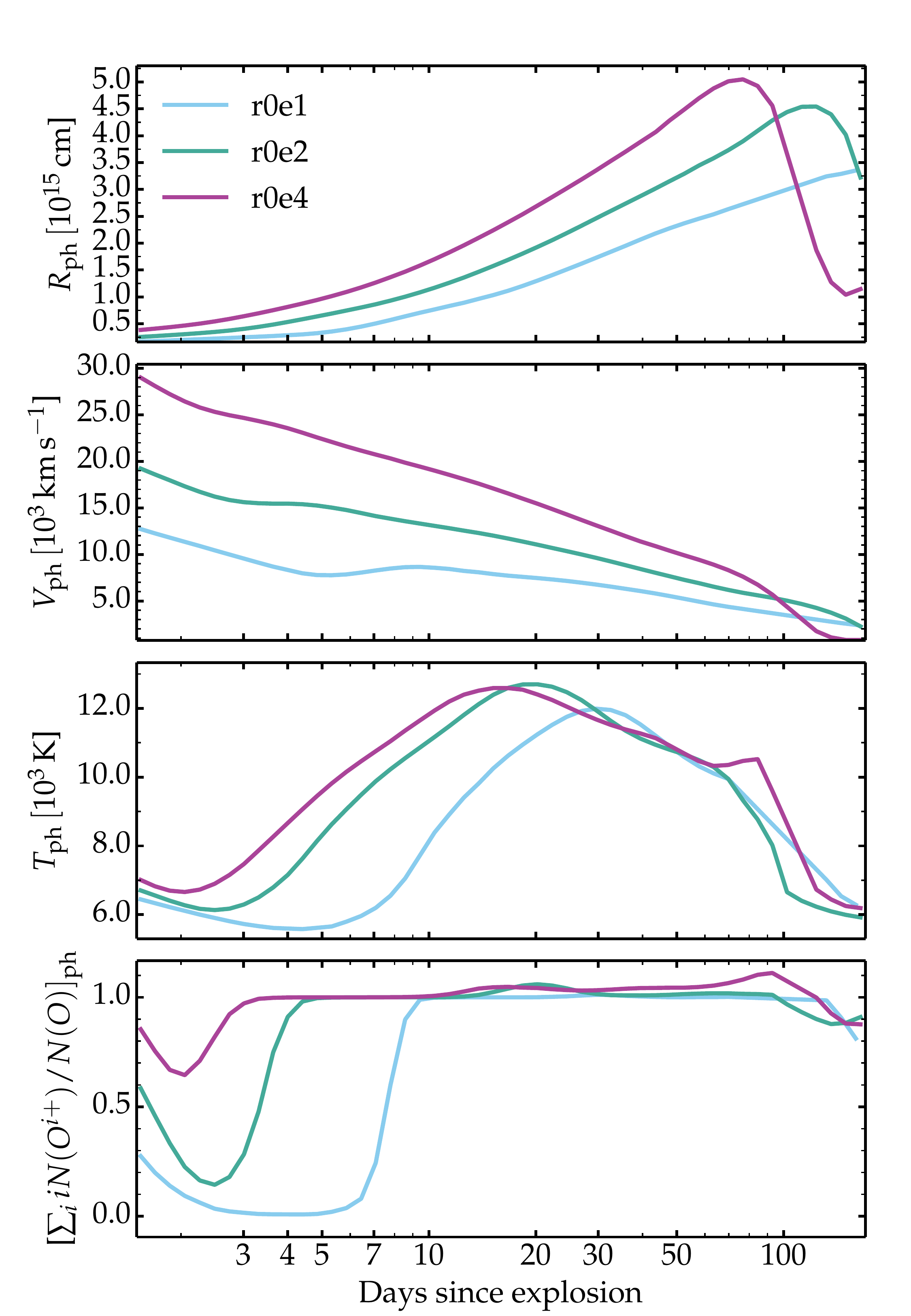}
   \caption{Same as Fig.~\ref{fig_YN_phot}, but now for models r0e1, r0e2, and r0e4 (models that differ in ejecta kinetic energy).
\label{fig_ekin_phot}
}
\end{figure}

The bottom panel of Fig.~\ref{fig_edep_spec} illustrates the spectral features in model r0e2s at 427\,d after explosion. The spectrum is very ionized relative to a core collapse SN. Here, the main lines are due to O\,\two\,7324.3\,\AA, O\three\ at 4363.2, 4958.9, 5006.8\,\AA,  S\three\ at 6312.0, 9068.6, and 9530.6\,\AA, Ar\three\,7135.8\,\AA, Ca\two\ H \& K, Fe\two\, at 4923.9, 5018.4, 5169.0\,\AA, and Fe\three\ at 4658.0, 4701.5, and 5011.3\,\AA. All these lines are also present in model r0e2 (with different relative strength), but they are more easily identified here with the reduced line overlap.

The magnetar-power deposition profile therefore impacts the SN radiation in many ways. It controls the timing for the influence felt at the photosphere as well as the different reaction of the gas for a given power (i.e., whether the power is injected at higher or lower density). These changes cause various quantitative shifts but do not change the qualitative picture. Overall, the primary impact is on the total heat content, which is controlled by the volume-integrated magnetar-power. The exact distribution is a secondary effect.

\begin{figure}
  \includegraphics[width=\hsize]{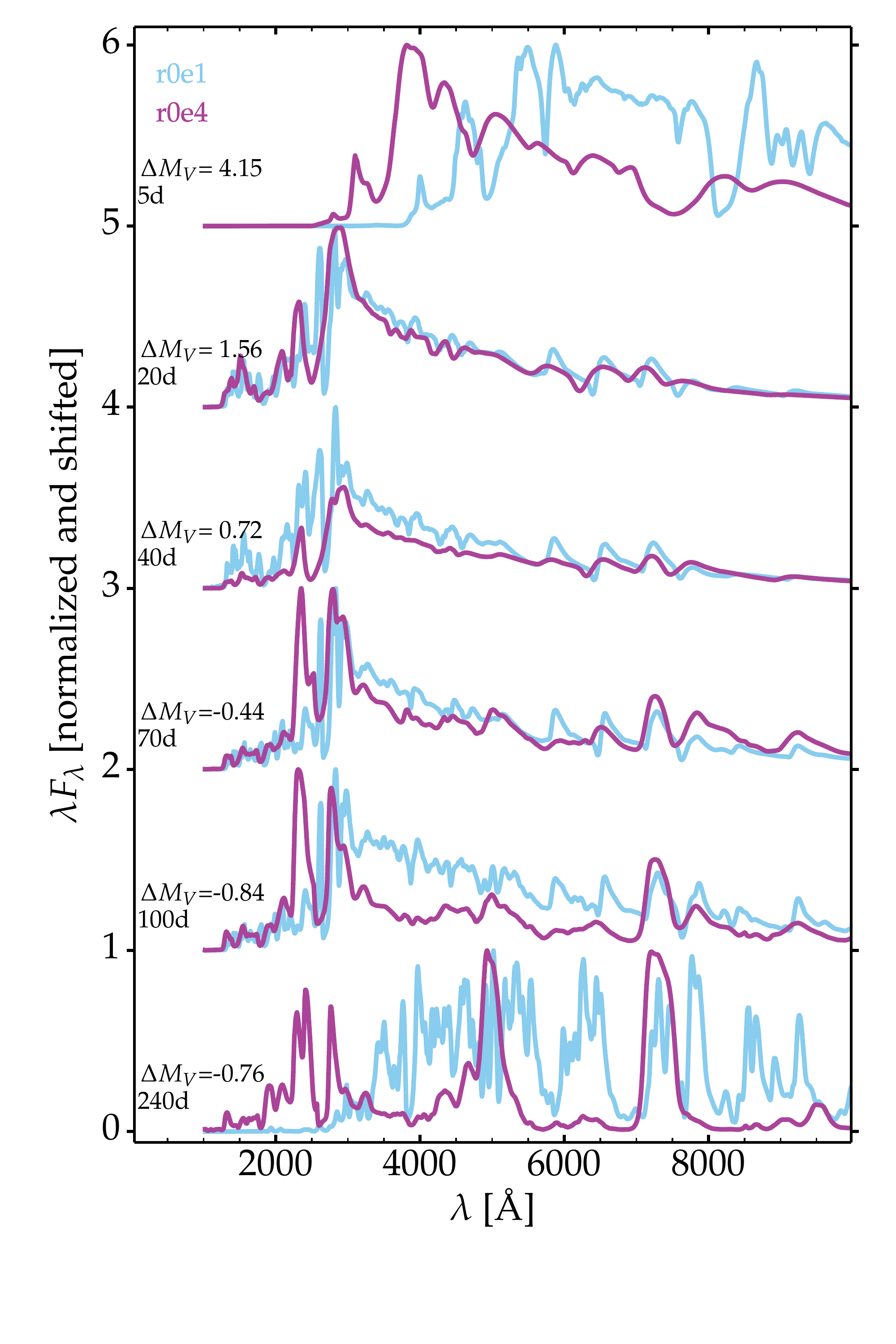}
\vspace{-1.5cm}
  \caption{Comparison of multi-epoch spectra for models r0e1 and r0e4, which are identical to  model r0e2 except for the lower and higher ejecta kinetic energy. Model r0e2 is discussed in Section~\ref{sect_r0e2}. Its spectral properties are intermediate (e.g., in SED color, line widths, or ionization) between those of models r0e1 and r0e4.
  \label{fig_ekin_spec}
  }
\end{figure}

\section{Influence of ejecta kinetic energy}
\label{sect_ekin}

This section presents the results for counterparts of model r0e2 for which the ejecta kinetic energy is lower  (model r0e1) and higher (model r0e4; see Section~\ref{sect_setup}). By adopting the same magnetar properties in all three models, one can test the influence of the ejecta kinetic energy on the SLSN Ic properties. For the magnetar power deposition profile, model r0e1 uses $V_0 = 2700$\,\kms\ and $dV=1350$\,\kms, model r0e2 uses $V_0 = 5700$\,\kms\ and $dV=2850$\,\kms, and model r0e4 uses $V_0 = 10,000$\,\kms\ and $dV=5000$\,\kms\ (see Section~\ref{sect_setup} and Fig.~\ref{fig_edep_profile}).

\begin{figure}
\includegraphics[width=\hsize]{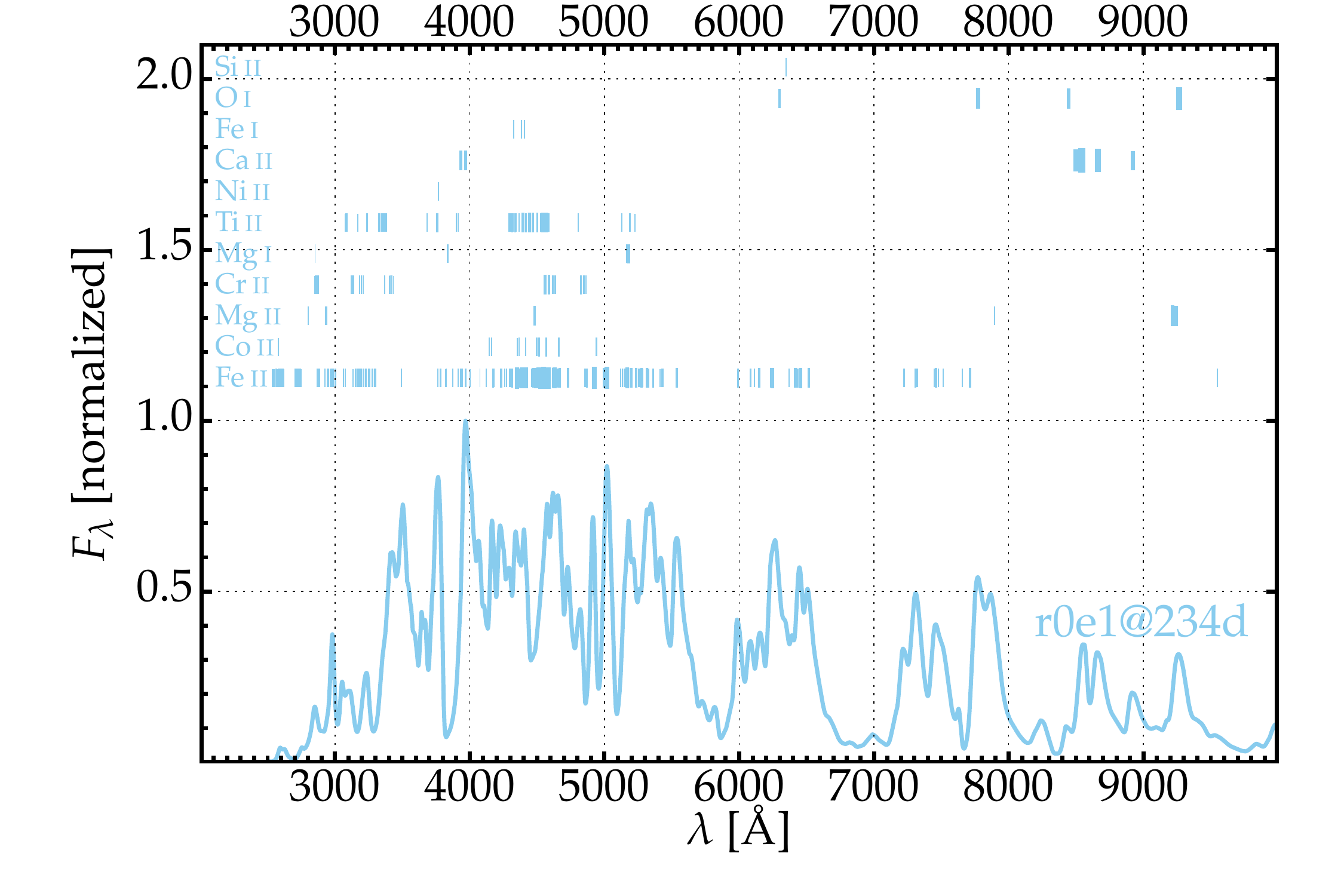}
\includegraphics[width=\hsize]{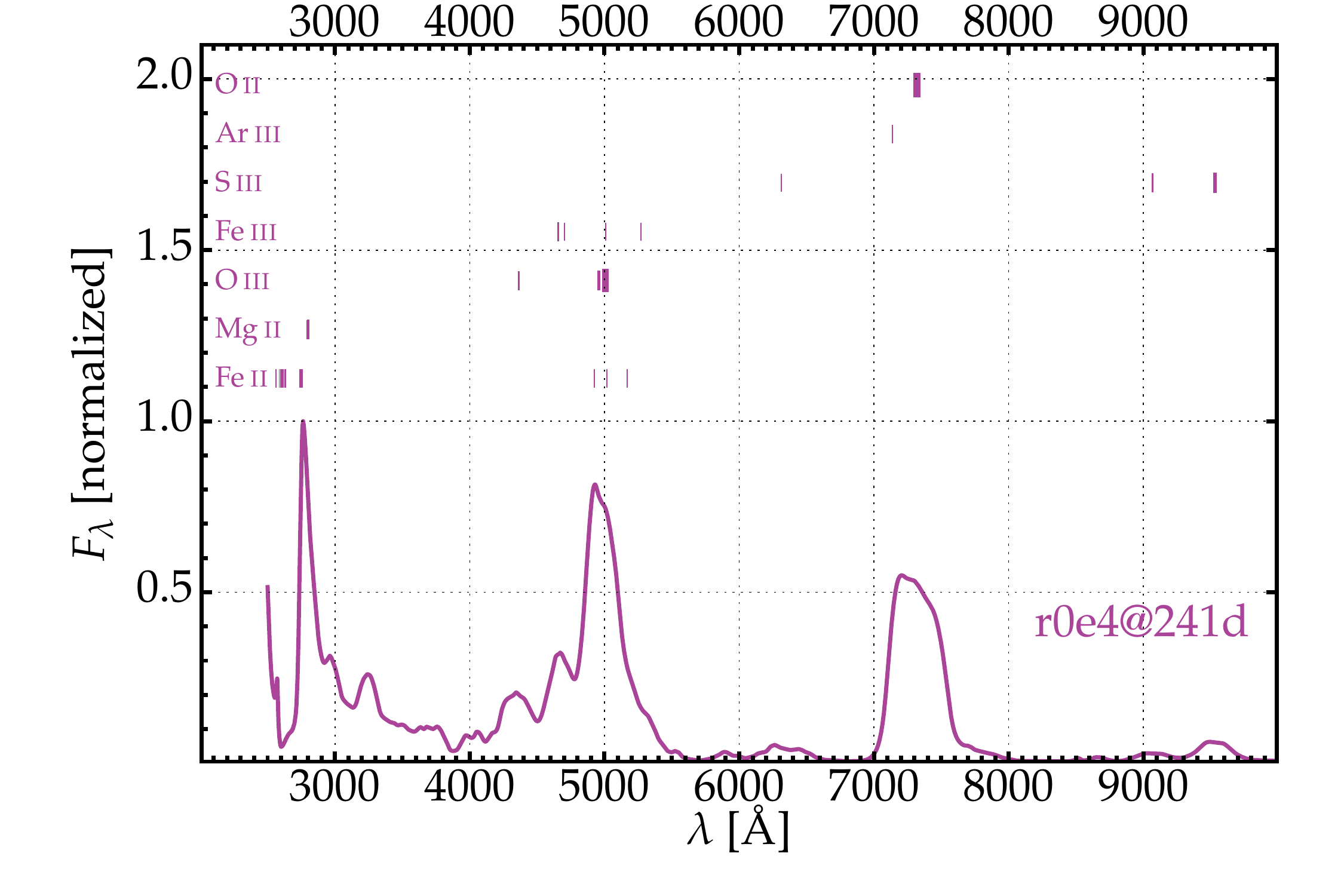}
\caption{Same as the bottom panel of Fig.~\ref{fig_r0e2_spec_early_late}, but now for models r0e1 and r0e4 at 234 and 241\,d after explosion.
\label{fig_ekin_spec_240d}
}
\end{figure}

Figure~\ref{fig_ekin} shows the bolometric light curve, the $V$-band light curve, and the $U-V$ and $V-I$ color curves for models r0e1, r0e2, and r0e4. The effect seen here is similar to employing a broader deposition profile (see previous section). With higher ejecta kinetic energy, the luminosity (or optical brightness) increases earlier (because of the lower ejecta optical depth and the broader deposition profile), peaks to a greater maximum (because it peaks earlier, when the magnetar power is greater), and is bluer early on. With higher ejecta kinetic energy, the photosphere is located further out (larger radii and velocity), although the ejecta becomes optically thin earlier (Fig.~\ref{fig_ekin_phot}). At the time of bolometric maximum, model r0e4 ($E_{\rm kin} = 1.23 \times 10^{52}$\,erg) has a photospheric velocity of $\sim$\,14000\,\kms, while model r0e1 ($E_{\rm kin} = 1.14 \times 10^{51}$\,erg) has a photospheric velocity of $\sim$\,5800\,\kms. This reflects closely the difference in $\sqrt{2E_{\rm kin}/M_{\rm ej}}$. The photospheric temperature is greater because of the earlier influence of the magnetar, but all models have a similar $T_{\rm ph}$ around the phase of maximum.

\begin{figure}
  \includegraphics[width=\hsize]{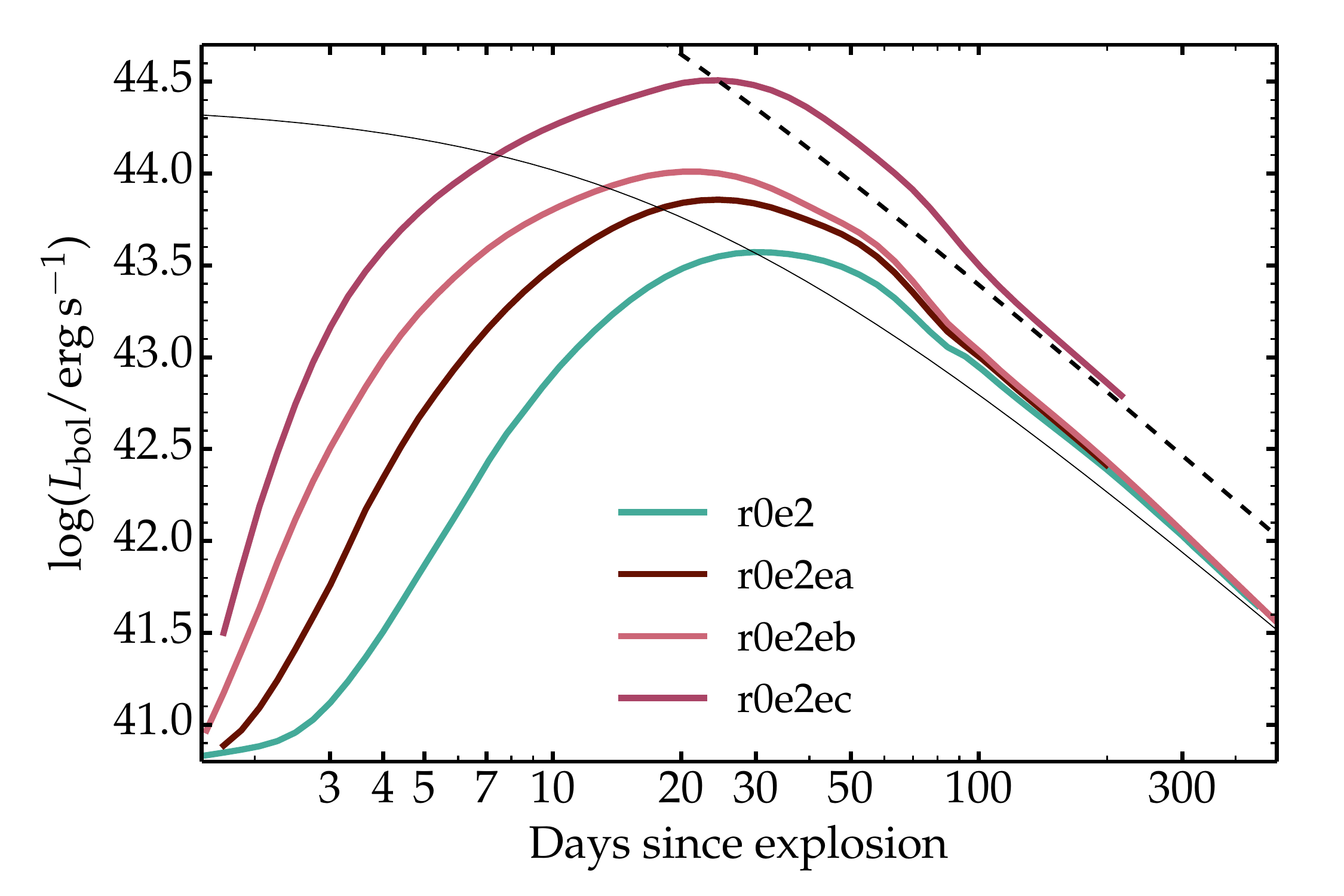}
  \includegraphics[width=\hsize]{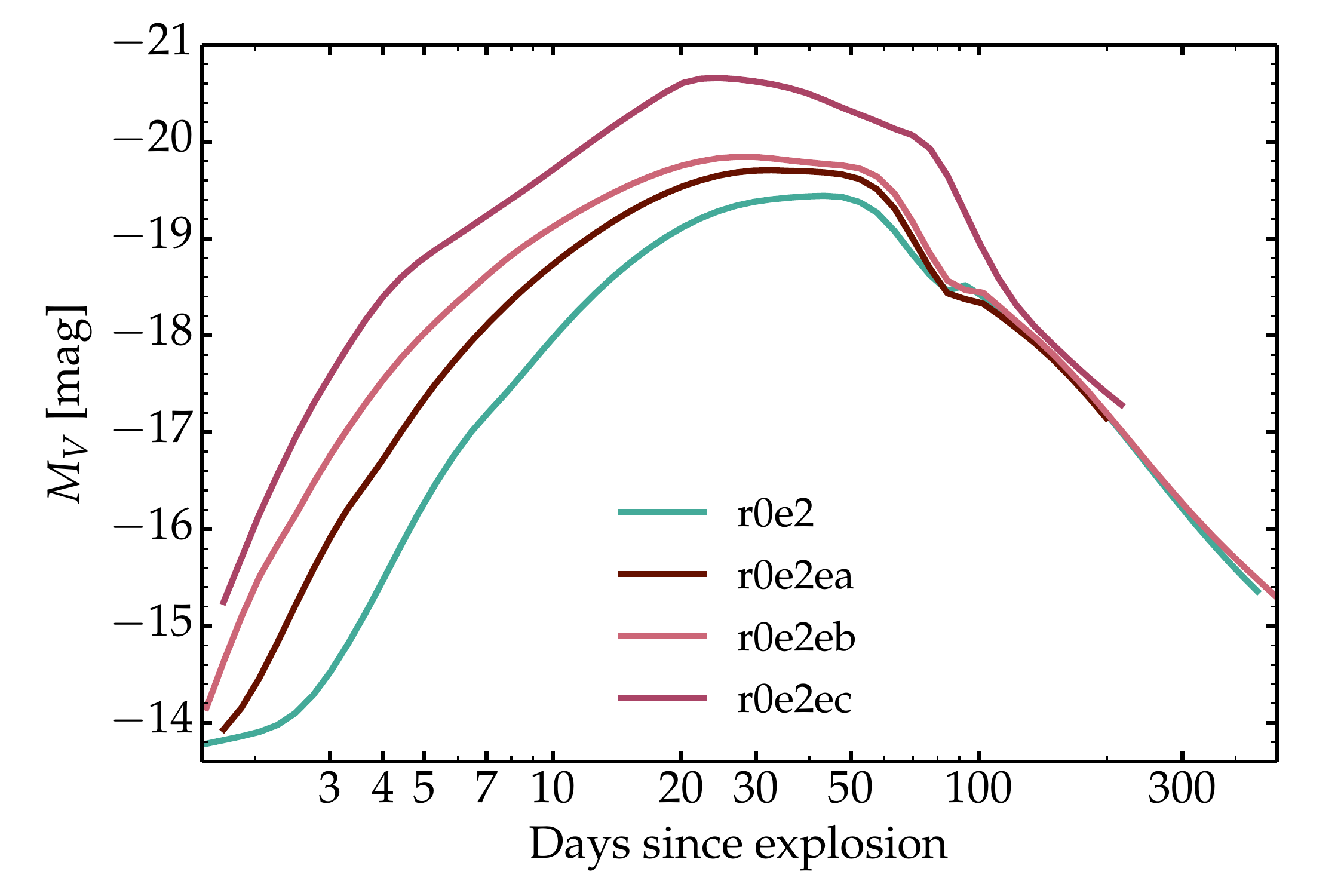}
  \includegraphics[width=\hsize]{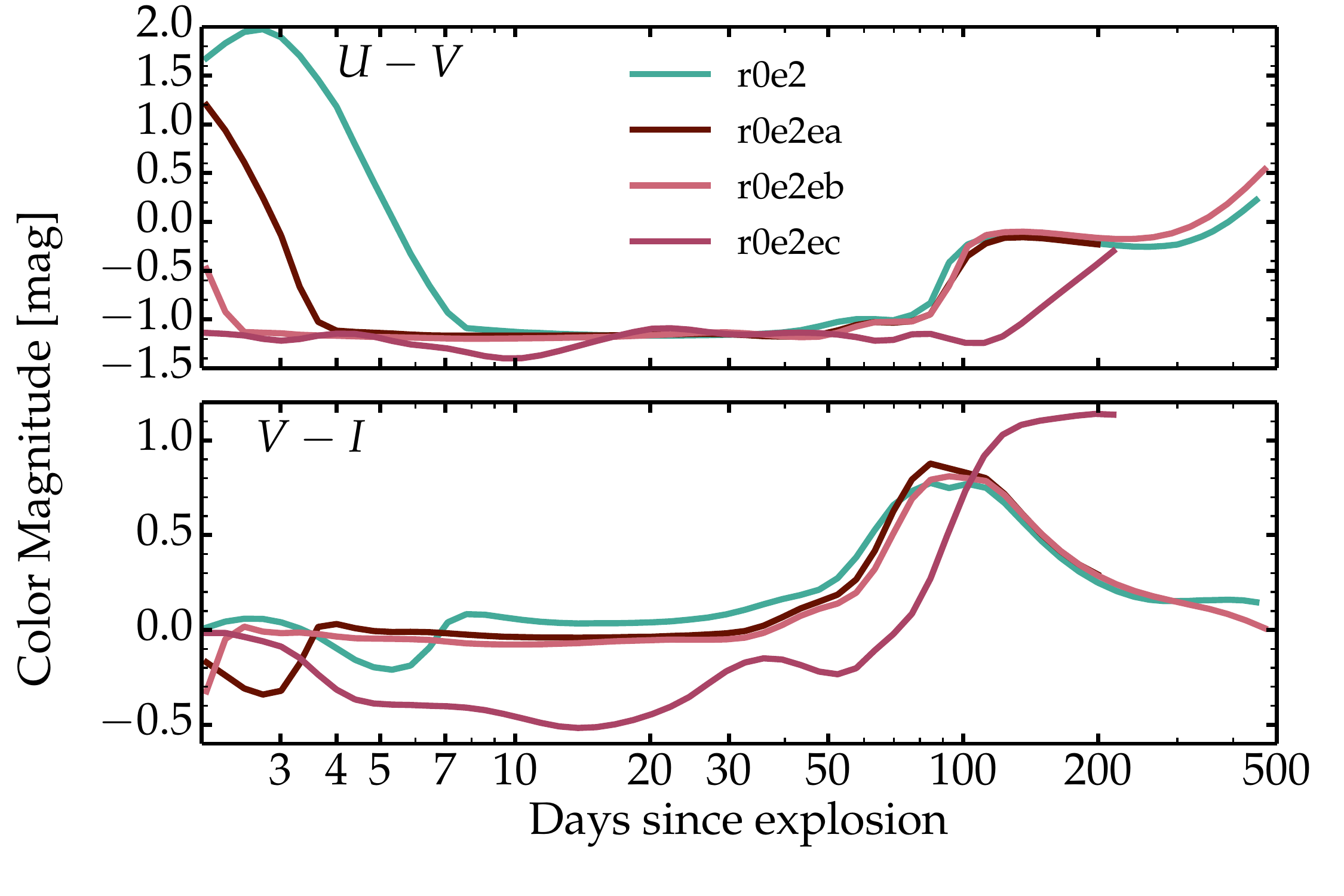}
  \caption{Same as Fig.~\ref{fig_YN_mag}, but now for models r0e2, r0e2ea, r0e2eb, and r0e2ec, which differ in magnetar rotational energy (i.e., initial spin period).
    \label{fig_epm}
  }
\end{figure}

Figure~\ref{fig_ekin_spec} shows the spectral evolution for models r0e1, r0e2, and r0e4. With higher ejecta kinetic energy, the SED becomes bluer earlier, the lines are broader and weaker, and the appearance of forbidden lines occurs earlier (e.g., O\,\two\,7324.3\,\AA\ and O\three\ at 4363.2, 4958.9, 5006.8\,\AA). While the ionization conditions are similar around maximum (dominance of permitted transitions from C\two\, and O\two), the conditions at late times are different. Model r0e4 shows nearly exclusively lines of O\two\ and O\three\ while model r0e1 shows lines primarily from neutral and once-ionized elements (e.g.,  O\one, Si\two, and a forest of Fe\two\ lines). Figure~\ref{fig_ekin_spec_240d} illustrates this contrast for models r0e1 and r0e4 at about 240\,d after explosion.

Hence, for the same magnetar properties, the nebular phase spectra can drastically differ. The origin of this feature is that in a higher energy explosion, the inner ejecta is less dense (the density profile is flatter, with more mass at large velocity) so for a given power one obtains a greater temperature and ionization at nebular times. This is exacerbated by the fact that the deposition profile is broader in model r0e4, leading to very broad O\two\ and O\three\ lines. In model r0e1, the lines are from neutral or once-ionized species like O\one\ or Fe\two\ and they are narrower.

\section{Influence of magnetar rotation energy}
\label{sect_epm}

This section explores the impact of varying the magnetar initial rotational energy (or spin) for the same ejecta model. Using model r0e2 as a reference, the value $E_{\rm pm}$ is  increased from 0.4 (r0e2) to 0.8 (r0e2ea), 1.2 (r0e2eb), and $5.0 \times 10^{51}$\,erg (r0e2ec; in this last model, $B_{\rm pm}$ is decreased from 3.5 to $2.0 \times 10^{14}$\,G).  The change in initial spin also affects the spin-down timescale since it scales as $1/\omega_{\rm pm}^2$ (or $1/E_{\rm pm}$). Hence, $t_{\rm pm}$ drops from 19.1\,d in model r0e2, to 9.6, 6.4, and 4.7\,d (model r0e2ec, with the fastest initial spin). This causes a boost of about a factor of ten in peak bolometric luminosity (top panel of Fig.~\ref{fig_epm}), while the $V$-band maximum increases by less than 2 magnitudes. This results from the shift to bluer colors in the highest energy model (bottom panel of Fig.~\ref{fig_epm}). The rise time to maximum is essentially the same because all models have the same expansion time scales and diffusion times scales (here the magnetar does not affect the dynamics) --- this is modulated in part by the fact that a broader deposition profile is used in higher energy models).

\begin{figure}
   \includegraphics[width=\hsize]{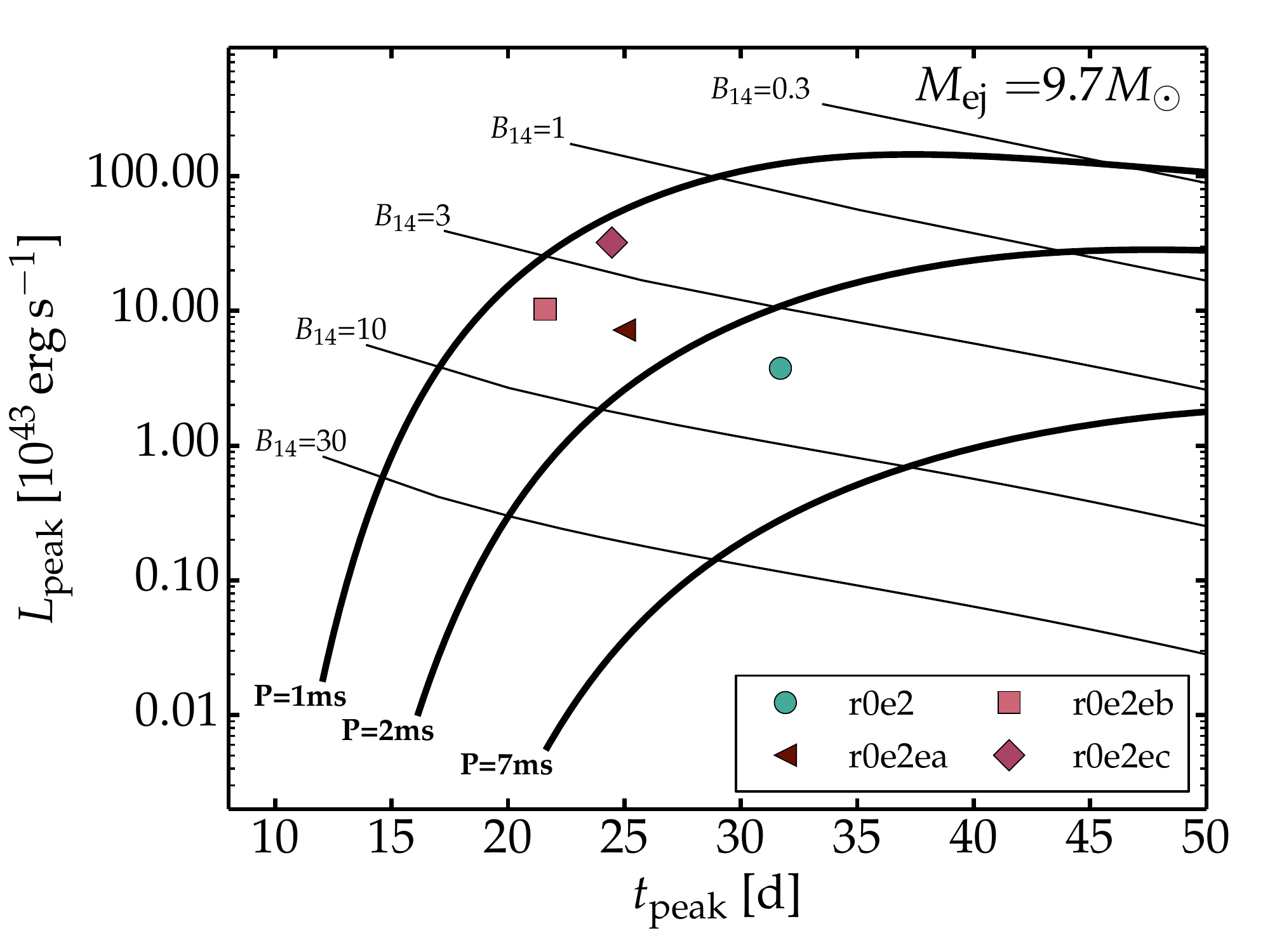}
\caption{Same as Fig.~\ref{fig_kb10}, but now adapted for the ejecta properties of model r0e2 ($M_{\rm ej}=$\,9.69\,\msun\ and $E_{\rm kin} = 4.12 \times 10^{51}$\,erg). Over-plotted are the results for models r0e2, r0e2ea, r0e2eb, and r0e2ec, which differ primarily in the initial magnetar spin period. The initial magnetar properties for the models are $B_{\rm pm}=3.5\times 10^{14}$\,G and $P_{\rm pm}=7.0$\,ms (r0e2),  $B_{\rm pm}=3.5\times 10^{14}$\,G and $P_{\rm pm}=5.0$\,ms (r0e2ea), $B_{\rm pm}=3.5\times 10^{14}$\,G and $P_{\rm pm}=4.1$\,ms (r0e2eb), $B_{\rm pm}=2.0\times 10^{14}$\,G and $P_{\rm pm}=2.0$\,ms (r0e2ec). The value of the rise time is affected by the prescribed deposition profile (see Section~\ref{sect_edep}) and probably by the neglect of dynamical effects as well (more problematic in more energetic explosions; but see ~Appendix~\ref{sect_comp_code}).
\label{fig_kb10_epm}
}
\end{figure}

As shown in Fig.~\ref{fig_kb10_epm}, the rise time to maximum and the luminosity at maximum obtained here are in rough in agreement with the analytic predictions of \citet{KB10}, who use a one-zone model (compared to an extended deposition profile here) and neglect dynamics (although they account for the magnetar energy in estimating the mean ejecta velocity).

\begin{figure}
  \vspace{-0.32cm}
  \includegraphics[width=\hsize]{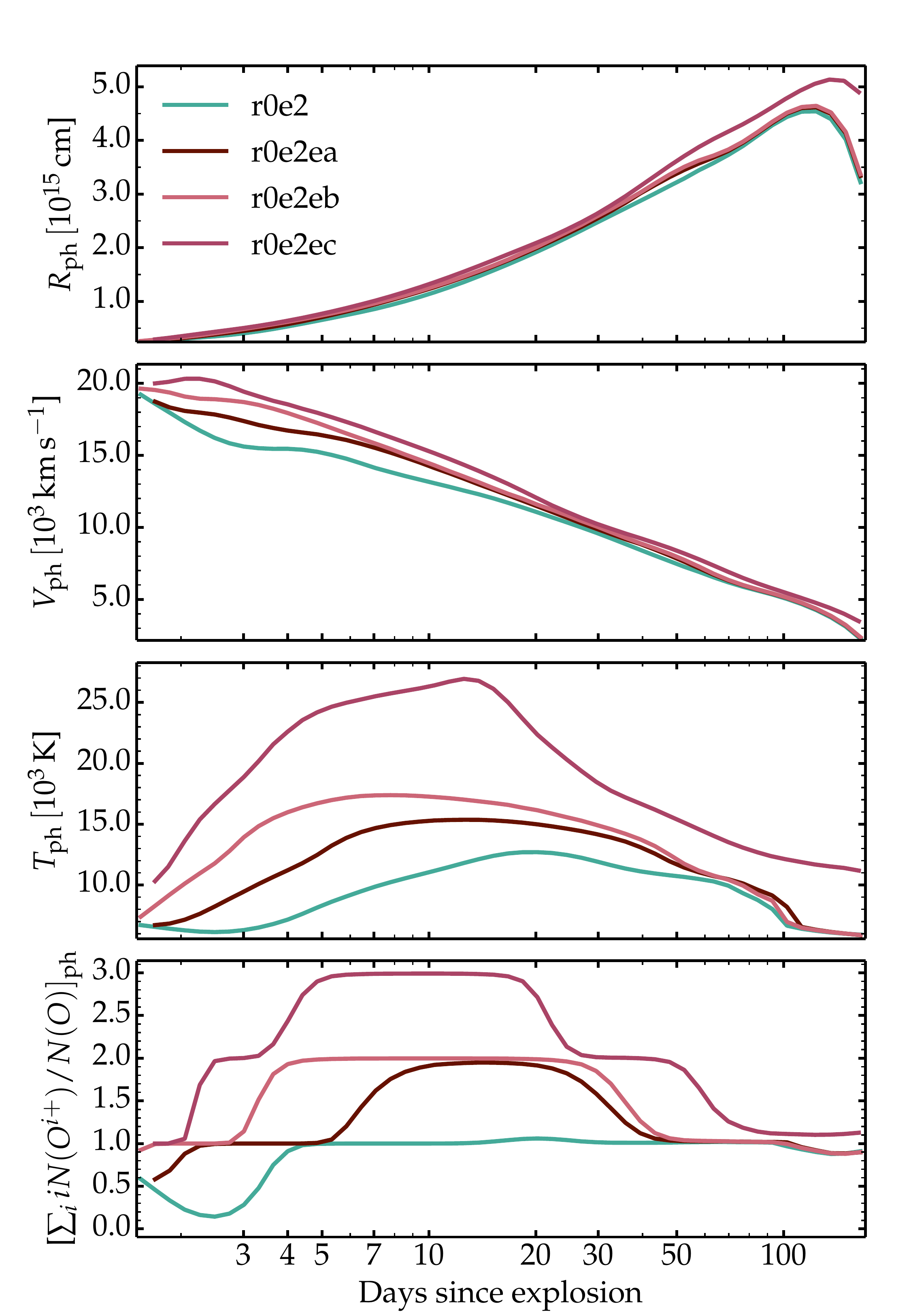}
  \caption{Same as Fig.~\ref{fig_YN_phot}, but now for models r0e2, r0e2ea, r0e2eb, and r0e2ec, which differ in magnetar rotational energy (i.e., initial spin period).
    \label{fig_epm_phot}
  }
\end{figure}

\begin{figure}
  \includegraphics[width=\hsize]{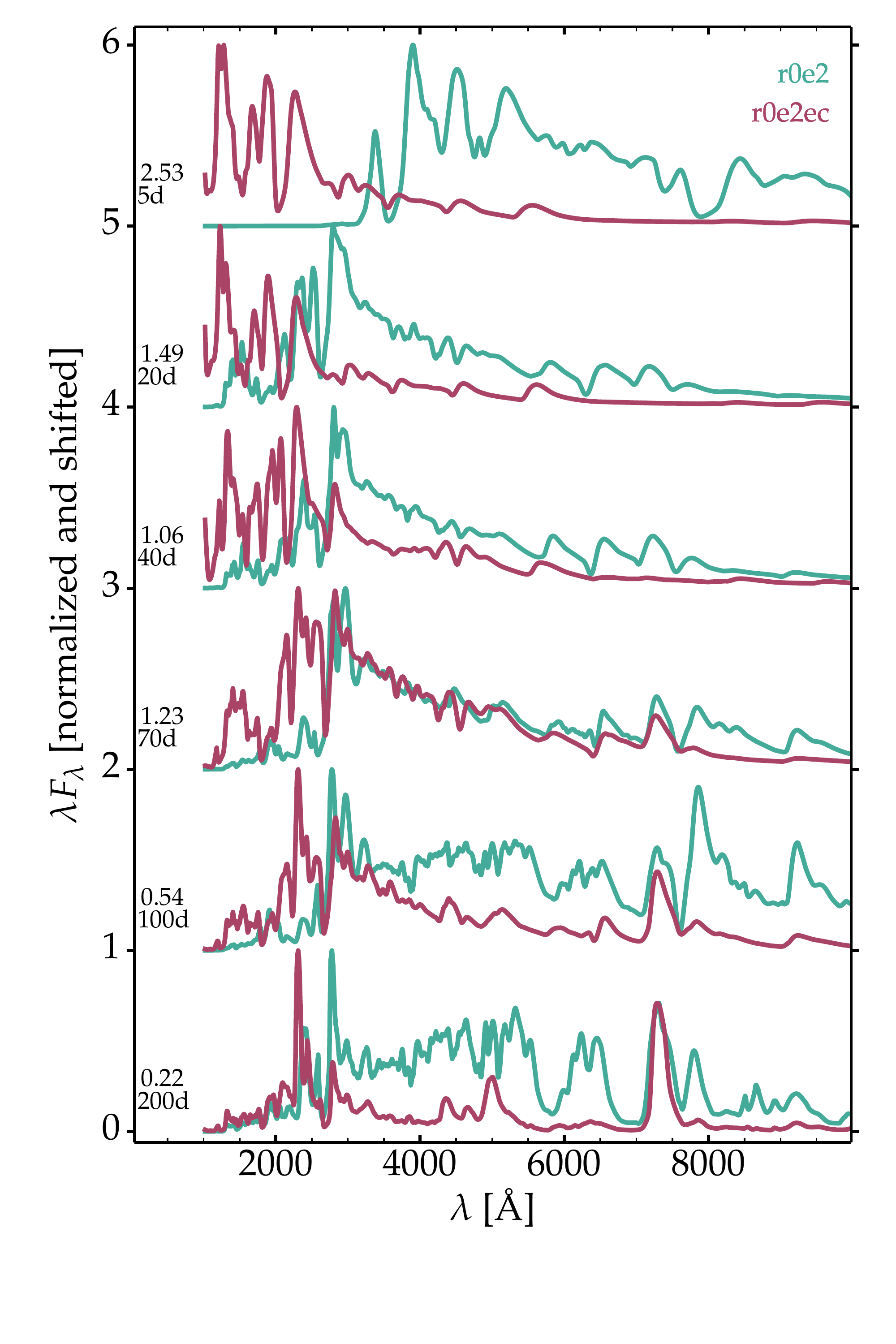}
  \includegraphics[width=\hsize]{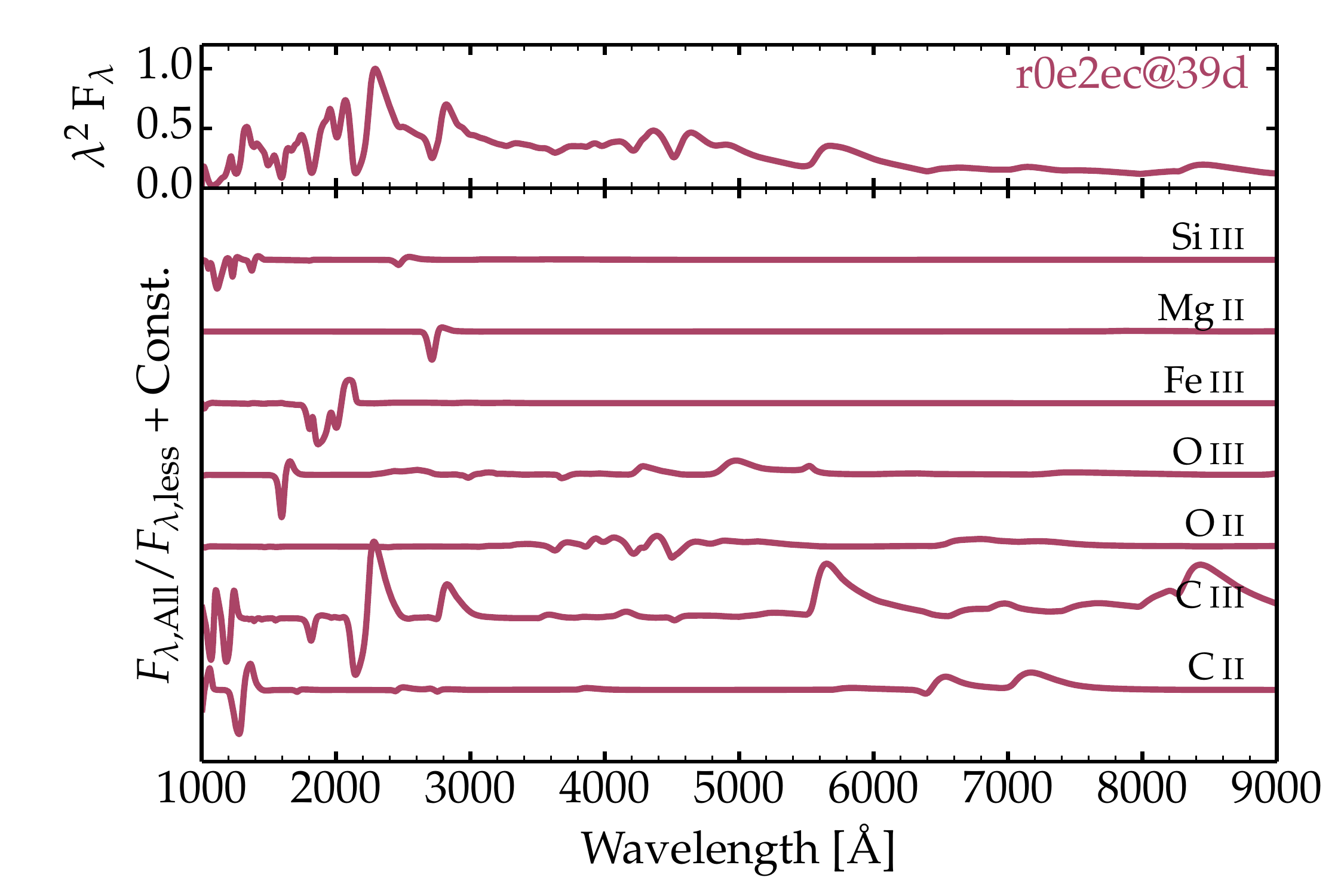}
  \caption{Top: Comparison of multi-epoch spectra for models r0e2 and r0e2ec, which differ in magnetar rotational energy (the label at left gives the $V$-band magnitude difference between the two models and the post explosion epoch). Bottom: Ladder plot for model r0e2ec at 39\,d after explosion (the upper part shows the quantity $\lambda^2 F_\lambda$).
    \label{fig_epm_spec}
  }
\end{figure}

The boost to the luminosity and brightness obtained with more energetic magnetars is associated here with a boost in the photospheric temperature (Fig.~\ref{fig_epm_phot}). Indeed, the photosphere is located in the same mass shells (i.e., same radius or velocity), but the temperature rises to a maximum of $\sim$\,12000\,K (at bolometric maximum) in model r0e2 and to $\sim$\,27000\,K (about 10\,d before bolometric maximum) in model r0e2ec. In model r0e2ec, the photospheric temperature is always greater than in model r0e2, by 2 to 10\,kK. This temperature contrast is associated with a change in ionization. While O is once ionized in model r0e2 throughout the simulation, it evolves from neutral to 3-times ionized on the way to maximum in model r0e2ec, before progressively dropping to being once-ionized at late times.

Models with different magnetar rotational energy show very different spectral properties, reflecting in part the contrast in temperature and ionization discussed above. The top panel of Fig.~\ref{fig_epm_spec} shows multi-epoch spectra (the quantity displayed is $\lambda F_\lambda$) for models r0e2 and r0e2ec (which differ in $E_{\rm pm}$ by a factor of 50). The higher energy model has an SED shifted to the blue, peaking in the UV at all times (the flux bias toward the UV region would look more extreme if showing $F_\lambda$). Because of this higher ionization, model r0e2ec shows lines of C\three\ (e.g.,  1175.6, 1247.4, 1894.3,  2296.9, 2844.1, 4650.2, 5695.9, and 8500.3\,\AA) and O\three\ (e.g., at 4363.2 and 5006.8\AA), in addition to the C\two\ and O\two\ lines that dominate the optical range in model r0e2 (see Section~\ref{sect_r0e2}). In the UV, apart from the strong C\three\ lines, Fe\three\ causes blanketing in the range $1700-2100$\,\AA. The bottom panel of Fig.~\ref{fig_epm_spec} illustrates the line contributions in model r0e2ec around the time of maximum.

\begin{figure}
   \includegraphics[width=\hsize]{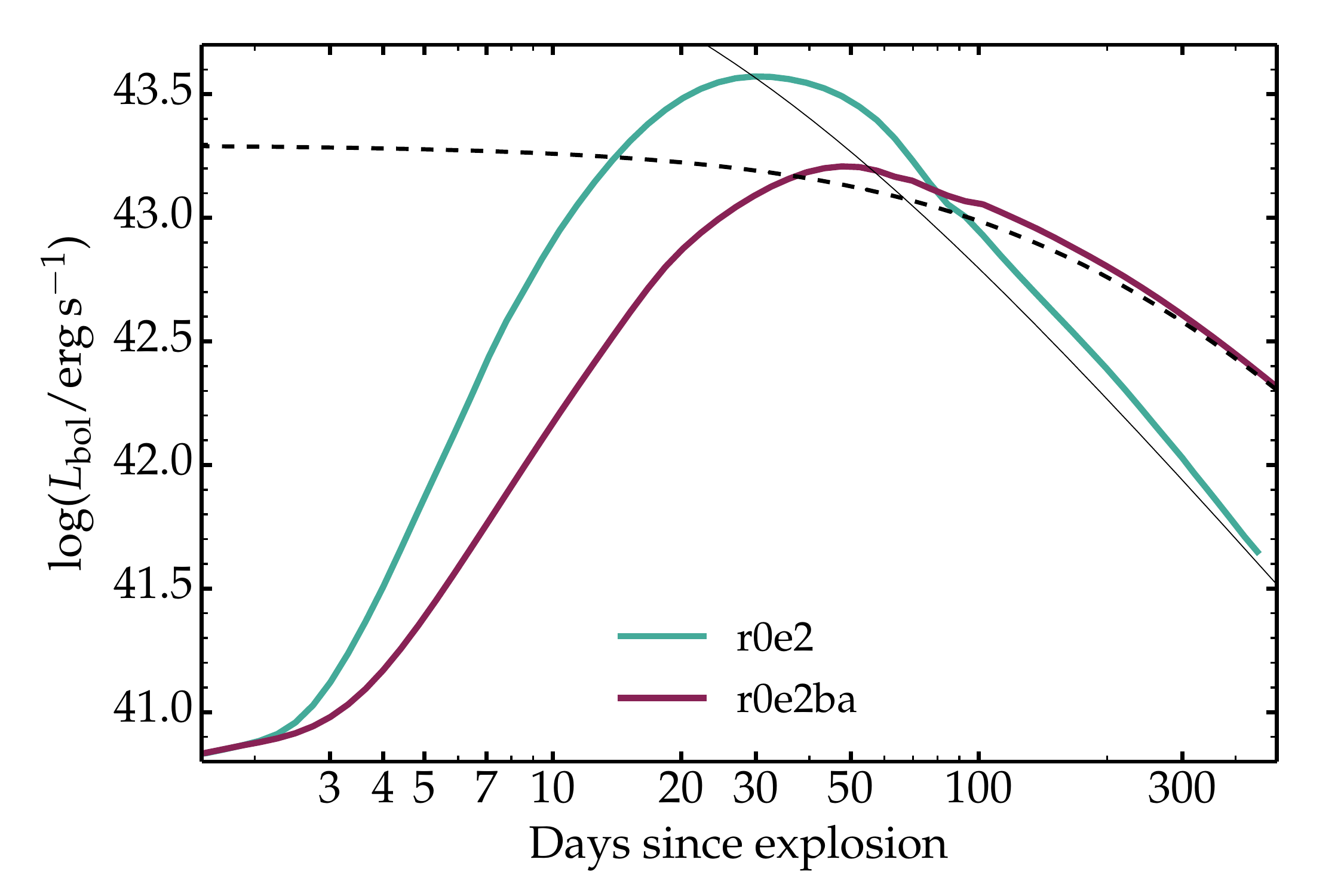}
\caption{Bolometric light curves for models r0e2 and r0e2ba, which differ only in magnetar field strength. The thin solid and dashed lines correspond to the instantaneous magnetar power absorbed by each ejecta. The contribution from 0.13\,\msun\ of \nifs\ is in part responsible for the offset after 100\,d. This offset is small in the model r0e2ba due to the much greater magnetar power at late times.
\label{fig_bpm}
}
\end{figure}

\section{Influence of magnetar field strength}
\label{sect_bpm}

    Model r0e2ba explores the effect of a weaker magnetic field relative to model r0e2 ($10^{14}$\,G compared to $3.5 \times 10^{14}$\,G). This yields a greater magnetar spin-down timescale (which scales as $1/B_{\rm pm}^2$), increasing from 19.1 in model r0e2 to 234.4\,d in model r0e2ba. The magnetar power is thus smaller early on but greater after about 100\,d for the weaker field case. This causes the bolometric and visual (not shown) light curves to peak later and at a fainter maximum in model r0e2ba (Fig.~\ref{fig_bpm}), and the color shift to the blue is also delayed (not shown). Compared to model r0e2, the photosphere in model r0e2ba heats up later and becomes more ionized later, but then the greater power at late times maintains a higher temperature and a higher ionization for longer (not shown).

Spectroscopically (no figure provided), model r0e2ba shows a similar evolution to model r0e2s early on (because the influence of the magnetar is delayed due to the smaller power initially).  At late times, the enhanced power in model r0e2ba yields a highly ionized spectrum, similar to what is found in model r0e4 but with narrower lines. In model r0e4, the high ionization comes from the deposition of power in a lower density ejecta (because of the large explosion energy relative to model r0e2) while in model r0e2ba, the density is higher (the explosion energy is smaller, equal to that in model r0e2) but the magnetar power is greater.

 \citet{d18_iptf14hls} used a similar approach to that presented here, but for a blue-supergiant star explosion model, and showed results for a SN ejecta influenced by an even weaker magnetar field ($7.0 \times 10^{13}$\,G). That model predicts a sustained luminosity, optical brightness, and blue color for 2 years, as observed in iPTF14hls \citep{arcavi_iptf14hls}.  For these weaker magnetic fields, the longer spin down timescale implies that the dynamical effects are weaker and a smaller fraction of the magnetar energy is released at early times, making the present assumptions in \cmfgen\ more suitable.

\section{Influence of non-thermal processes}
\label{sect_nonth}

Model 5p11Bx2 is used to explore the influence of non-thermal effects on the SN ejecta and radiation. So, a counterpart is run (model 5p11Bx2th; the sequence covers from 3 to 100\,d only) in which all the magnetar and decay energy is treated as heat (i.e. non-thermal effects are ignored).

\begin{figure}
  \includegraphics[width=\hsize]{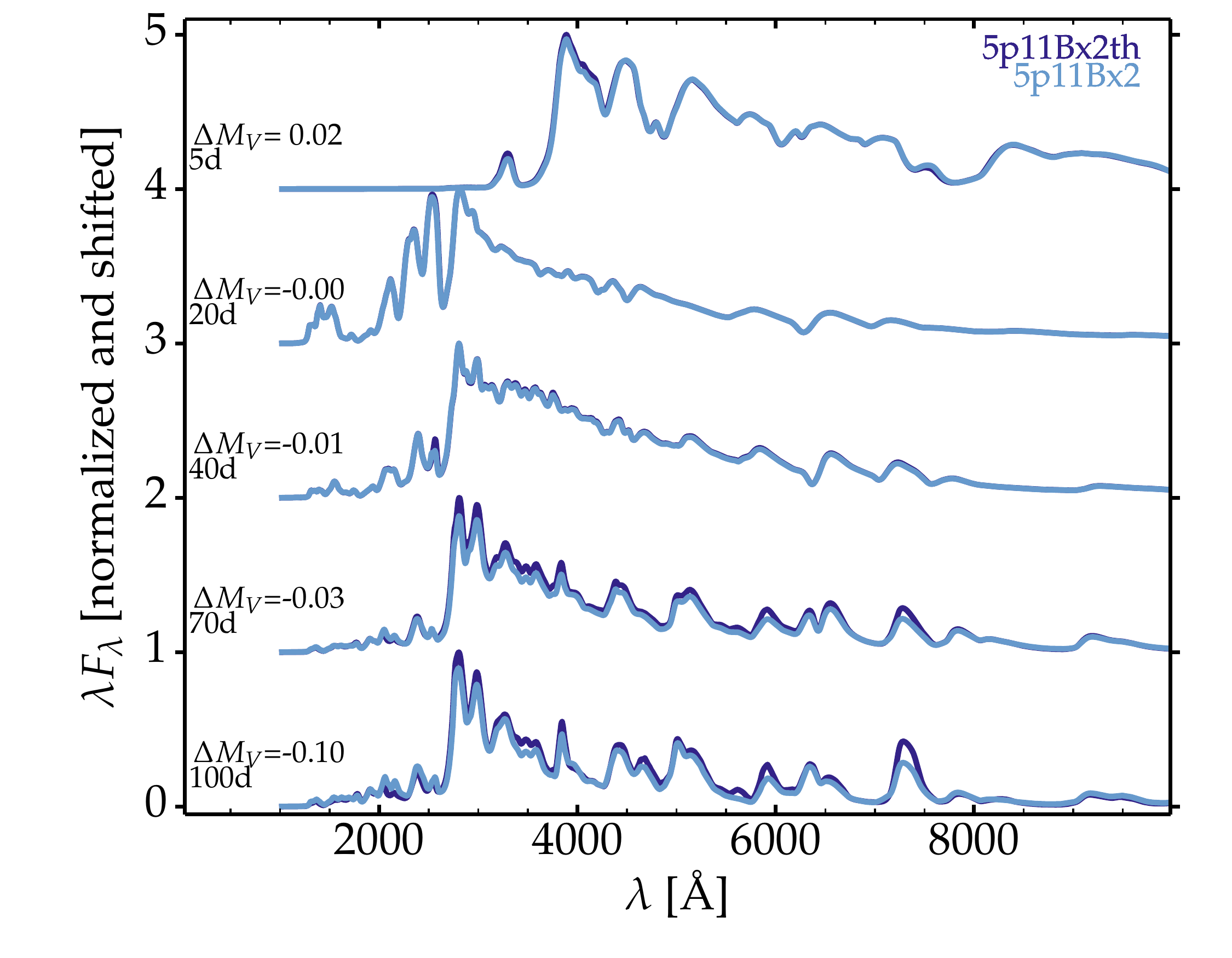}
  \caption{Comparison of multi-epoch spectra for model 5p11Bx2, which treats non-thermal processes, and model 5p11Bx2th, which ignores them (in this case, the decay and magnetar powers are treated as a thermal source). The quantity shown is $\lambda F_\lambda$.
    \label{fig_nth_th}
      }
\end{figure}

Figure~\ref{fig_nth_th} shows multi-epoch spectra for the two model sequences at 5, 20, 40, 70, and 100\,d. As can be seen, the impact of non-thermal processes throughout the photospheric phase (i.e., the high brightness phase) is negligible. A slight difference is visible at 100\,d in the O\two\ feature at  about 7324.3\,\AA\ (composed of several lines), and arises from the higher ionization in the non-thermal model (a mix of O$^+$ and O$^{2+}$ is present at the photosphere), while the thermal model is less ionized (O$^+$ dominates at the photosphere). The photospheric properties (not shown) are essentially identical in both models, with only a slight over-ionization in the non-thermal model after bolometric maximum. At later times, the offset between the two models grows.

In model 5p11Bx2, all elements are at least once ionized throughout the ejecta at all times so the electron density is very large. This is known to quench non-thermal effects. Indeed, \citet{KF92} find that as the ionization level increases, a growing fraction of the decay energy is channeled into heat (i.e., heating of thermal electrons), largely irrespective of the composition. In \cmfgen\ simulations of Type II SN\,1987A, \citet{li_etal_12_nonte} find that in the recombined layers of the H-rich ejecta ($x_{\rm e}\sim$\,0.01, where $x_{\rm e}$ is the ratio of the total ion population and the total element [atom and ions] population), 20\% of the decay energy goes into heat, 40\% goes into excitation, and 40\% into ionization of the elements. In the partial ionization conditions of Type Ibc photosphere  around bolometric maximum ($x_{\rm e}\sim$\,0.3), \citet{d12_snibc} find that 80\% of the decay energy goes into heat, and 10\% goes into excitation and 10\% into ionization of the elements. In the magnetar-powered model 5p11Bx2 at 30\,d (close to bolometric maximum), $x_{\rm e}$ is 0.95 at the photosphere and drops to about 0.6 within a few 1000\,\kms\ above it. In this spectrum formation region, 85 to 91\% of the magnetar power is channeled into heat, with the remaining $\sim$\,10\% are channeled equally between non-thermal excitation and ionization. With the increase in ionization and temperature, the radiation field is also much stronger in the UV (the SED peaks in the UV in SLSNe Ic), allowing stronger photo-ionization than typically possible in cool plasmas like standard SNe Ibc photospheres. In early-time Type II (or IIb) SNe, emission from lines of elements in a relatively high ionization state have been observed [O\five\ and O\six\ in SN\,2013fs \citep{yaron_13fs_17};  N\three\ and N\four\ in SN\,2013cu \citep{galyam_13cu_14}; He\two\ in SN\,2006bp \citep{quimby_06bp_07}]. Rather than non-thermal processes, these emission lines are caused by the high temperature in the spectrum formation region \citep{dessart_05cs_06bp,grafener_vink_13cu_16,groh_13cu,d18_13fs}. The combination of a high ionization (which favors heating of thermal electrons) and the large gas temperature (which boosts the UV flux) explain why non-thermal processes have no visible impact on the spectra. This also confirms the earlier spectral simulations of SLSNe Ic by \citet{d12_magnetar}, which ignored non-thermal processes, but nonetheless reproduced the salient features of SLSNe Ic like PTF09atu \citep{quimby_slsnic_11}. Although this depends on the exact level of ionization of the gas, this seems in contradiction with \citet{mazzali_slsn_16}, who argue that non-thermal effects are important for the spectrum formation of superluminous SNe Ic, including for the formation of He\one\ lines. The present simulations, which solve explicitly for non-thermal processes, predict that ``thermal'' processes dominate (this does not mean that the conditions are in LTE).

He\one\ lines are present in the synthetic spectra computed here but generally limited to the transitions at 5875.7 and 10830.2\,\AA. In models 5p11Bx2 and r0e2, He\one\,5875.6\,\AA\ represents about 1/5 of the 5890\,\AA\ feature, whose strength is dominated instead by C\two. This (and other) C\two\ features are present for as long as He\one\,5875.6\,\AA, and the carbon contribution always dominates in the present models. In the near-IR, He\one\,10830.2\,\AA\ overlaps with O\one, O\two, and Mg\two\ lines. The helium contribution is weak early on, but it strengthens after maximum, and at late times non-thermal effects influence that line. The photospheric temperature of about 10000\,K during the high brightness phase is, however, the main reason for the presence of these lines at that time. The weakness of He\one\ lines is probably due to the low helium abundance in the outer ejecta ($5-10$\,\% in model r0e2, and 30\% in model 5p11Bx2, the rest being primarily carbon and oxygen). Given the large CO core in model r0e2, the helium is located too far out in the ejecta, yielding low line optical depths and consequently weak features. The progenitor CO core mass is a fundamental quantity controlling the production of a Type Ib or a Type Ic SN \citep{d12_snibc}.

\section{Influence of clumping}
\label{sect_cl}

This section presents an exploration of the effect of ejecta clumping. The treatment of clumping is identical to that in \citet{d18_fcl}. Clumping is treated through a volume-filling factor approach which does not affect the radial column density, hence it does not facilitate the escape of photons, whether of low energy (optical photons) or high energy ($\gamma$-ray photons). A value of $f_{\rm vol}$ equal to 1 corresponds to a smooth ejecta, and a value of 0.1 corresponds to 10\% of the volume being filled with material (the rest being vacuum). Whatever the level of clumping, the composition is considered homogeneous at a given velocity (i.e., chemical segregation allowing for clumps of distinct composition is not considered). In \citet{d18_fcl}, the main effect of clumping was to enhance the recombination efficiency of the gas at and above the photosphere, leading to a faster recession of the photosphere and a boost to the luminosity (in the context of a Type II-pec or a Type II-P SN). Here, clumping is studied for its influence on the ionization state of the gas. Indeed, \citet{jerkstrand_slsnic_17} found that a strong clumping was required to  explain the nebular-phase spectra of superluminous SNe. Specifically, a volume-filling factor of 0.1\% for the ONeMg rich material was required to reproduce, for example, the O\one\ line doublet at 6316.0\,\AA\ and quench the emission from O\two\ and O\three\ lines (i.e., shift the ionization from ionized oxygen to neutral oxygen).

The next section explores the impact of various levels of clumping in model r0e1, r0e2, and r0e4 at a single nebular-phase epoch (i.e., 240\,d after explosion). The subsequent section presents a time-dependent simulation for model r0e2cl, which is identical to model r0e2 but differs in that the ejecta is clumped uniformly with a 10\% volume-filling factor.

\begin{figure}
   \includegraphics[width=\hsize]{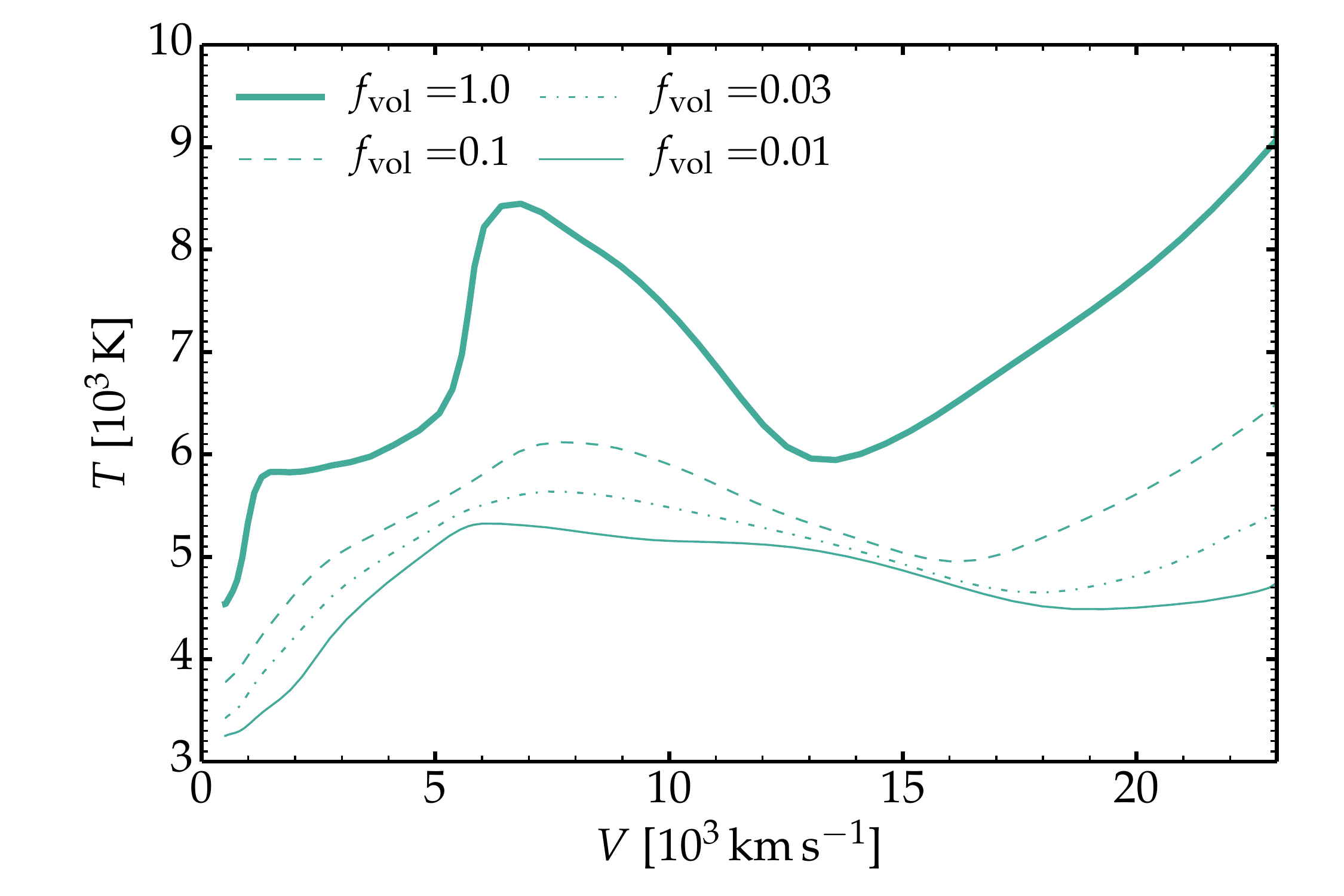}
   \includegraphics[width=\hsize]{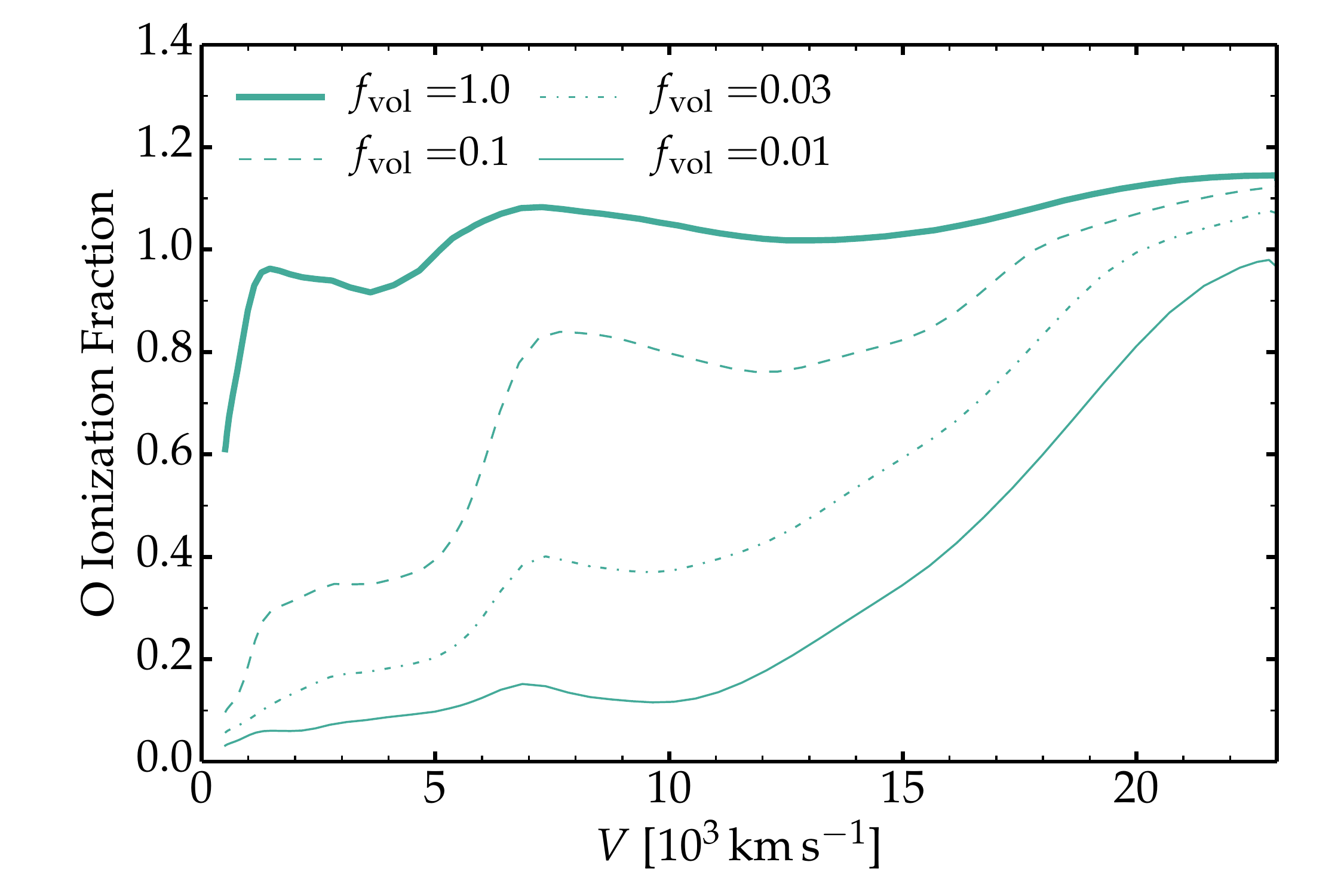}
\caption{
Illustration of the impact of clumping on the gas temperature (top) and the oxygen ionization (bottom) in the magnetar-powered SN model r0e2 at 240\,d after explosion, when conditions are nebular. Together with the smooth model ($f_{\rm vol}=$\,1.0) discussed in previous sections (and in particular in Section~\ref{sect_r0e2}), the figure shows the results for clumped models with $f_{\rm vol}$ of 0.1, 0.03, and 0.01. In each case, the adopted clumping is constant with depth in the ejecta. Models with higher clumping correspond to ejecta that are cooler and less ionized, and with optical spectra that are consequently redder and with lines from ions with a lower ionization (Fig.~\ref{fig_clumping_r0e2_57}).
\label{fig_clumping_phot}
}
\end{figure}

\begin{figure}
   \includegraphics[width=\hsize]{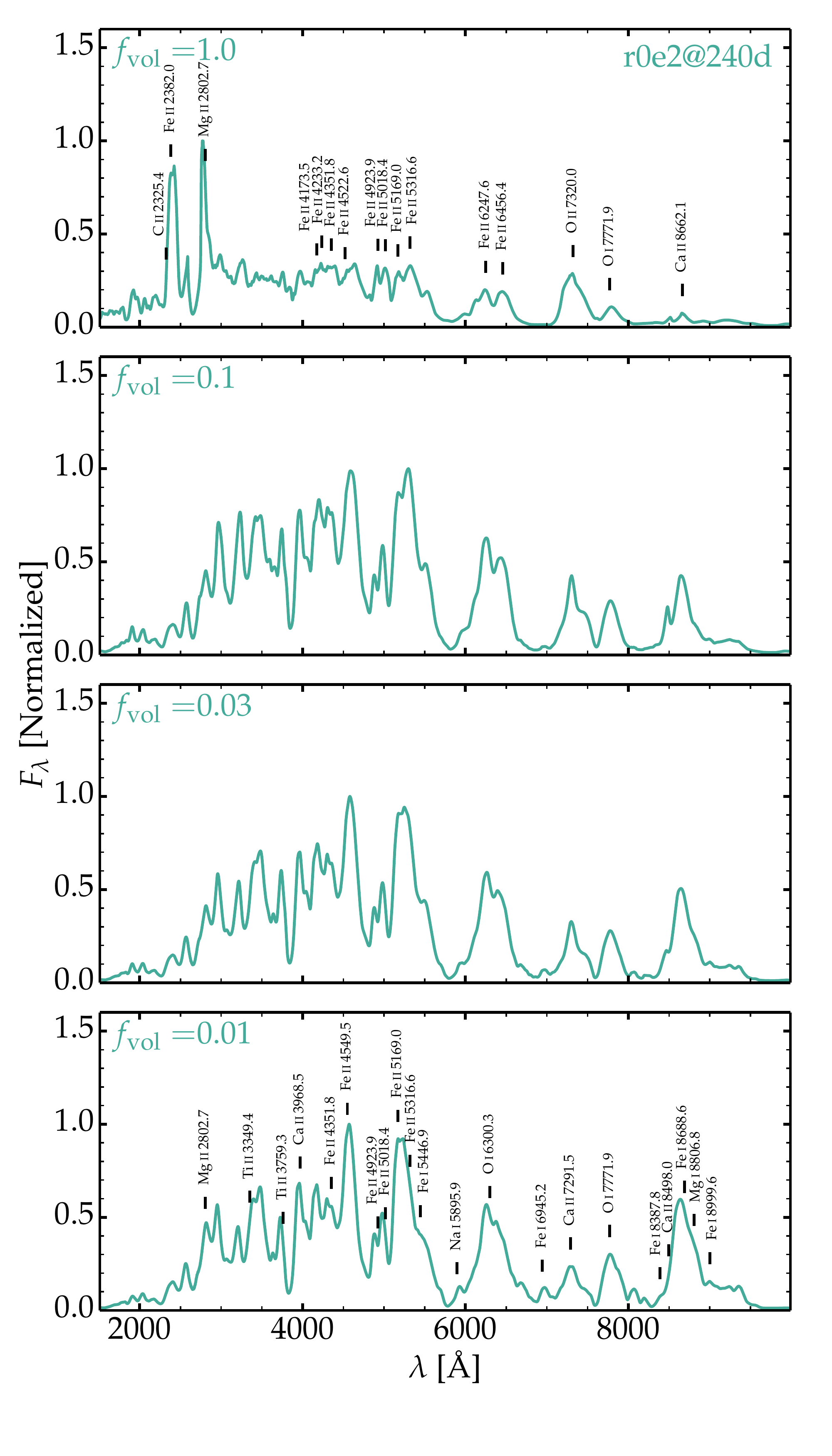}
       \vspace{-1cm}
\caption{Comparison of nebular-phase spectra for model r0e2 but adopting a volume filling factor $f_{\rm vol}$ of 1 (smooth ejecta), 0.1, 0.03, and 0.01. As the ejecta clumping is enhanced, the spectra become redder and exhibit lines from ions with a lower ionization state (see for example, the shift from O\two\,7324.3\,\AA\ to Ca\two\,7307.6\,\AA\, and from Fe\two\,6247.6\,\AA\ to O\one\,6316.0\,\AA; only the strongest transitions are labeled).
\label{fig_clumping_r0e2_57}
}
\end{figure}

\subsection{Comparison of models at nebular times for different clumping factors}

Figures~\ref{fig_clumping_phot} and \ref{fig_clumping_r0e2_57} show the impact of various levels of clumping ($f_{\rm vol}$ of 1, 0.1, 0.03, and 0.01) on the ejecta temperature and oxygen ionization, as well as on the spectrum, for model r0e2. For a greater clumping (smaller value of $f_{\rm vol}$), the gas is cooler and more recombined (oxygen is a good represent of the ionization state of the ejecta material since it dominates the composition).  The smooth model is more ionized and cools through emission in O\two\ or  Fe\two\ lines. The clumped models are more recombined (already for $f_{\rm vol}=0.1$), with a dominance of neutral oxygen over ionized oxygen. For example, the doublet line of O\one\ at 6316.0\,\AA\ is strong in the three clumped models, but absent in the smooth model, where the 6300\,\AA\ feature is due instead to Fe\two\ lines. A similar shift is seen for the near-IR Ca\two\ triplet, which is stronger with increasing clumping as the Ca ionization shifts to being primarily Ca$^+$. The SED is markedly altered in clumped models, which appear redder.

\begin{figure*}
   \includegraphics[width=0.5\hsize]{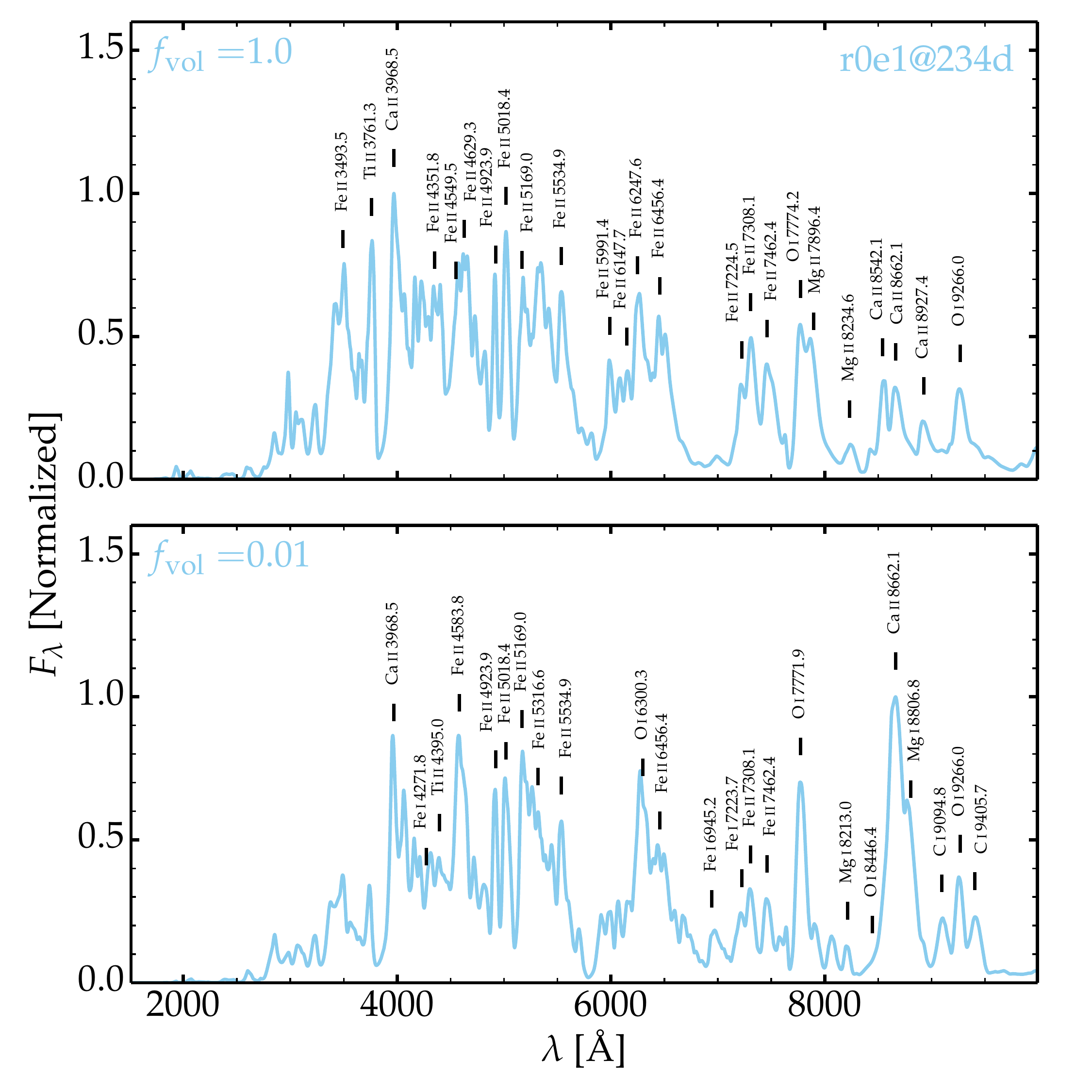}
   \includegraphics[width=0.5\hsize]{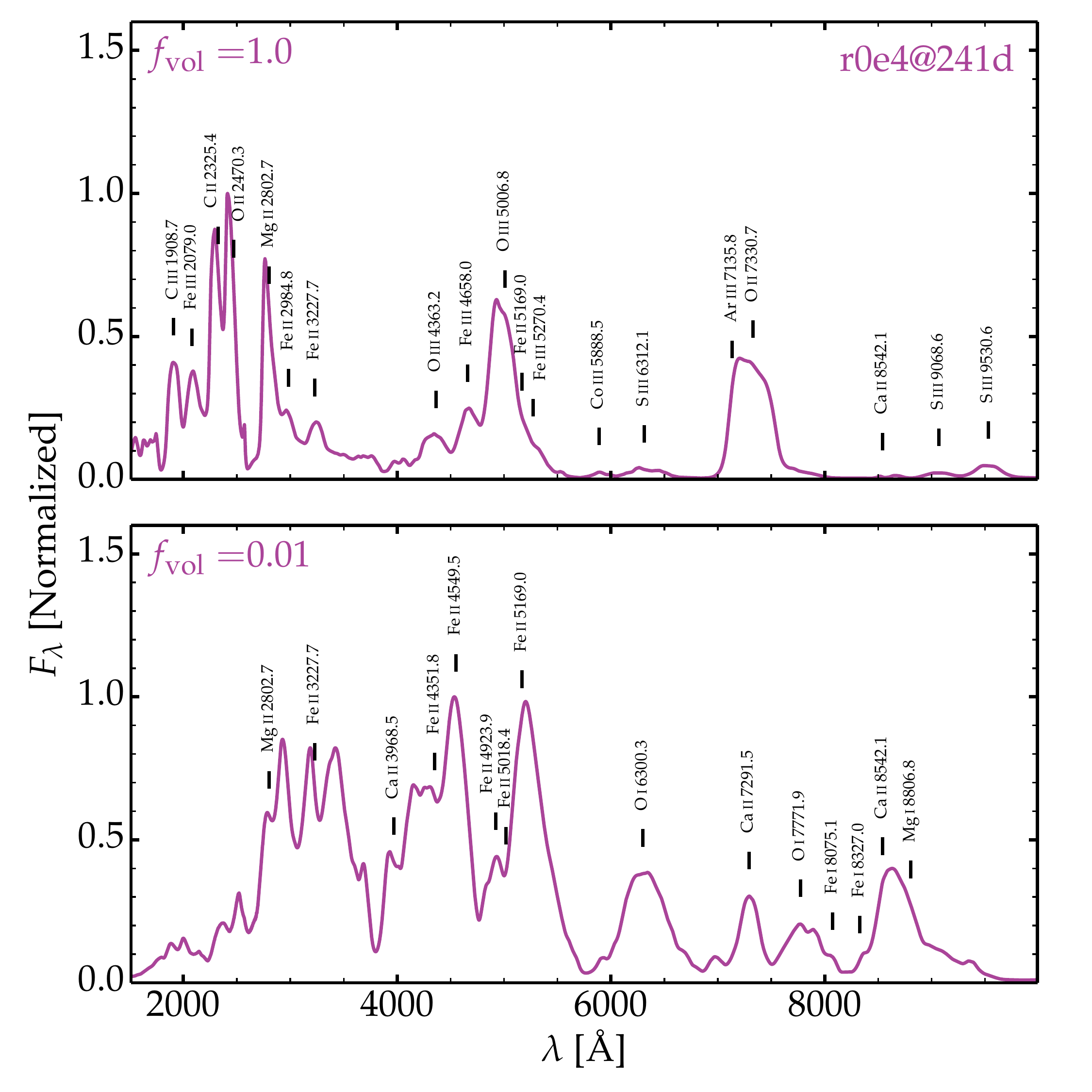}
\caption{Comparison of normalized spectra at about 240\,d after explosion for model r0e1 (left) and r0e4 (right) for two different levels of clumping ($f_{\rm vol}$ of 1 in the top row and 0.01 in the bottom row). Clumping has a weak impact on the  lower energy model (which has a denser inner ejecta) but a drastic influence on the higher energy model (which has a lower-density inner ejecta), associated with a strong reduction in ionization.
\label{fig_clumping_r0e1_r0e4}
}
\end{figure*}

\begin{figure}
  \includegraphics[width=\hsize]{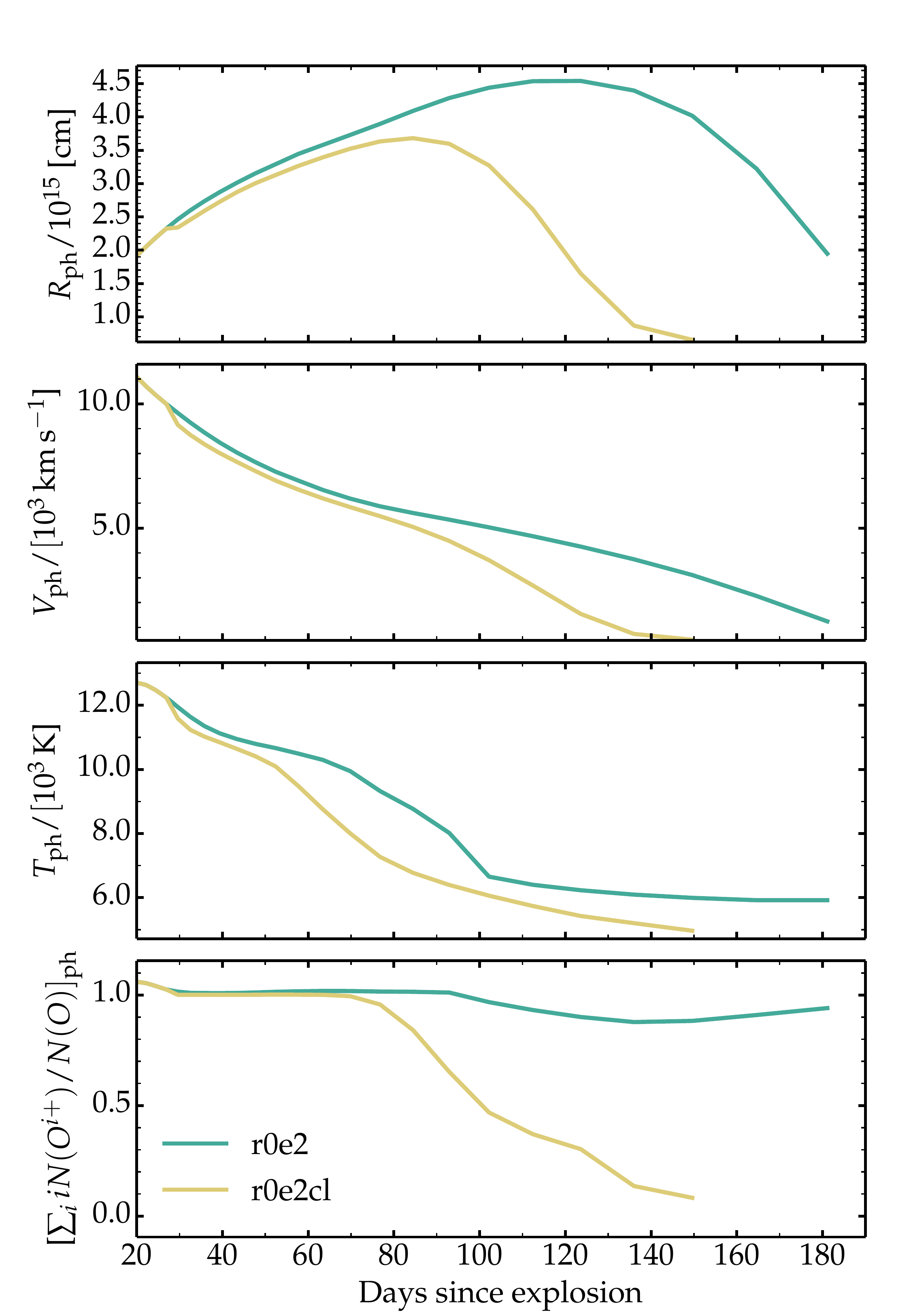}
  \caption{Same as Fig.~\ref{fig_YN_phot}, but now comparing the photospheric properties for models r0e2 (smooth ejecta) and r0e2cl (same as model r0e2 but the ejecta is clumped, with a 10\% volume filling factor). The clumped model is initially similar to the smooth counterpart, but as time progresses after bolometric maximum (which occurs in both at about 30\,d), the clumped ejecta cools and recombines faster, entering the nebular phase 150\,d after  explosion, which is 30\,d earlier than in model r0e2.
    \label{fig_sm_cl_phot}
  }
\end{figure}

\begin{figure}
  \includegraphics[width=\hsize]{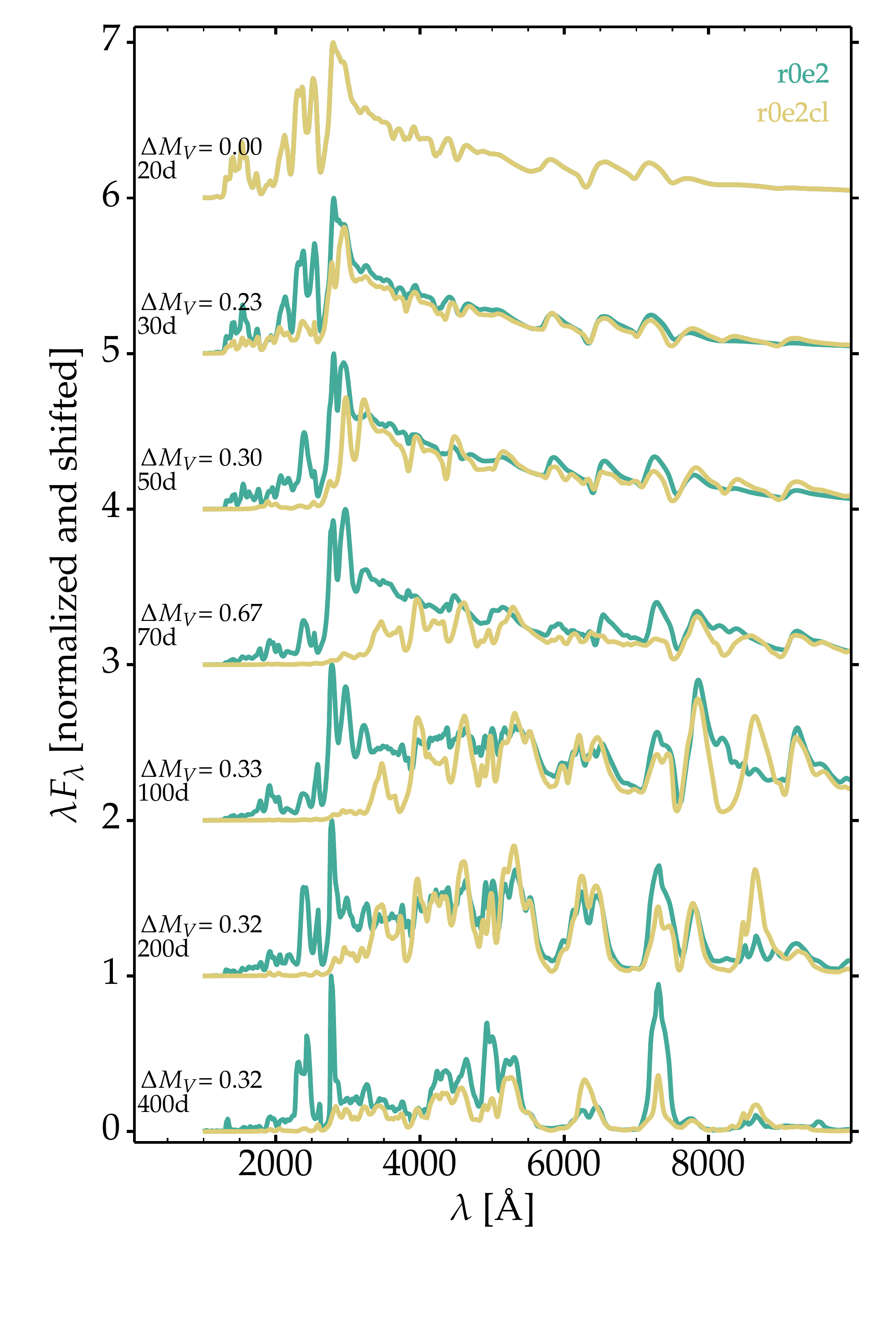}
  \includegraphics[width=\hsize]{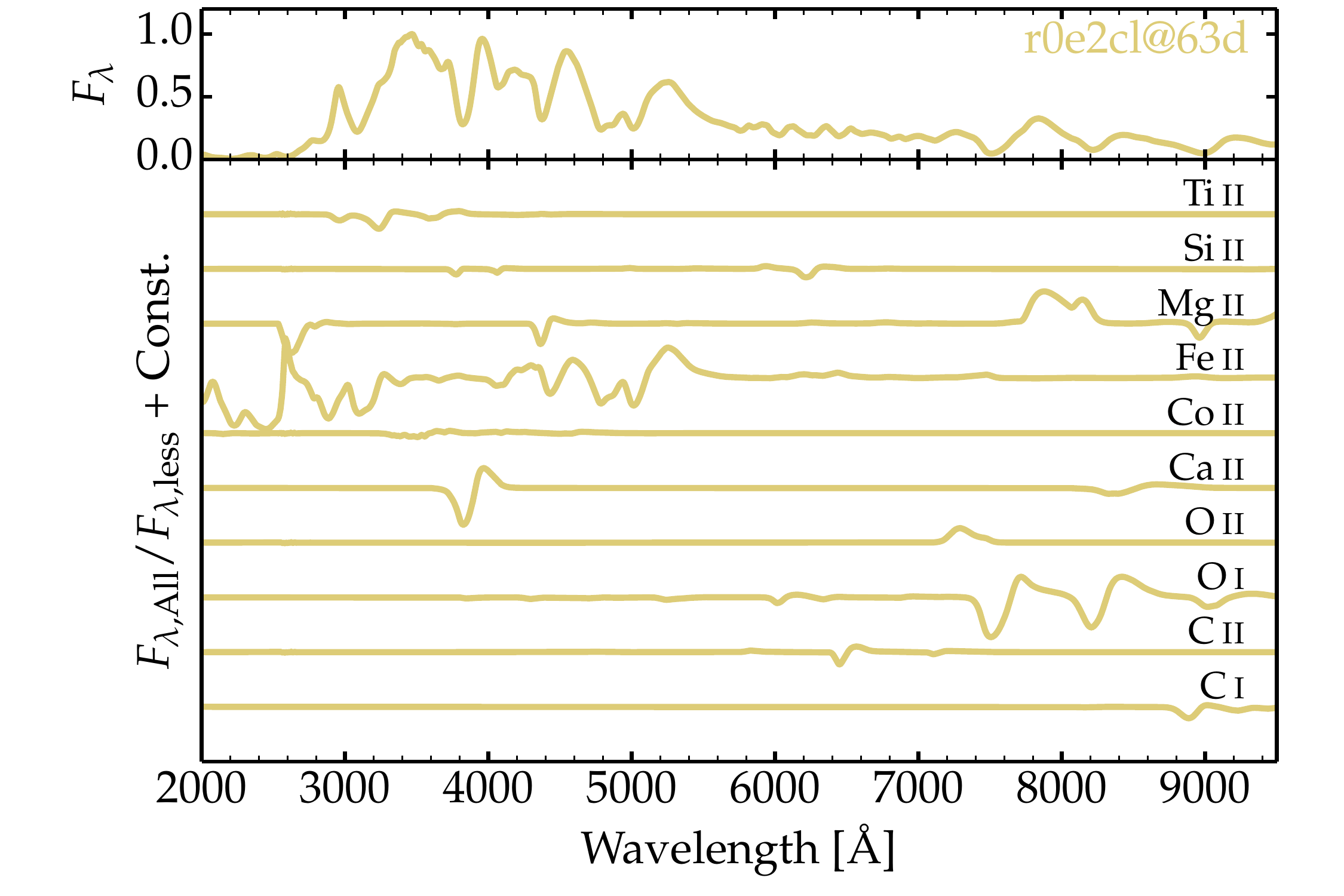}
  \caption{Top: Multi-epoch spectra for models r0e2 (smooth) and r0e2cl (clumped, with a uniform volume-filling factor of 10\%; the quantity $\lambda F_{\lambda}$ is shown). With clumping, the spectra are much redder, with lines from ions of a lower ionization state (neutral or once ionized). Bottom: Illustration of the line contributions for the most important ions in model r0e2cl at 63\,d after explosion.
    \label{fig_sm_cl_spec}
  }
\end{figure}

In model r0e2, the effect of clumping is strong for $f_{\rm vol}=0.1$, and not much enhanced as the clumping factor if further enhanced. The value of $f_{\rm vol}$ needed to cause the ionization shift depends on the SN age, the ejecta properties (primarily density, which depends on the progenitor structure, the explosion energy), and the magnetar properties (power, deposition profile).  \citet{jerkstrand_slsnic_17} performed one-zone models for various compositions and found that $f_{\rm vol}$ needed to be as low as 0.001 to yield a satisfactory correspondence to observed spectra of SLSNe Ic. This seems extreme since the O-rich material represents most of the ejecta mass (which would imply that most of the ejecta volume is essentially vacuum), but it is clear that clumping can have a profound impact on the resulting SN spectrum.

The impact of clumping depends on the ejecta properties. Indeed, in general, the ionization state of the gas for a given power depends on the density so one expects a different result for clumping in a magnetar power acting on a low or a high energy explosion. Figure~\ref{fig_clumping_r0e1_r0e4} shows the impact of clumping in models r0e1 and r0e4. Evidently, the impact is rather modest in model r0e1, because this lower energy model is characterized by a dense inner ejecta. The ejecta ionization is rather low even in the smooth case. Introducing clumping does not alter this much, although one sees that the O\one\,6316.0\,\AA\ line is stronger in the clumped model. In the higher energy model r0e4, the situation is completely different. In this case, the ionization of the smooth ejecta is very high at 240\,d (see also Section~\ref{sect_ekin}). In this case, clumping leads to a drastic change in ejecta ionization and spectral morphology. The strong and broad O\two\ and O\three\ lines present for $f_{\rm vol}=1$ (top right panel of Fig.~\ref{fig_clumping_r0e1_r0e4}) are absent in the clumped model (bottom right panel of Fig.~\ref{fig_clumping_r0e1_r0e4}) with $f_{\rm vol}=0.01$, which shows lines of O\one\ (as well as Fe\two\ and Ca\two).

In model r0e4, if the ejecta is smooth, a large fraction of the flux is emitted in O\two\, 7323.4\,\AA, O\three\,4363.2, and O\three\,5006.8\,\AA. In the clumped ejecta model with $f_{\rm vol}=$\,0.01, the same amount of magnetar power is absorbed by the oxygen rich material (because the deposition profile is independent of clumping), but the oxygen line flux is much smaller. Instead, the bulk of the flux comes out in Fe\two\ lines. The reason for this feature is probably that the main O\one\ forbidden line at 6316.0\,\AA\ is a poor coolant compared to O\two\, 7323.4\,\AA\ or O\three\,5006.8\,\AA. The oscillator strengths are 10 to 100 times greater for these last two lines. So, if  O$^+$ and O$^{2+}$ are present, the oxygen-rich material cools primarily through O\two\ and O\three\ lines, but if neutral O dominates, the cooling here is then performed in a large part by iron lines. This vividly illustrates that for the same oxygen mass and the same power absorbed by oxygen rich material, very different line strengths of oxygen lines can result, strong if O\two\ and O\three\ lines are present, weak if O\one\ lines are present. In other words, the strength of oxygen lines is not a robust and direct measure of oxygen abundance. Allowing for chemical segregation might alter this behavior.

As discussed in Section~\ref{sect_edep}, the nebular-phase properties also depend on the deposition profile, but all else being the same, this experiment demonstrates that clumping alone can completely change the spectral appearance of lower density ejecta (characteristic of higher energy explosions) in the nebular phase. In higher density ejecta, clumping is less influential because the density is higher and of low ionization even if the ejecta is smooth. Clumping has a greater impact only on smooth ejecta that are ionized. Cool recombined ejecta are less sensitive.

\subsection{Evolution of the impact of clumping in a time sequence, from maximum light to late times}
\label{sect_cl_seq}

This section compares the evolution of two magnetar powered models that are identical, except that one is smooth (r0e2) and the other is clumped (r0e2cl). In practice, model r0e2cl starts off from model r0e2 at 29.6\,d, but with a uniform volume filling factor of 10\%. The time sequence for model r0e2cl is them computed as for r0e2, and until 400\,d after explosion. This model shows how a fixed clumping level impacts the ejecta and the radiation as time proceeds.

Figure~\ref{fig_sm_cl_phot} shows the evolution of the photospheric properties for models r0e2 and r0e2cl. When they start off, the two models are identical, by design. But as the clumped model evolves, the enhanced density boosts the recombination efficiency of the gas. Initially, the effect is quite weak so the photosphere in both models have similar properties (although $T_{\rm ph}$ is markedly lower in the clumped model early on). But after 80\,d, the ionization in the clumped model suddenly deviates from the smooth model, meaning that the gas recombines faster and the photosphere recedes in the ejecta (at earlier times, $T_{\rm ph}$ was dropping but not enough to cause a change in ionization). The entrance to the nebular phase occurs 30\,d earlier in the clumped model as a result of the lower ionization of the gas. As shown here, oxygen, which is the dominant element in the ejecta, is recombined at 150\,d but once ionized in the smooth model. Clumping has therefore a similar effect on the photosphere in this model as discussed for a Type II SN in \citet{d18_fcl}. However, in contrast to the Type II SN ejecta, the bolometric light curve for models r0e2 and r0e2cl are identical to within a few percent (thus not shown). The reason is that in the Type II SN case, clumping immediately impacted the ionization and speeded up the photosphere recession (and the release of stored energy) whereas here, the choice of clumping has very little impact on the ionization up until 80\,d after explosion. At that time, the ejecta has a small optical depth and the luminosity is close to the magnetar and decay power absorbed by the ejecta. A higher level of clumping would perhaps influence the ionization earlier on and impact the light curve, as obtained in \citet{d18_fcl}. Quantifying the magnitude of clumping versus depth in magnetar powered SNe is left to future work.

The contrast in photospheric properties is reflected in the different spectral evolution of models r0e2 and r0e2cl (Fig.~\ref{fig_sm_cl_spec}). The impact on the r0e2cl model spectrum is already visible at 30\,d after explosion: the spectral features are essentially the same but the UV flux is reduced. As time proceeds, the SED becomes redder and eventually, the enhanced recombination is directly visible through the shift, for example, of O\two\ lines to O\one\ lines. The bottom panel of Fig.~\ref{fig_sm_cl_spec} illustrates the line contributions for model r0e2cl at 63\,d after explosion, with the preponderance of O\one, Ca\two, or Fe\two\ rather than O\two\ and C\two\ in the smooth model r0e2 (see also Section~\ref{sect_r0e2}). At nebular times, the results for r0e2cl are similar to those discussed in the previous section and are therefore not repeated.

Although the above results are function of the adopted clumping level and profile, they demonstrate that in a magnetar-powered SN, clumping may have a week effect up to maximum, but soon after that start to influence the ionization level and the spectrum appearance. When clumping is inferred from nebular phase spectra, it implies that clumping affects the SN radiation already soon after maximum because the ejecta optical depth is not large around maximum -- the spectrum forms over the entire ejecta soon after maximum.

\begin{table}
\caption{Characteristics of the observed SLSNe Ic used in this paper, including the time of $r$-band maximum, the redshift, the distance, the reddening, and the reference from where these quantities and observational data were taken.
\label{tab_obs}
}
\begin{center}
\begin{tabular}{l@{\hspace{3mm}}c@{\hspace{3mm}}c@{\hspace{3mm}}
c@{\hspace{3mm}}c@{\hspace{3mm}}c@{\hspace{3mm}}
}
\hline
                             &   $t_{\rm max}(r)$    &   $z$     &     $d$       &       $E(B-V)$       &     Ref.   \\
                             &         [MJD]                 &              &    [Mpc]    &           [mag]        &                \\
\hline
SN\,2007bi       &    54160.0     & 0.1279  & 591.6        &         0.03        &   a \\
SN\,2010gx       &   55283.0     & 0.2297  & 1181.3      &          0.04       & b      \\
PTF12dam          &   56095.0   &    0.1073 &   462.8     &          0.0          & c\\
PTF12gty            &   56143.4  & 0.176     &  750.0      &           0.058     & d \\
LSQ14an             &   56595.0    & 0.1637   &   766.0     &           0.07      &  e,f \\
SN\,2015bn       & 57108.0  & 0.1136   & 513.0           &   0.0                  & g,f   \\
Gaia16apd         &  57554.0  & 0.1018  & 481.93  &  0.013 &  h \\
\hline
\end{tabular}
\end{center}
\flushleft
Notes: The references used are \citet[a]{galyam_07bi_09}, \citet[b]{pasto_10gx_10}, \citet[c]{nicholl_slsn_13}, \citet[d]{de_cia_slsn_ic_17}, \citet[e]{inserra_slsn_ic_17}, \citet[f]{jerkstrand_slsnic_17}, \citet[g]{nicholl_15bn_long_16}, and \citet[h]{yan_Gaia16apd_17}.
\end{table}

\section{Comparison to observations}
\label{sect_obs}

This section presents some comparison to observations. The next section first gives a preamble on the scope of these comparisons. Section~\ref{sect_data} summarizes the observational data used. Section~\ref{sect_comp_obs} presents a comparison of light curves and spectra to a few SLSNe Ic and in particular SN\,2007bi, SN\,2010gx, PTF12dam, PTF12gty, LSQ14an, SN\,2015bn, and Gaia16apd.

\begin{figure*}
 \begin{center}
 \includegraphics[width=0.9\hsize]{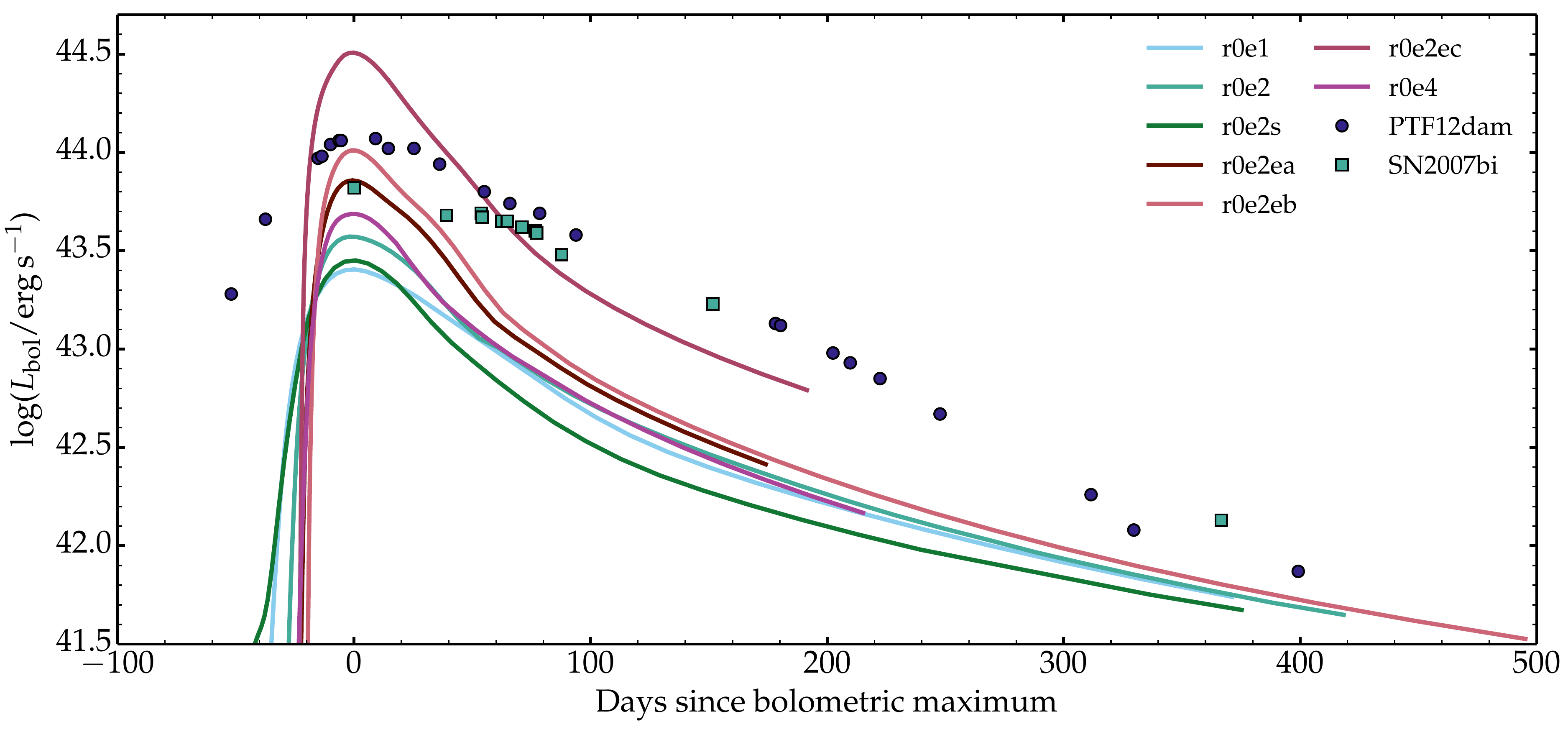}
  \includegraphics[width=0.9\hsize]{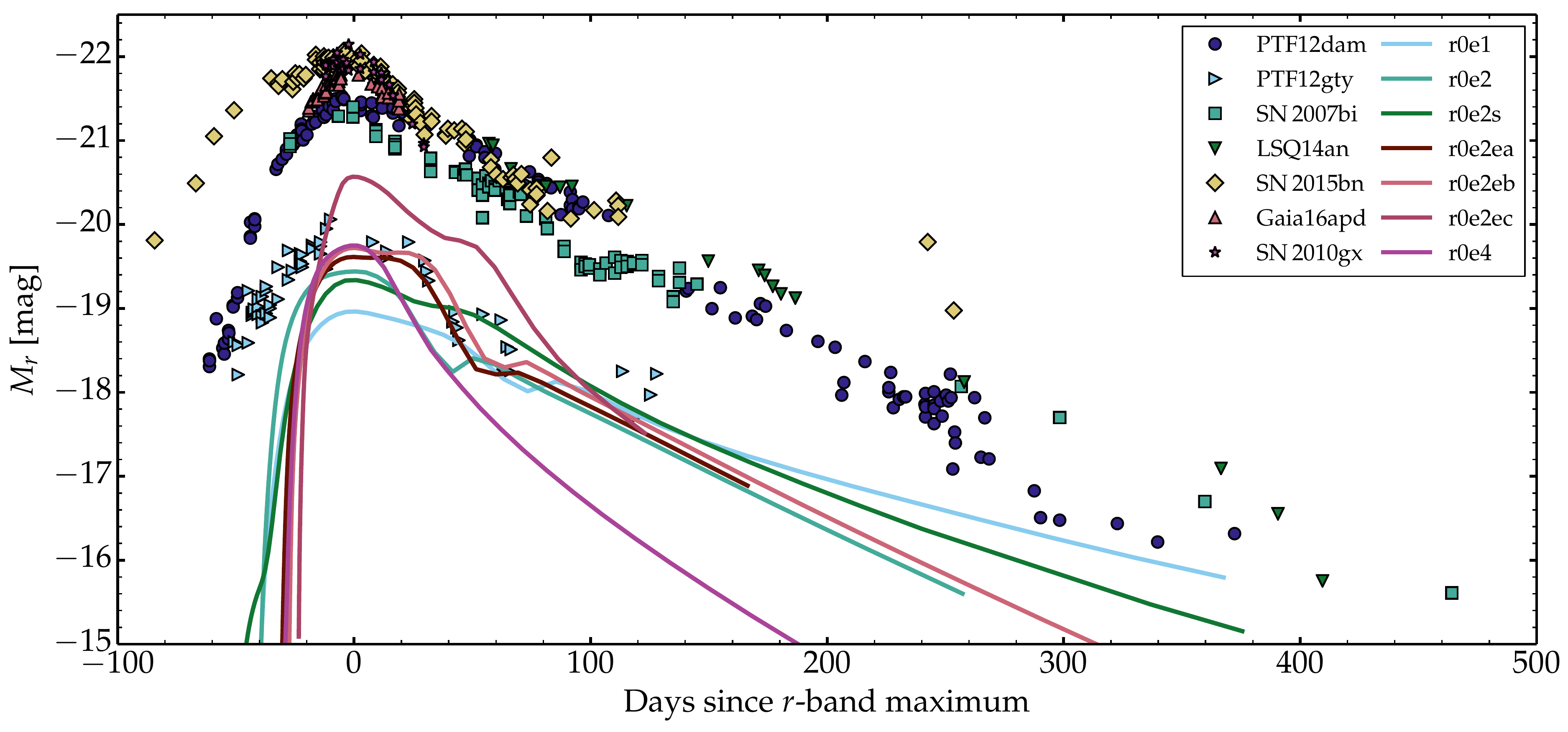}
\end{center}
  \caption{Top: Comparison of the inferred bolometric light curves of PTF12dam and SN\,2007bi \citep{chen_slsn_15} with a subset of models. Bottom: Same as top, but for the $r$-band light curve. A sizable offset between the top and bottom panels can arise if the bolometric correction is large (as for model r0e2ec). Time dilation, reddening, and distance are corrected for.
    \label{fig_comp_lc_obs}
  }
\end{figure*}

\subsection{Preamble}

The radiative properties of a magnetar-powered SN are controlled by numerous processes such as the magnetar field strength and its initial spin, or the ejecta mass and its kinetic energy \citep{KB10}. They also depend on how the magnetar power is deposited in the ejecta (Section~\ref{sect_edep}; see also \citealt{dessart_audit_18}), and whether the ejecta is clumped (Section~\ref{sect_cl}; see also \citealt{jerkstrand_slsnic_17}). Even in its simplest form, there are a handful of parameters that impact the resulting SN photometric and spectroscopic evolution.

The present approach ignores dynamics and assumes a spherical ejecta and magnetar-power deposition. Given the results in Section~\ref{sect_edep}, the model predictions on the rise to maximum (photometry, rise time, spectral evolution) are not robust. Similarly, little is known about the clumping, which implies that the color and spectral evolution after bolometric maximum is uncertain. At present, the treatment of clumping is 1D and neglects chemical segregation. The diversity in the properties of SLSNe Ic may come in part from asymmetry.  Overall, it seems that the present simulations cannot set strong constraints on the progenitor stars (mass and composition) or how they exploded (energy). But the models give a robust qualitative description of the influence of a magnetar on an H-deficient and He-poor ejecta. Further improvements await a better knowledge of the 3D hydrodynamic properties of magnetar-powered SNe.

\begin{figure*}
\begin{center}
   \includegraphics[width=0.9\hsize]{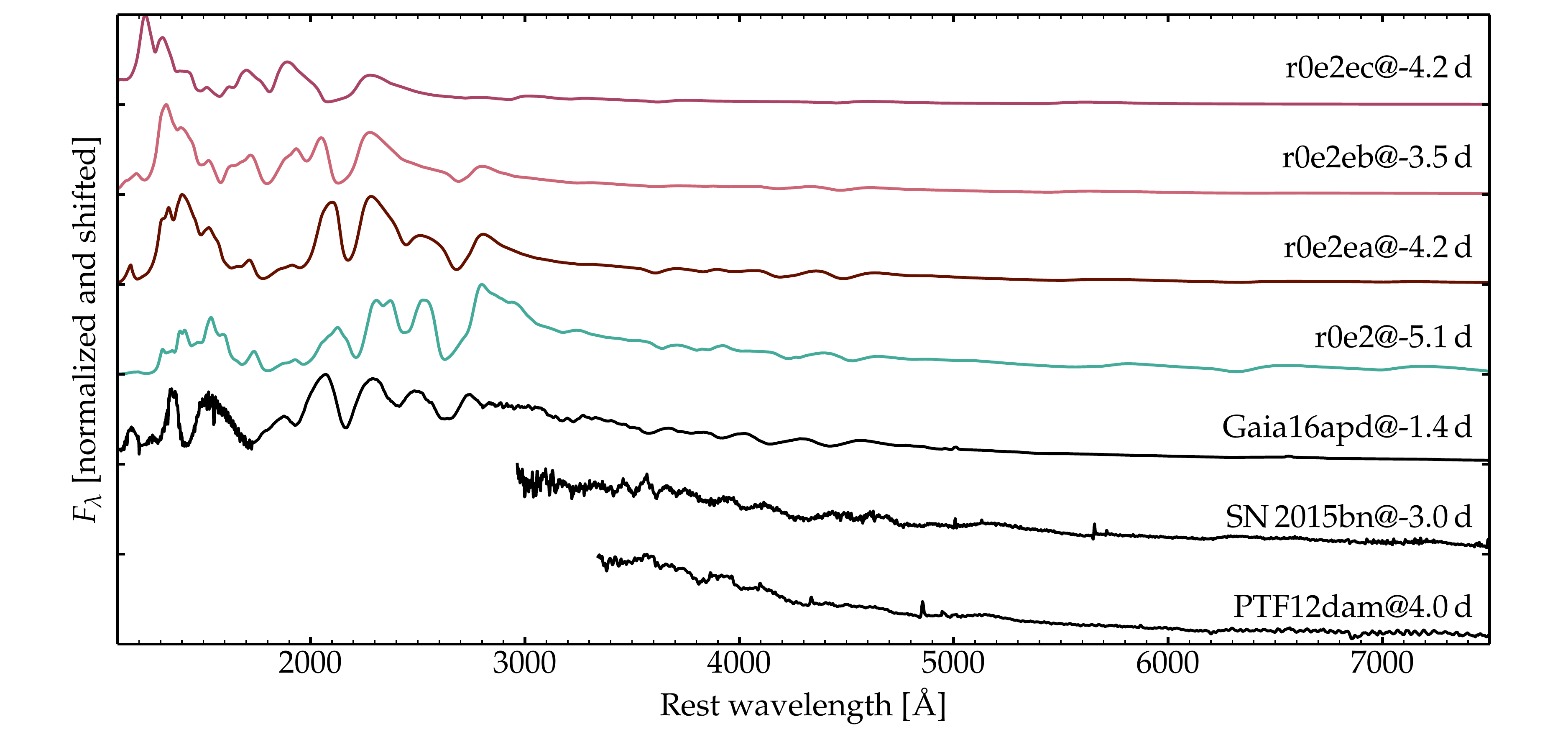}
   \includegraphics[width=0.9\hsize]{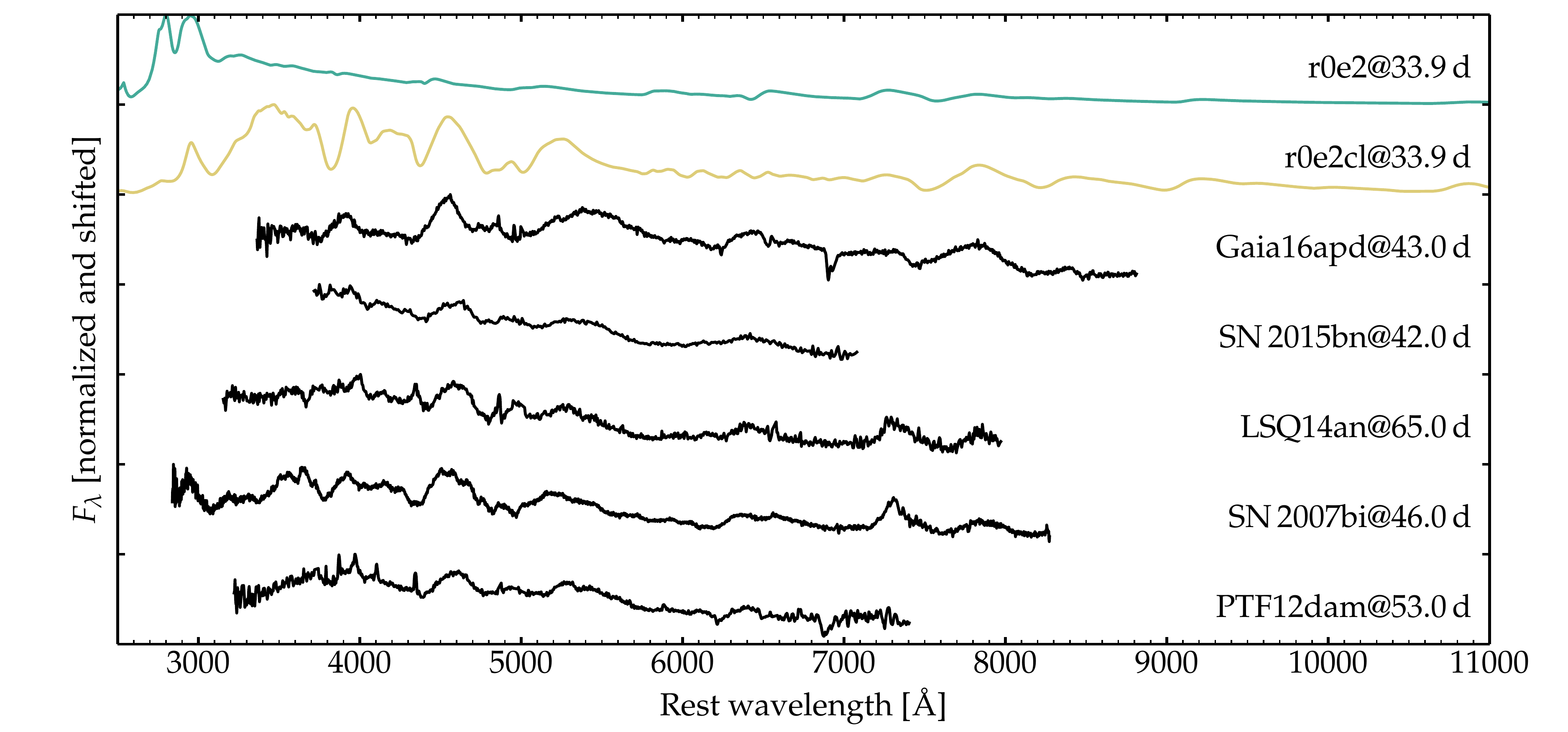}
   \end{center}
       \vspace{-0.5cm}
\caption{Montage of optical spectra (and UV when available) for a few observed SLSNe Ic and a subset of models. For Gaia16apd, the spectrum combines the UV and optical spectra, which were taken two days apart. Labels indicate the SN name and the rest-frame post-maximum epoch. The top panel corresponds to $1-4$\,d prior to maximum. The bottom panel corresponds to a phase about $35-65$\,d after maximum. At the time of maximum, most models reproduce the O\two\ and C\two\ lines typical of observed SLSNe Ic. However, a month later, only the clumped ejecta model r0e2cl reproduces the observations, which exhibit uniform spectral properties.
\label{fig_comp_spec_obs_photospheric}
}
\end{figure*}

\subsection{Dataset}
\label{sect_data}

There is now a vast collection of photometric and spectroscopic data for SLSNe Ic (see e.g., \citealt{quimby_slsnic_11}, \citealt{inserra_slsn_13}, \citealt{nicholl_slsn_13}, \citealt{de_cia_slsn_ic_17}, \citealt{lunnan_slsn_ic_18}). The sample selected for the comparisons shown in the next section is limited to SN\,2007bi, SN\,2010gx, PTF12dam, PTF12gty, LSQ14an, SN2015bn, and Gaia16apd. Comparing to more SNe does not add anything substantial, since there is a lot of degeneracy in the observations and because of the limitations of the numerical approach. A summary of the relevant characteristics for the SN sample is given in Table~\ref{tab_obs}. In practice, the photometric and spectroscopic data were retrieved from the SN catalog \citep{sn_catalog} at \href{https://sne.space}{https://sne.space}, except the maximum-light spectrum of Gaia16apd (Lin Yan, priv. comm.).

In addition, for SN\,2007bi, the photometric data of \citet{chen_slsn_15} is used to subtract the flux contamination from the host galaxy, as discussed in \citet{jerkstrand_slsnic_17}.The inferred bolometric light curves of SN\,2007bi and PTF12dam are from \citet{chen_slsn_15}. For LSQ14an,  the nebular phase spectrum (already corrected for host contamination) is taken from WISEREP \citep{wiserep}.

\subsection{Comparison of models to data}
\label{sect_comp_obs}

The top panel of Fig.~\ref{fig_comp_lc_obs} compares the inferred bolometric luminosity (with respect to the time of maximum) of a fraction of the model sample with the SLSNe Ic PTF12dam and SN\,2007bi \citep{chen_slsn_15}. This is done merely to show that the range of peak bolometric luminosities overlaps with that inferred from observations. However, in the model set, most of the light curves peak to a relatively faint maximum, exhibit a fast rise to maximum and a narrow light curve width. This results primarily from the large magnetic field used (1 to $3.5 \times 10^{14}$\,G) and the relatively small initial spins (most models have $P_{\rm ms}=$\,7.0\,ms). Only the models r0e2ea ($P_{\rm ms}=$\,5.0\,ms), r0e2eb ($P_{\rm ms}=$\,4.1\,ms), and r0e2ec ($P_{\rm ms}=$\,2.0\,ms) stretch to larger peak luminosities. In this paper, the goal was to understand the basic (non-dynamical) effects on SN ejecta and radiation for a given magnetar and investigate dependencies. The goal was not to match any specific SLSN Ic.

\begin{figure*}
   \includegraphics[width=0.5\hsize]{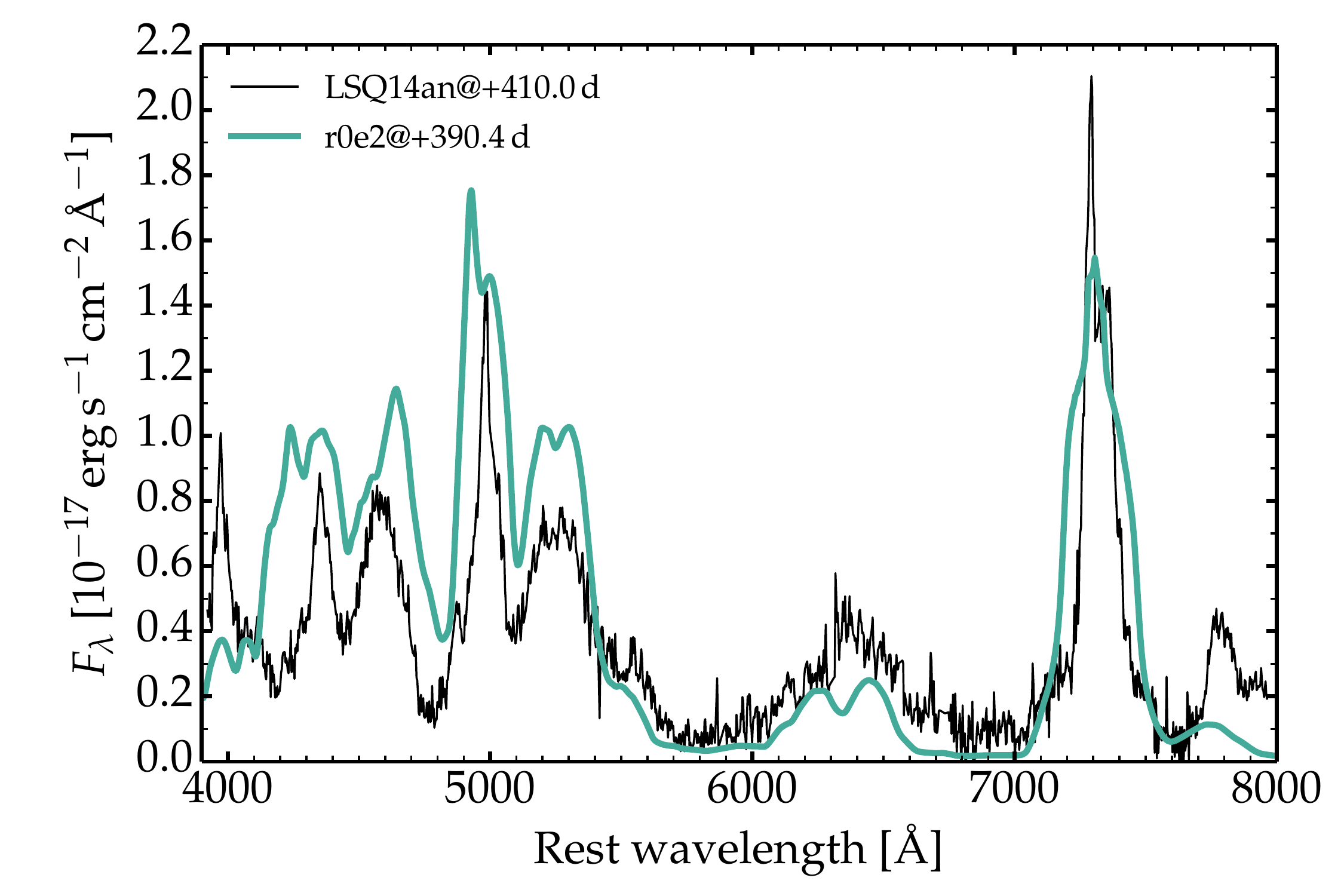}
   \includegraphics[width=0.5\hsize]{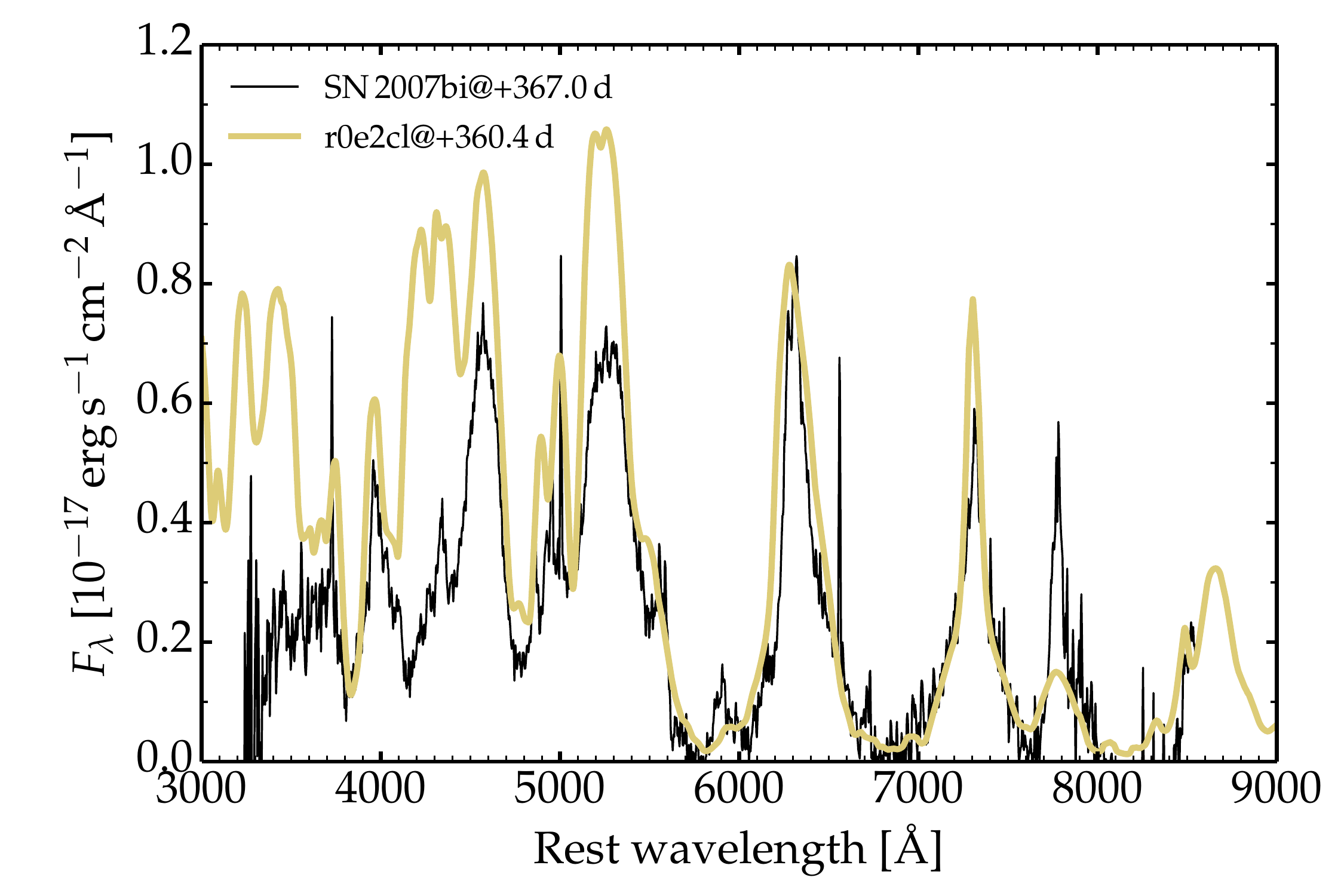}
\caption{Left: Spectral comparison of LSQ14an at +410.0\,d with model r0e2 at +390.4\,d. Right: Spectral comparison of SN\,2007bi at +367.0\,d with model r0e2cl at +360.6\,d (times are with respect to rest-frame days since maximum). The observations are corrected for redshift and reddening, as well as for the light contamination from the host galaxy (see Section~\ref{sect_data}). The models are normalized to the observations to facilitate the comparison.
\label{fig_comp_spec_obs_neb}
}
\end{figure*}

The bottom panel of Fig.~\ref{fig_comp_lc_obs} compares the $r$-band light curve (with respect to the time of $r$-band maximum) for a subset of models with a sample of observed SLSNe Ic. Models with faint bolometric luminosities exhibit a faint $r$-band maximum. But a faint $r$-band maximum is also obtained for the very energetic and luminous (in a bolometric sense) model r0e2ec. In this model, the bulk of the flux comes out in the UV, producing a relatively low brightness in the optical. The $r$-band or bolometric light curves can be modulated through changes in the magnetar properties. For example, PTF12dam is compatible with a faster spinning moderately magnetized magnetar, allowing a long rise time, broad light curve, and sustained brightness at late times (see e.g., \citealt{chen_slsn_15,de_cia_slsn_ic_17,nicholl_mosfit_mag_17}). However, these works propose a factor of 5 range in magnetic field, a factor of $2-3$ in ejecta mass, as well as invoke a leakage of magnetar power at late times.

Figure~\ref{fig_comp_spec_obs_photospheric} compares spectroscopic information for a few SLSNe Ic and a sample of our models. The top panel gives a comparison around maximum light (1 to 4 days prior to $r$-band maximum) using the observations of Gaia16apd, SN\,2015bn, and PTF12dam. The UV and optical ranges are shown, since HST data is available for Gaia16apd and the UV range is always computed by \cmfgen. The idea here is not to be quantitative. However, most of our model set, with the exception of the most energetic magnetars, predicts the O\two\ and C\two\ lines observed in SLSNe Ic (as revealed already in the first observations of these events; \citealt{pasto_10gx_10}; \citealt{quimby_slsnic_11}). Model r0e2ea is in good agreement with Gaia16apd from the UV to the optical, in the sense that the same line features appear in both spectra and with a location of the maximum absorptions at comparable wavelengths. As discussed earlier, the O\two\ lines are numerous and blended, while the C\two\ lines are only doublets and triplets and less affected by line overlap. The models tend to overestimate the observed strength of optical C\two\ lines, with the exception of SN\,2018bsz \citep{anderson_18bsz}.

The bottom panel of Fig.~\ref{fig_comp_spec_obs_photospheric} shows a spectral comparison at $1-2$ months after maximum. In models that assume a smooth ejecta, the spectrum is still very blue, with the persistence of O\two\ and C\two\ lines, as around the time of maximum. In contrast, the observations show a drastic evolution in spectral properties, with a dominance of lines from neutral and once-ionized species, in particular O\one, Ca\two, and Fe\two. The only model in the present set that reproduces these features is r0e2cl, which assumes a clumped ejecta (see Fig.~\ref{fig_sm_cl_spec} for the full spectral evolution and the line identifications at 63\,d after explosion).  The disagreement with models is more severe for lower-density ejecta (i.e., higher $E_{\rm kin}/M_{\rm ej}$, as in model r0e4) or faster-spinning magnetars (i.e., higher $E_{\rm pm}$, as in model r0e2ec). In model r0e1, the $E_{\rm kin}/M_{\rm ej}$ is the lowest in our set and the ionization is lower than in model r0e2 after maximum, but not as low as in the clumped model r0e2cl. Here, clumping is not a tuning parameter for line strength. It is instead a fundamental tuning process for the ejecta ionization. Indeed, with clumping (here with a uniform volume filling factor of 10\%), the ejecta ionization is low and suitable to produce most of the lines observed in the selected sample of observed SLSNe Ic (despite the diversity in their light curve properties, and hence in ejecta and magnetar properties). It thus appears that clumping is not just important at nebular times \citep{jerkstrand_slsnic_17,nicholl_slsn_neb_18}. Clumping is also influential early after maximum (see Section~\ref{sect_cl_seq}). Physically, this makes sense since the total ejecta electron-scattering optical depth $\tau_{\rm es}$ is usually only a few tens at maximum. For example, in model r0e2, $\tau_{\rm es}$ is 65 at maximum and only 7 at about 40\,d after maximum, so that the SN radiation already forms in the inner parts of the ejecta at that time. Thus, it is the same clumped inner ejecta that influences the SN radiation at nebular times and at early times after maximum. This clumping results from the dynamical influence of the magnetar, which operates most effectively early on, when the ejecta is compact, dense, and hot and the magnetar power  is at its greatest. The clumpy structure of the ejecta is probably frozen within at most a few days after explosion. These results also suggest that SN\,2007bi would have looked like, for example, PTF12dam, at maximum light, with a blue SED containing O\two\ and C\two, lines.

Figure~\ref{fig_comp_spec_obs_neb} compares the observations of LSQ\,14an and SN\,2007bi with the models r0e2 and r0e2cl at about one year after maximum light. LSQ14an shows the presence of O\three\,5006.8\,\AA\ \citep{inserra_slsn_ic_17}, while SN\,2007bi appears similar to GRB SN\,1998bw \citep{jerkstrand_slsnic_17}, with the presence of O\one\ and Ca\two\ lines. Model r0e2 exhibits lines that are somewhat too broad (as a result of an adopted deposition profile that may be too broad) but all features present in LSQ14an are predicted by the model.  The main contributions are from a forest of Fe\two\ lines up to about 5500\,\AA, followed by  more isolated Fe\two\ lines at 6247.6 or 6456.4\,\AA, and then contributions from oxygen with O\one\ at 7773.4\,\AA\ (the O\one\ doublet at 6316.0\,\AA\ is a very small contribution to the Fe\two\ blend),  O\two\,7324.3\,\AA, and O\three\,5006.8\,\AA. In LSQ14an, O\one\ at 7773.4\,\AA\ is underestimated, primarily because oxygen is once ionized (see also Section~\ref{sect_r0e2}). The O\three\ line is narrower and therefore forms over a smaller ejecta volume (smaller velocities) than in the model. Model r0e2 also predicts a strong Ca\two\ NIR triplet.

The clumped model r0e2cl predicts most of the features observed in SN\,2007bi at 367\,d after maximum. The contribution in the blue from Fe\two\ lines is somewhat overestimated, but important lines in the red part of the optical are well matched. The main contributions in the model at that time are O\one\,6300.0--6363.8\,\AA, Ca\two\,7307.6\,\AA\ (with overlapping but weaker contributions from Fe\two\ and O\two), O\one\,7773.4\,\AA, and the Ca\two\ NIR triplet (at the edge of the CCD so the observed flux is probably uncertain). In the optical, magnesium lines are weaker than numerous overlapping contributions from Fe\two\ and Ti\two\ lines (for example, Mg\one\,4571.1\,\AA\ has the same EW as Co\two\,4569.3\,\AA, and is about a tenth of the contribution from Fe\two\,4549.5\,\AA\  and a fifth from Ti\two\,4572.0\,\AA). \citet{jerkstrand_slsnic_17} modeled the nebular phase spectrum of SLSNe Ic with line contributions exclusively from O and Mg, but the present model suggests that the spectrum may contain lines from other metals (this may depend on the assumption of microscopic mixing made here). In the observed spectra, there are narrow peaks corresponding to O\three\ lines but weaker than in LSQ14an. It is not possible to estimate from observations alone the contribution of O\two\ (if any) to the 7300\,\AA\ emission (since it overlaps with Ca\two), nor the contribution of Fe\two\ to the 6300\,\AA\ emission (since it overlaps with O\one). The exact level of clumping depends on the density in the smooth ejecta, which is dependent on ejecta mass and kinetic energy, but clumping seems fundamental to reproducing the nebular-phase spectrum of SN\,2007bi. This agrees with the findings of \citet{jerkstrand_slsnic_17}, although the clumping of 0.1 (corresponding to a 10\% volume filling factor) is sufficient here to cause an ionization shift in the present model set, while  \citet{jerkstrand_slsnic_17} invokes a clumping 100 times greater. Physically, it seems extreme to fit about 90\% of the ejecta mass in 0.1\% of the ejecta volume, in particular if at the same time one assumes that the amount of magnetar power deposited in the ONeMg clumps is constant. Instead, as the clumps become smaller and smaller, one eventually enters a regime where little magnetar power gets absorbed by the clumps. Here, the magnetar power deposition is also assumed to be independent of clumping, but the clumping factor invoked is lower.

Magnetar power may lead to the co-existence of low-density hot ionized regions that would produce O\two\ and O\three\ lines,  and dense, cool, recombined, clumps that would produce O\one, Ca\two, and  Fe\two\ lines. This seems to be the case in SN\,2007bi, which exhibits weak O\three\ lines (and probably an O\two\ contribution to the 7300\,\AA). This co-existence may be complex. For example, this density and ionization structure may vary with radius and angle. A radial stratification could be easily discerned by a dichotomy in line width between lines from more ionized and less ionized ions. With the current treatment in \cmfgen, the inter-clump medium is considered as void so that only the clumps can absorb or emit.

\section{Summary and conclusions}
\label{sect_conc}

   This study presented non-LTE time-dependent radiative transfer simulations for magnetar-powered SNe. Using two solar-metallicity carbon-rich Wolf-Rayet progenitors and four ejecta in total (all endowed with \nifs\ at the time of explosion), additional models are computed using a range of initial spins and magnetic field for the magnetar. The simulations are started at 1\,d and continued until 200 to 600\,d after explosion, depending on the model. The magnetar power deposition is prescribed using a simple, time-independent, analytical form function of the density and velocity structure. In each ejecta shell, magnetar power is treated the same way as radioactive decay, and the associated non-thermal effects are solved for.

   Because the dynamical effects from the magnetar are ignored in \cmfgen, the quantitative results are expected to be a little off, and the more so for faster-spinning magnetar models. For example, rise times or peak luminosities may be underestimated or overestimated, but the qualitative aspects and the trends should hold. The neglect of dynamics is quantified for models r0e2 and r0e2ec with the code \heracles\ (see discussion and illustrations in Appendix~\ref{sect_comp_code}). The \cmfgen\ and \heracles\ results compare well for bolometric light curves (with slight offsets in rise times and light curve width) while the ejecta density structure is, as expected, most different in the case of the faster-spinning magnetar.

 With magnetar fields in the range of 1.0 to $3.5 \times 10^{14}$\,G and initial rotational energies in the  range of 0.4 to  $5.0 \times 10^{51}$\,erg (spin periods in the range from 7.0 to 2.0\,ms), the grid of magnetar-powered ejecta yield Type Ic SNe with rise times in the range of 21.6 to 51.7\,d to a bolometric maximum in the range of 0.16 to $3.2 \times 10^{44}$\,\ergs. The model counterpart without magnetar power peaks roughly at the same time, but with a luminosity ten times smaller and a narrower light curve. The main effect of the magnetar power is to raise the internal energy of the SN ejecta. This boosts the SN luminosity but also causes a drastic shift in temperature and ionization. In a standard SN Ic, oxygen is neutral or partially ionized at the photosphere at all times, while in a magnetar-powered SN Ic, oxygen is once-ionized throughout the high-brightness phase. Consequently, the ejecta optical depth is enhanced and the photosphere is located further out in radius (or velocity). The light curve is broader, the optical color is bluer, and the maximum-light optical spectra exhibit lines of O\two\ and C\two\ (instead of, for example, O\one, Na\one, and Fe\two\ in a standard SN Ic).

This temperature and ionization shift occurs in all the models presented here. In other words, even the weaker magnetar model can sizably influence the thermodynamics of the gas, alter the SN radiation, and produce the spectral appearance of a SLSN Ic. For the highest magnetar-energy model, the shift can be larger, with the O and C twice or three times ionized. This is possible in cases where the magnetar spin down timescale is shorter, allowing the temperature rise to occur while the ejecta is still relatively compact. Many different combinations of ejecta and magnetar properties are possible but the present grid of model reflects well the relative uniformity of observed properties of SLSNe Ic around maximum light.

Varying the extent of the magnetar-power deposition introduces shifts in the evolution of the SN radiation, while qualitatively, the SN model behaves in a similar fashion. With a narrower profile, the model takes longer to rise to a fainter maximum; the photospheric temperature and ionization increase later; the SED is at all times redder with a strong deficit of flux in the UV; the maximum-light optical spectra are essentially undistinguishable apart from the narrower line profiles; the late time spectra show lines from more ionized species. The differences at late times may not hold with a more realistic treatment of the magnetar power since then, the associated energy deposition may be extended in velocity space, and the more so as time progresses.

For a given ejecta mass and magnetar properties, increasing the ejecta kinetic energy (models r0e1, r0e2, and r0e4) by a factor of ten shifts the rise time from 43.0 to 25.8\,d, and raises the peak luminosity from 2.53 to $4.86 \times 10^{43}$\,\ergs. This occurs at constant time integrated luminosity since an increase in $E_{\rm kin}/M_{\rm ej}$ decreases the ejecta diffusion time (or more physically its ability to trap radiation). Around the time of maximum, the spectra are very similar, except that at higher $E_{\rm kin}$, the lines exhibit a greater velocity at maximum absorption and their emission component is weaker. However, at late times, the difference between models r0e1, r0e2, and r0e4 is large, reflecting the range in density in the inner ejecta for the different models. At higher $E_{\rm kin}$, the ejecta density is lower so that for a given magnetar power, the temperature and ionization are higher. As a result, in the nebular phase, model r0e4 exhibits lines of O\two\ and O\three\ while model r0e1 shows lines of O\one, Ca\two, and Fe\two.

Varying the magnetar properties causes a change in the spin-down time scale (which goes as $P_{\rm pm}^2/B_{\rm pm}^2$), the total energy liberated (which goes as $1/P_{\rm pm}^2$), and the power at late times (which goes as $1/t^2B_{\rm pm}^2$). These introduce quantitative offsets but bring no new qualitative behavior. Reducing the spin period  shortens the rise to maximum (from 31.7 to 21.6 from models r0e2 to r0e2eb) while lowering the magnetic field delays it. For sufficiently small spin periods, the SN can attain a very large temperature and ionization at the photosphere, making the SED bluer (with most of the flux falling in the UV) and strengthening O\three\ and C\three\ lines in the optical range. For a lower magnetic field, the power absorbed by the ejecta at late times is greater, boosting the ionization so that the ejecta cools mostly through forbidden-line emission from O\two\ and O\three\ transitions.

Non-thermal processes are explicitly treated in all simulations except one (model 5p11Bx2th) which is used to quantify the magnitude of these effects. The result shows that throughout the photospheric phase, the escaping radiation is essentially identical between model 5p11Bx2th and its non-thermal counterpart, that is, non-thermal processes have no visible impact on the plasma for this particular choice of ejecta and magnetar (in contradiction with \citealt{mazzali_slsn_16}, but in agreement with expectations from previous studies, such as \citealt{KF92}, \citealt{d12_snibc} and \citealt{li_etal_12_nonte}). The interpretation is that the ionization is large (all the oxygen is generally once-ionized in all models), so that the free-electron density is also large, causing non-thermal electrons to lose their energy through Coulomb scattering, thereby heating the thermal bath. Since all SLSNe Ic exhibit lines of O\two\ during the high brightness phase, their photospheres are generally hot and ionized, and thus non-thermal effects should be small at those times. Instead, the high photospheric temperature and strong UV flux boost photoionization processes. At late times, non-thermal effects should strengthen if or when the ionization drops. In model 5p11Bx2, the He\one\,10830 line is indeed stronger than in the thermal model counterpart 5p11Bx2th at 100\,d after explosion (early nebular phase). Overall, non-thermal effects thrive under low-ionization conditions, which are not representative of SLSNe Ic at maximum.

Following the approach presented in \citet{d18_fcl}, a number of simulations are performed to test the effect of clumping on the SN radiation and gas properties. Clumping can impact the ionization level of the gas and all the above simulations for smooth ejecta (with the exception of model r0e1) exhibit a high level of ionization (with lines of O\two\ and O\three) until late times. There is also evidence from multi-dimensional simulations \citep{chen_pm_2d_16,suzuki_pm_2d_17} and nebular-phase spectra \citep{jerkstrand_slsnic_17} that clumping is present. Here, the few simulations carried out show that the post-maximum bolometric light curve is unaffected but the UV-optical color of the clumped model is much redder. In model r0e2cl, clumping has a visible impact even at maximum light, although the greatest impact occurs a few weeks after maximum with a rapid drop in photospheric temperature and a rapid recombination of the ejecta to a partial ionization at +40\,d. At this time, since the ejecta optical depth is only a few, the flux is roughly equal to the power absorbed and no longer scales with $R_{\rm ph}^2 T_{\rm ph}^4$. At nebular times, the clumped model shows stronger O\one\ and weaker O\two. Because the spectrum formation region encompasses the entire ejecta at 40\,d after maximum, a clumped ejecta should exhibit distinct signatures from a smooth ejecta soon after maximum light -- the effect of clumping is not confined to the nebular phase. Simulations at nebular times also show that as clumping is enhanced, the SN temperature and ionization drop as long as the conditions are ionized. Once partially ionized, a further increase in clumping has little impact. This is likely because the recombination rate drops due to the lower free-electron density as well as the strengthening of non-thermal effects at lower ionization.  The effect of clumping is stronger in ejecta with a higher $E_{\rm kin}/M_{\rm ej}$.

The grid of models presented here was not designed to match any particular SLSN Ic. The choice of a relatively high magnetic field and slow spin implies that the models are located at the low-brightness and the fast-rise end of the distribution of potential SLSNe Ic. Nonetheless, the model properties reproduce the systematic presence of O\two\ and C\two\ lines in SLSNe Ic. Model r0e2 matches closely the maximum-light spectrum of Gaia16apd in the UV and optical range. The simulations suggest that clumping is essential to reproduce the colors and spectral lines observed within a month of maximum light. Without clumping, the models remain blue and persist in showing O\two\ and C\two\ lines for months. With clumping, a lower ionization is eventually enforced. For example, model r0e2cl matches well the observed spectra of the selected sample of SLSNe Ic, including SN\,2007bi at +46\,d or Gaia16apd at +43\,d. The smooth ejecta model r0e2 reproduces some of the salient features of LSQ14an at +410\,d (which exhibits a strong O\three\,5006.8\,\AA\ line) and the clumped ejecta model r0e2cl reproduces well the observations of SN\,2007bi at +367\,d (characterized by lines primarily from neutral and once-ionized species). While clumping is supported by the present simulations, it is 100 times weaker than proposed by \citet{jerkstrand_slsnic_17}. Here, the bulk of the effect of clumping is already present when adopting a 10\% volume filling factor -- enhancing the clumping further causes little change. In the future, this study will be extended to include a bigger set of progenitors covering from subsolar to solar metallicities, and also to improve the physical consistency with a better, less ad-hoc, treatment of magnetar-power deposition.

\begin{acknowledgements}

LD thanks Lin Yan for providing the HST and optical maximum-light spectra of Gaia16apd. LD thanks ESO-Vitacura for their hospitality. This work utilized computing resources of the mesocentre SIGAMM, hosted by the Observatoire de la C\^ote d'Azur, Nice, France, as well as computing resources of the ``Maison de la simulation'', CEA, Gif-sur-Yvette, France. This research was supported by the Munich Institute for Astro- and Particle Physics (MIAPP) of the DFG cluster of excellence ``Origin and Structure of the Universe".

\end{acknowledgements}

\appendix

\section{Comparison between \cmfgen\ and \heracles\ results}
\label{sect_comp_code}

\cmfgen\ is a non-LTE time-dependent radiative transfer code with no treatment of dynamics. The ejecta must be in homologous expansion and each mass shell moves at constant velocity. The density structure evolves as $1/t^3$. Hence, even in the presence of strong pressure gradients, as for example caused by energy deposition by a magnetar, there is no influence on the velocity or density structure. In reality, the magnetar influence on the ejecta may be strong, both on the internal energy budget (which may come out as radiation) and on the density structure, for example through a snow-plow effect \citep{KB10}.

This section compares the results with \cmfgen\ for models r0e2 and r0e2ec with the results from radiation-hydrodynamics simulations with \heracles. The setup in \heracles\ is the same as in \citet{dessart_audit_18}. The code starts from the \cmfgen\ ejecta properties at 1.4\,d. For simplicity, \heracles\ uses a fixed composition with 99.9\% oxygen and 0.1\% iron when estimating the opacity of the material. A low floor opacity of 0.0003\,cm$^2$\,g$^{-1}$ is used although it is in effect only at very low optical depth and thus has no influence on the light curve. The gray approximation is used. The code assumes an ideal gas equation of state, with $\gamma=$\,5/3, and a mean molecular weight of 10 (this is a rough estimate for an oxygen-dominated composition with partial ionization). The choice of molecular weight has no impact on the light curve (or dynamics; the plasma is radiation dominated) but it influences the gas temperature (where the albedo is high or where the optical depth is low). Radioactive decay is ignored in the \heracles\ simulation.

\begin{figure}
   \includegraphics[width=0.9\hsize]{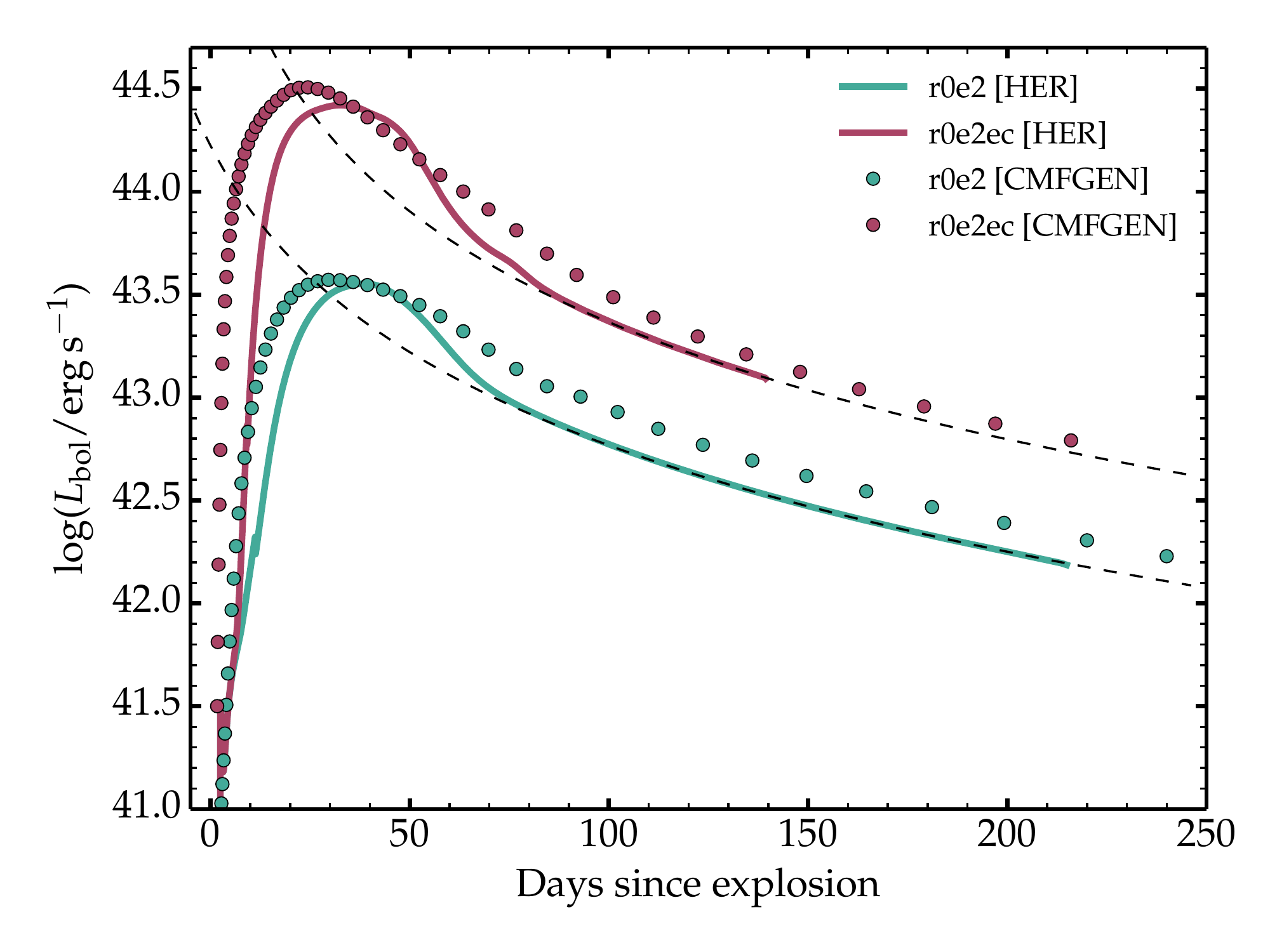}
\caption{Comparison of the bolometric light curve computed with \heracles\ (solid line; the simulation neglects radioactive decay power) and with \cmfgen\ (dots) for models r0e2 and r0e2ec. The same color coding is used as before. The total magnetar power in each model is shown as a dashed line.
\label{fig_comp_lbol_her_cmfgen}
}
\end{figure}

Figure~\ref{fig_comp_lbol_her_cmfgen} compares the bolometric light curves computed with \cmfgen\ and \heracles\ for models r0e2 and r0e2ec. A time shift is applied to the \heracles\ light curve to account for the light-travel time to the outer boundary at 10$^{16}$\,cm (which corresponds to 3.86\,d). The light curves are in rough agreement. The main difference is in the early-time behavior and rise time, which arises from the different profile for the magnetar-power deposition.  The light curve width differs also from the greater opacity returned by \cmfgen\, which accounts accurately for the effect of lines. At late times, the \cmfgen\ luminosity is offset from the magnetar power absorbed for a number of reasons, including the additional contribution from 0.13\,\msun\ of \nifs, the larger ejecta optical depth in \cmfgen\ (which treats the opacity accurately), and also because the magnetar-power deposition is broader than in the \heracles\ simulation (which caused a small boost in the observer's frame during the photospheric phase).

\begin{figure*}
   \includegraphics[width=0.5\hsize]{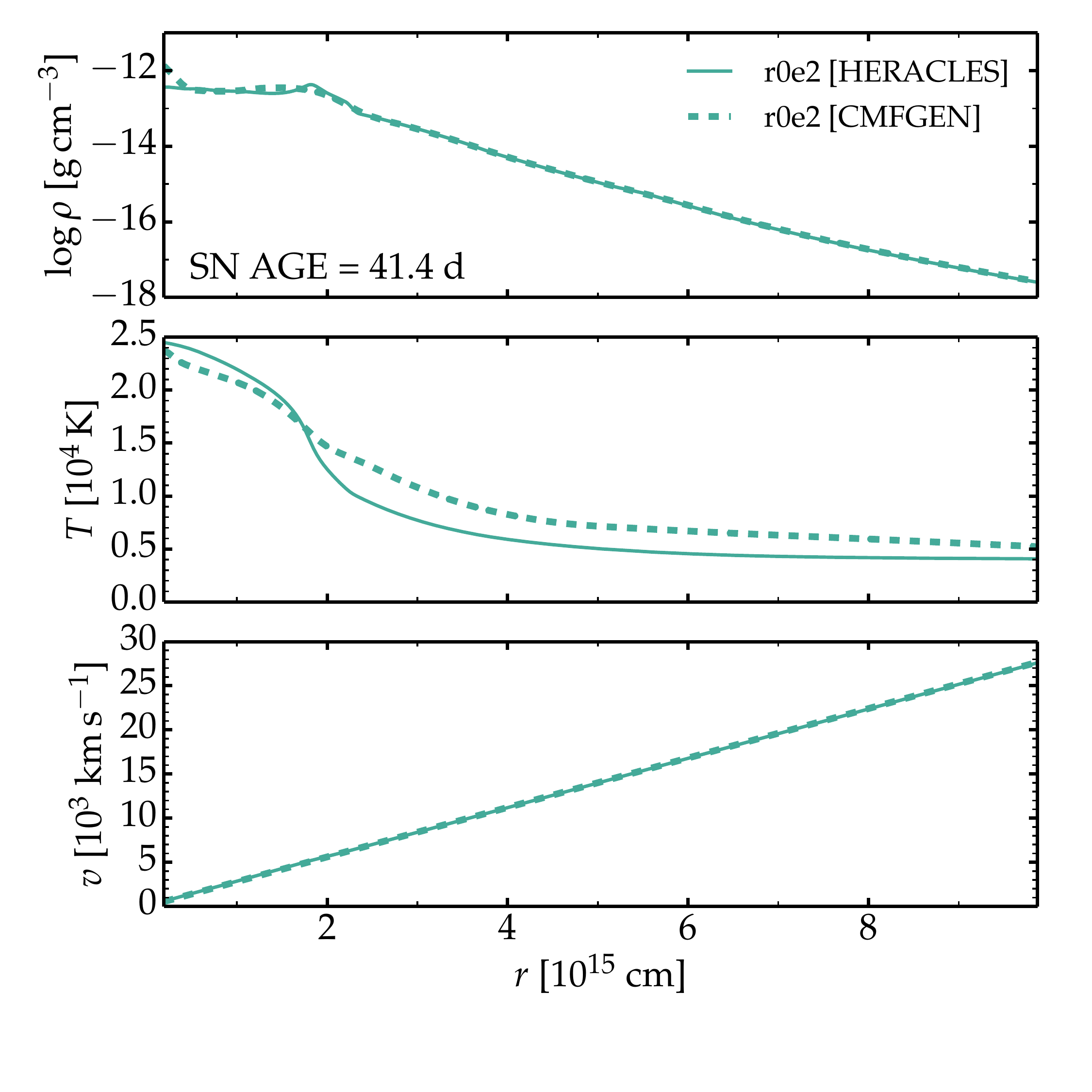}
   \includegraphics[width=0.5\hsize]{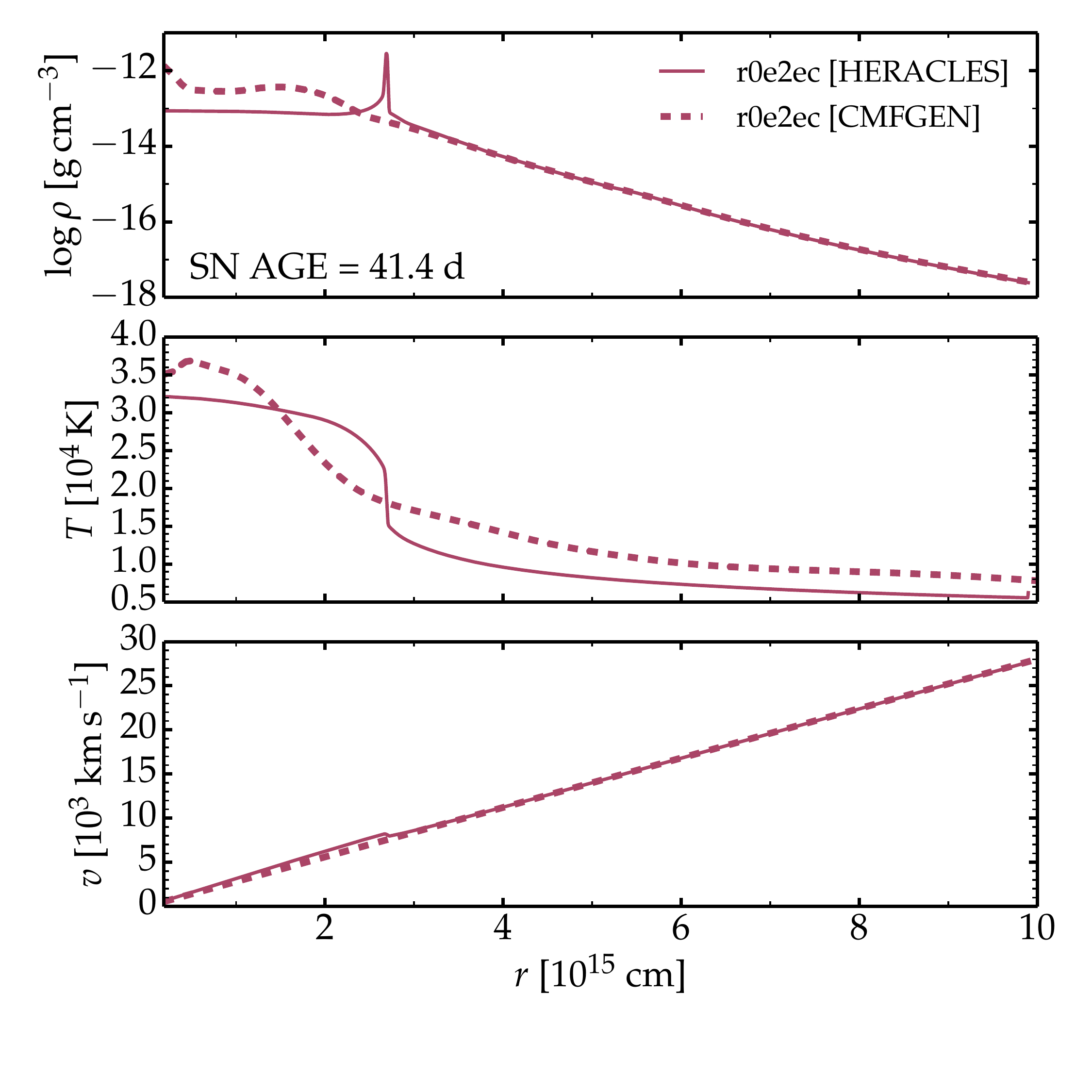}
\caption{Comparison of some ejecta properties computed with \heracles\ (solid line) and with \cmfgen\ (dashed line) for models r0e2 (left) and r0e2ec (right) at 41.4\,d after explosion.
\label{fig_comp_ejecta_her_cmfgen}
}
\end{figure*}

Figure~\ref{fig_comp_ejecta_her_cmfgen} compares the ejecta properties computed by \cmfgen\ and \heracles\ at 41.4\,d after explosion. In the regions where the magnetar power does not affect the dynamics, the density structure is identical in both models, because it results exclusively from mass conservation. Deeper in the ejecta, the density structure differs in both models, and more strongly in the model r0e2ec characterized by a large magnetar energy ($E_{\rm pm} = 5 \times 10^{51}$\,erg). As expected, the magnetar power has a strong dynamical effect (even with the relatively broad deposition profile used in \heracles), and a snow-plow effect is visible through the presence of a dense shell at about 8000\,\kms. The snow-plow effect in model r0e2 is weak. The impact on the temperature (and the different results between the codes) result from the different dynamical evolution (in particular the resulting density), as well as the fact that the temperature is computed in non-LTE in \cmfgen. The bottom panel of Fig.~\ref{fig_comp_ejecta_her_cmfgen} shows that the velocity can deviate from homologous expansion for a large magnetar power and energy. For the model r0e2ec, the ejecta kinetic energy is roughly equal to the magnetar energy at birth ($E_{\rm pm} = 5.0 \times 10^{51}$\,erg compared to $E_{\rm kin} = 4.1 \times 10^{51}$\,erg).

Overall, despite the neglect of hydrodynamics in \cmfgen, the resulting ejecta and radiation properties (for what can be compared) are similar to those produced by \heracles. This suggests that the approach used for the main part of this paper is adequate.  Furthermore, it is clear that the 1-D hydrodynamics treatment of magnetar-powered  ejecta is not ideal since it overestimates the snow-plow effect and ignores multi-dimensional fluid instabilities, which are known to prevent the formation of a dense swept-up shell \citet{chen_pm_2d_16,suzuki_pm_2d_17}. With the present approach, \cmfgen\ allows for detailed non-LTE and time dependence and mimics the influence of some multi-dimensional effects. Further work is needed to improve the physical consistency of the approach.


\begin{thebibliography}{80}
\expandafter\ifx\csname natexlab\endcsname\relax\def\natexlab#1{#1}\fi

\bibitem[{{Akiyama} {et~al.}(2003){Akiyama}, {Wheeler}, {Meier}, \&
  {Lichtenstadt}}]{akiyama_mri_03}
{Akiyama}, S., {Wheeler}, J.~C., {Meier}, D.~L., \& {Lichtenstadt}, I. 2003,
  \apj, 584, 954

\bibitem[{{Anderson} {et~al.}(2018){Anderson}, {Pessi}, {Dessart}, {Inserra},
  {Hiramatsu}, {Taggart}, {Smartt}, {Leloudas}, {Chen}, {M{\"o}ller}, {Roy},
  {Schulze}, {Perley}, {Selsing}, {Prentice}, {Gal-Yam}, {Angus}, {Arcavi},
  {Ashall}, {Bulla}, {Bray}, {Burke}, {Callis}, {Cartier}, {Chang}, {Chambers},
  {Clark}, {Denneau}, {Dennefeld}, {Flewelling}, {Fraser}, {Galbany},
  {Gromadzki}, {Guti{\'e}rrez}, {Heinze}, {Hosseinzadeh}, {Howell}, {Hsiao},
  {Kankare}, {Kostrzewa-Rutkowska}, {Magnier}, {Maguire}, {Mazzali}, {McBrien},
  {McCully}, {Morrell}, {Lowe}, {Onken}, {Onori}, {Phillips}, {Rest},
  {Ridden-Harper}, {Ruiter}, {Sand}, {Smith}, {Smith}, {Stalder},
  {Stritzinger}, {Sullivan}, {Tonry}, {Tucker}, {Valenti}, {Wainscoat},
  {Waters}, {Wolf}, \& {Young}}]{anderson_18bsz}
{Anderson}, J.~P., {Pessi}, P.~J., {Dessart}, L., {et~al.} 2018, ArXiv e-prints
  [\eprint[arXiv]{1806.10609}]

\bibitem[{{Arcavi} {et~al.}(2017){Arcavi}, {Howell}, {Kasen}, {Bildsten},
  {Hosseinzadeh}, {McCully}, {Wong}, {Katz}, {Gal-Yam}, {Sollerman}, {Taddia},
  {Leloudas}, {Fremling}, {Nugent}, {Horesh}, {Mooley}, {Rumsey}, {Cenko},
  {Graham}, {Perley}, {Nakar}, {Shaviv}, {Bromberg}, {Shen}, {Ofek}, {Cao},
  {Wang}, {Huang}, {Rui}, {Zhang}, {Li}, {Li}, {Zhang}, {Valenti}, {Guevel},
  {Shappee}, {Kochanek}, {Holoien}, {Filippenko}, {Fender}, {Nyholm}, {Yaron},
  {Kasliwal}, {Sullivan}, {Blagorodnova}, {Walters}, {Lunnan}, {Khazov},
  {Andreoni}, {Laher}, {Konidaris}, {Wozniak}, \& {Bue}}]{arcavi_iptf14hls}
{Arcavi}, I., {Howell}, D.~A., {Kasen}, D., {et~al.} 2017, \nat, 551, 210

\bibitem[{{Bersten} {et~al.}(2016){Bersten}, {Benvenuto}, {Orellana}, \&
  {Nomoto}}]{bersten_15lh_16}
{Bersten}, M.~C., {Benvenuto}, O.~G., {Orellana}, M., \& {Nomoto}, K. 2016,
  \apjl, 817, L8

\bibitem[{{Bodenheimer} \& {Ostriker}(1974)}]{bodenheimer_ostriker_74}
{Bodenheimer}, P. \& {Ostriker}, J.~P. 1974, \apj, 191, 465

\bibitem[{{Chen} {et~al.}(2016){Chen}, {Woosley}, \&
  {Sukhbold}}]{chen_pm_2d_16}
{Chen}, K.-J., {Woosley}, S.~E., \& {Sukhbold}, T. 2016, \apj, 832, 73

\bibitem[{{Chen} {et~al.}(2015){Chen}, {Smartt}, {Jerkstrand}, {Nicholl},
  {Bresolin}, {Kotak}, {Polshaw}, {Rest}, {Kudritzki}, {Zheng}, {Elias-Rosa},
  {Smith}, {Inserra}, {Wright}, {Kankare}, {Kangas}, \&
  {Fraser}}]{chen_slsn_15}
{Chen}, T.-W., {Smartt}, S.~J., {Jerkstrand}, A., {et~al.} 2015, \mnras, 452,
  1567

\bibitem[{{De Cia} {et~al.}(2018){De Cia}, {Gal-Yam}, {Rubin}, {Leloudas},
  {Vreeswijk}, {Perley}, {Quimby}, {Yan}, {Sullivan}, {Fl{\"o}rs}, {Sollerman},
  {Bersier}, {Cenko}, {Gal-Yam}, {Maguire}, {Ofek}, {Prentice}, {Schulze},
  {Spyromilio}, {Valenti}, {Arcavi}, {Corsi}, {Howell}, {Mazzali}, {Kasliwal},
  {Taddia}, \& {Yaron}}]{de_cia_slsn_ic_17}
{De Cia}, A., {Gal-Yam}, A., {Rubin}, A., {et~al.} 2018, \apj, 860, 100

\bibitem[{{Dessart}(2018)}]{d18_iptf14hls}
{Dessart}, L. 2018, \aap, 610, L10

\bibitem[{{Dessart} \& {Audit}(2018)}]{dessart_audit_18}
{Dessart}, L. \& {Audit}, E. 2018, \aap, 613, A5

\bibitem[{{Dessart} {et~al.}(2008){Dessart}, {Blondin}, {Brown}, {Hicken},
  {Hillier}, {Holland}, {Immler}, {Kirshner}, {Milne}, {Modjaz}, \&
  {Roming}}]{dessart_05cs_06bp}
{Dessart}, L., {Blondin}, S., {Brown}, P.~J., {et~al.} 2008, \apj, 675, 644

\bibitem[{{Dessart} {et~al.}(2012{\natexlab{a}}){Dessart}, {Hillier}, {Li}, \&
  {Woosley}}]{d12_snibc}
{Dessart}, L., {Hillier}, D.~J., {Li}, C., \& {Woosley}, S. 2012{\natexlab{a}},
  \mnras, 424, 2139

\bibitem[{{Dessart} {et~al.}(2012{\natexlab{b}}){Dessart}, {Hillier},
  {Waldman}, {Livne}, \& {Blondin}}]{d12_magnetar}
{Dessart}, L., {Hillier}, D.~J., {Waldman}, R., {Livne}, E., \& {Blondin}, S.
  2012{\natexlab{b}}, \mnras, 426, L76

\bibitem[{{Dessart} {et~al.}(2018){Dessart}, {Hillier}, \& {Wilk}}]{d18_fcl}
{Dessart}, L., {Hillier}, D.~J., \& {Wilk}, K.~D. 2018, \aap, 619, A30

\bibitem[{{Dessart} {et~al.}(2015){Dessart}, {Hillier}, {Woosley}, {Livne},
  {Waldman}, {Yoon}, \& {Langer}}]{D15_SNIbc_I}
{Dessart}, L., {Hillier}, D.~J., {Woosley}, S., {et~al.} 2015, \mnras, 453,
  2189

\bibitem[{{Dessart} {et~al.}(2016){Dessart}, {Hillier}, {Woosley}, {Livne},
  {Waldman}, {Yoon}, \& {Langer}}]{D16_SNIbc_II}
{Dessart}, L., {Hillier}, D.~J., {Woosley}, S., {et~al.} 2016, \mnras, 458,
  1618

\bibitem[{{Dessart} {et~al.}(2017{\natexlab{a}}){Dessart}, {John Hillier}, \&
  {Audit}}]{d18_13fs}
{Dessart}, L., {John Hillier}, D., \& {Audit}, E. 2017{\natexlab{a}}, \aap,
  605, A83

\bibitem[{{Dessart} {et~al.}(2017{\natexlab{b}}){Dessart}, {John Hillier},
  {Yoon}, {Waldman}, \& {Livne}}]{dessart_98bw_17}
{Dessart}, L., {John Hillier}, D., {Yoon}, S.-C., {Waldman}, R., \& {Livne}, E.
  2017{\natexlab{b}}, \aap, 603, A51

\bibitem[{{Dessart} {et~al.}(2010{\natexlab{a}}){Dessart}, {Livne}, \&
  {Waldman}}]{dlw10b}
{Dessart}, L., {Livne}, E., \& {Waldman}, R. 2010{\natexlab{a}}, \mnras, 408,
  827

\bibitem[{{Dessart} {et~al.}(2010{\natexlab{b}}){Dessart}, {Livne}, \&
  {Waldman}}]{dlw10a}
{Dessart}, L., {Livne}, E., \& {Waldman}, R. 2010{\natexlab{b}}, \mnras, 405,
  2113

\bibitem[{{Dessart} {et~al.}(2013){Dessart}, {Waldman}, {Livne}, {Hillier}, \&
  {Blondin}}]{d13_pisn}
{Dessart}, L., {Waldman}, R., {Livne}, E., {Hillier}, D.~J., \& {Blondin}, S.
  2013, \mnras, 428, 3227

\bibitem[{{Gal-Yam} {et~al.}(2014){Gal-Yam}, {Arcavi}, {Ofek}, {Ben-Ami},
  {Cenko}, {Kasliwal}, {Cao}, {Yaron}, {Tal}, {Silverman}, {Horesh}, {De Cia},
  {Taddia}, {Sollerman}, {Perley}, {Vreeswijk}, {Kulkarni}, {Nugent},
  {Filippenko}, \& {Wheeler}}]{galyam_13cu_14}
{Gal-Yam}, A., {Arcavi}, I., {Ofek}, E.~O., {et~al.} 2014, \nat, 509, 471

\bibitem[{{Gal-Yam} {et~al.}(2009){Gal-Yam}, {Mazzali}, {Ofek}, {Nugent},
  {Kulkarni}, {Kasliwal}, {Quimby}, {Filippenko}, {Cenko}, {Chornock},
  {Waldman}, {Kasen}, {Sullivan}, {Beshore}, {Drake}, {Thomas}, {Bloom},
  {Poznanski}, {Miller}, {Foley}, {Silverman}, {Arcavi}, {Ellis}, \&
  {Deng}}]{galyam_07bi_09}
{Gal-Yam}, A., {Mazzali}, P., {Ofek}, E.~O., {et~al.} 2009, \nat, 462, 624

\bibitem[{{Georgy} {et~al.}(2009){Georgy}, {Meynet}, {Walder}, {Folini}, \&
  {Maeder}}]{georgy_snibc_09}
{Georgy}, C., {Meynet}, G., {Walder}, R., {Folini}, D., \& {Maeder}, A. 2009,
  \aap, 502, 611

\bibitem[{{Glas} {et~al.}(2018){Glas}, {Just}, {Janka}, \&
  {Obergaulinger}}]{glas_ccsn_3d_18}
{Glas}, R., {Just}, O., {Janka}, H.-T., \& {Obergaulinger}, M. 2018, ArXiv
  e-prints [\eprint[arXiv]{1809.10146}]

\bibitem[{{Gr{\"a}fener} \& {Vink}(2016)}]{grafener_vink_13cu_16}
{Gr{\"a}fener}, G. \& {Vink}, J.~S. 2016, \mnras, 455, 112

\bibitem[{{Groh}(2014)}]{groh_13cu}
{Groh}, J.~H. 2014, \aap, 572, L11

\bibitem[{{Guillochon} {et~al.}(2017){Guillochon}, {Parrent}, {Kelley}, \&
  {Margutti}}]{sn_catalog}
{Guillochon}, J., {Parrent}, J., {Kelley}, L.~Z., \& {Margutti}, R. 2017, \apj,
  835, 64

\bibitem[{{Hillier} \& {Dessart}(2012)}]{HD12}
{Hillier}, D.~J. \& {Dessart}, L. 2012, \mnras, 424, 252

\bibitem[{{Hillier} {et~al.}(2013){Hillier}, {Dessart}, \&
  {Li}}]{hillier_hedp_13}
{Hillier}, D.~J., {Dessart}, L., \& {Li}, C. 2013, High Energy Density Physics,
  9, 297

\bibitem[{{Hirschi} {et~al.}(2004){Hirschi}, {Meynet}, \&
  {Maeder}}]{hirschi_rot_04}
{Hirschi}, R., {Meynet}, G., \& {Maeder}, A. 2004, \aap, 425, 649

\bibitem[{{Howell} {et~al.}(2013){Howell}, {Kasen}, {Lidman}, {Sullivan},
  {Conley}, {Astier}, {Balland}, {Carlberg}, {Fouchez}, {Guy}, {Hardin},
  {Pain}, {Palanque-Delabrouille}, {Perrett}, {Pritchet}, {Regnault}, {Rich},
  \& {Ruhlmann-Kleider}}]{howell_slsnic_13}
{Howell}, D.~A., {Kasen}, D., {Lidman}, C., {et~al.} 2013, \apj, 779, 98

\bibitem[{{Inserra} {et~al.}(2017){Inserra}, {Nicholl}, {Chen}, {Jerkstrand},
  {Smartt}, {Kr{\"u}hler}, {Anderson}, {Baltay}, {Della Valle}, {Fraser},
  {Gal-Yam}, {Galbany}, {Kankare}, {Maguire}, {Rabinowitz}, {Smith}, {Valenti},
  \& {Young}}]{inserra_slsn_ic_17}
{Inserra}, C., {Nicholl}, M., {Chen}, T.-W., {et~al.} 2017, \mnras, 468, 4642

\bibitem[{{Inserra} {et~al.}(2013){Inserra}, {Smartt}, {Jerkstrand}, {Valenti},
  {Fraser}, {Wright}, {Smith}, {Chen}, {Kotak}, {Pastorello}, {Nicholl},
  {Bresolin}, {Kudritzki}, {Benetti}, {Botticella}, {Burgett}, {Chambers},
  {Ergon}, {Flewelling}, {Fynbo}, {Geier}, {Hodapp}, {Howell}, {Huber},
  {Kaiser}, {Leloudas}, {Magill}, {Magnier}, {McCrum}, {Metcalfe}, {Price},
  {Rest}, {Sollerman}, {Sweeney}, {Taddia}, {Taubenberger}, {Tonry},
  {Wainscoat}, {Waters}, \& {Young}}]{inserra_slsn_13}
{Inserra}, C., {Smartt}, S.~J., {Jerkstrand}, A., {et~al.} 2013, \apj, 770, 128

\bibitem[{{Jerkstrand} {et~al.}(2017){Jerkstrand}, {Smartt}, {Inserra},
  {Nicholl}, {Chen}, {Kr{\"u}hler}, {Sollerman}, {Taubenberger}, {Gal-Yam},
  {Kankare}, {Maguire}, {Fraser}, {Valenti}, {Sullivan}, {Cartier}, \&
  {Young}}]{jerkstrand_slsnic_17}
{Jerkstrand}, A., {Smartt}, S.~J., {Inserra}, C., {et~al.} 2017, \apj, 835, 13

\bibitem[{{Kasen} \& {Bildsten}(2010)}]{KB10}
{Kasen}, D. \& {Bildsten}, L. 2010, \apj, 717, 245

\bibitem[{{Kasen} {et~al.}(2016){Kasen}, {Metzger}, \&
  {Bildsten}}]{kasen_pm_16}
{Kasen}, D., {Metzger}, B.~D., \& {Bildsten}, L. 2016, \apj, 821, 36

\bibitem[{{Kozma} \& {Fransson}(1992)}]{KF92}
{Kozma}, C. \& {Fransson}, C. 1992, \apj, 390, 602

\bibitem[{{Leloudas} {et~al.}(2015){Leloudas}, {Schulze}, {Kr{\"u}hler},
  {Gorosabel}, {Christensen}, {Mehner}, {de Ugarte Postigo}, {Amor{\'{\i}}n},
  {Th{\"o}ne}, {Anderson}, {Bauer}, {Gallazzi}, {He{\l}miniak}, {Hjorth},
  {Ibar}, {Malesani}, {Morell}, {Vinko}, \& {Wheeler}}]{leloudas_slsn_15}
{Leloudas}, G., {Schulze}, S., {Kr{\"u}hler}, T., {et~al.} 2015, \mnras, 449,
  917

\bibitem[{{Lentz} {et~al.}(2015){Lentz}, {Bruenn}, {Hix}, {Mezzacappa},
  {Messer}, {Endeve}, {Blondin}, {Harris}, {Marronetti}, \&
  {Yakunin}}]{lentz_ccsn_3d_15}
{Lentz}, E.~J., {Bruenn}, S.~W., {Hix}, W.~R., {et~al.} 2015, \apjl, 807, L31

\bibitem[{{Li} {et~al.}(2012){Li}, {Hillier}, \& {Dessart}}]{li_etal_12_nonte}
{Li}, C., {Hillier}, D.~J., \& {Dessart}, L. 2012, \mnras, 426, 1671

\bibitem[{{Liu} {et~al.}(2017){Liu}, {Modjaz}, \& {Bianco}}]{liu_slsnic_17}
{Liu}, Y.-Q., {Modjaz}, M., \& {Bianco}, F.~B. 2017, \apj, 845, 85

\bibitem[{{Livne}(1993)}]{livne_93}
{Livne}, E. 1993, \apj, 412, 634

\bibitem[{{Lunnan} {et~al.}(2018){Lunnan}, {Chornock}, {Berger}, {Jones},
  {Rest}, {Czekala}, {Dittmann}, {Drout}, {Foley}, {Fong}, {Kirshner},
  {Laskar}, {Leibler}, {Margutti}, {Milisavljevic}, {Narayan}, {Pan}, {Riess},
  {Roth}, {Sanders}, {Scolnic}, {Smartt}, {Smith}, {Chambers}, {Draper},
  {Flewelling}, {Huber}, {Kaiser}, {Kudritzki}, {Magnier}, {Metcalfe},
  {Wainscoat}, {Waters}, \& {Willman}}]{lunnan_slsn_ic_18}
{Lunnan}, R., {Chornock}, R., {Berger}, E., {et~al.} 2018, \apj, 852, 81

\bibitem[{{Lunnan} {et~al.}(2014){Lunnan}, {Chornock}, {Berger}, {Laskar},
  {Fong}, {Rest}, {Sanders}, {Challis}, {Drout}, {Foley}, {Huber}, {Kirshner},
  {Leibler}, {Marion}, {McCrum}, {Milisavljevic}, {Narayan}, {Scolnic},
  {Smartt}, {Smith}, {Soderberg}, {Tonry}, {Burgett}, {Chambers}, {Flewelling},
  {Hodapp}, {Kaiser}, {Magnier}, {Price}, \& {Wainscoat}}]{lunnan_slsn_14}
{Lunnan}, R., {Chornock}, R., {Berger}, E., {et~al.} 2014, \apj, 787, 138

\bibitem[{{Lunnan} {et~al.}(2016){Lunnan}, {Chornock}, {Berger},
  {Milisavljevic}, {Jones}, {Rest}, {Fong}, {Fransson}, {Margutti}, {Drout},
  {Blanchard}, {Challis}, {Cowperthwaite}, {Foley}, {Kirshner}, {Morrell},
  {Riess}, {Roth}, {Scolnic}, {Smartt}, {Smith}, {Villar}, {Chambers},
  {Draper}, {Huber}, {Kaiser}, {Kudritzki}, {Magnier}, {Metcalfe}, \&
  {Waters}}]{lunnan_slsnic_16}
{Lunnan}, R., {Chornock}, R., {Berger}, E., {et~al.} 2016, \apj, 831, 144

\bibitem[{{Maeda} {et~al.}(2007){Maeda}, {Tanaka}, {Nomoto}, {Tominaga},
  {Kawabata}, {Mazzali}, {Umeda}, {Suzuki}, \& {Hattori}}]{maeda_05bf_07}
{Maeda}, K., {Tanaka}, M., {Nomoto}, K., {et~al.} 2007, \apj, 666, 1069

\bibitem[{{Mazzali} {et~al.}(2016){Mazzali}, {Sullivan}, {Pian}, {Greiner}, \&
  {Kann}}]{mazzali_slsn_16}
{Mazzali}, P.~A., {Sullivan}, M., {Pian}, E., {Greiner}, J., \& {Kann}, D.~A.
  2016, \mnras, 458, 3455

\bibitem[{{Moriya} {et~al.}(2016){Moriya}, {Metzger}, \&
  {Blinnikov}}]{moriya_pm_bh_16}
{Moriya}, T.~J., {Metzger}, B.~D., \& {Blinnikov}, S.~I. 2016, \apj, 833, 64

\bibitem[{{M{\"u}ller} {et~al.}(2017){M{\"u}ller}, {Melson}, {Heger}, \&
  {Janka}}]{mueller_ccsn_3d_17}
{M{\"u}ller}, B., {Melson}, T., {Heger}, A., \& {Janka}, H.-T. 2017, \mnras,
  472, 491

\bibitem[{{Nicholl} {et~al.}(2018{\natexlab{a}}){Nicholl}, {Berger},
  {Blanchard}, {Gomez}, \& {Chornock}}]{nicholl_slsn_neb_18}
{Nicholl}, M., {Berger}, E., {Blanchard}, P.~K., {Gomez}, S., \& {Chornock}, R.
  2018{\natexlab{a}}, ArXiv e-prints [\eprint[arXiv]{1808.00510}]

\bibitem[{{Nicholl} {et~al.}(2016{\natexlab{a}}){Nicholl}, {Berger},
  {Margutti}, {Chornock}, {Blanchard}, {Jerkstrand}, {Smartt}, {Arcavi},
  {Challis}, {Chambers}, {Chen}, {Cowperthwaite}, {Gal-Yam}, {Hosseinzadeh},
  {Howell}, {Inserra}, {Kankare}, {Magnier}, {Maguire}, {Mazzali}, {McCully},
  {Milisavljevic}, {Smith}, {Taubenberger}, {Valenti}, {Wainscoat}, {Yaron}, \&
  {Young}}]{nicholl_15bn_16}
{Nicholl}, M., {Berger}, E., {Margutti}, R., {et~al.} 2016{\natexlab{a}},
  \apjl, 828, L18

\bibitem[{{Nicholl} {et~al.}(2016{\natexlab{b}}){Nicholl}, {Berger}, {Smartt},
  {Margutti}, {Kamble}, {Alexander}, {Chen}, {Inserra}, {Arcavi}, {Blanchard},
  {Cartier}, {Chambers}, {Childress}, {Chornock}, {Cowperthwaite}, {Drout},
  {Flewelling}, {Fraser}, {Gal-Yam}, {Galbany}, {Harmanen}, {Holoien},
  {Hosseinzadeh}, {Howell}, {Huber}, {Jerkstrand}, {Kankare}, {Kochanek},
  {Lin}, {Lunnan}, {Magnier}, {Maguire}, {McCully}, {McDonald}, {Metzger},
  {Milisavljevic}, {Mitra}, {Reynolds}, {Saario}, {Shappee}, {Smith},
  {Valenti}, {Villar}, {Waters}, \& {Young}}]{nicholl_15bn_long_16}
{Nicholl}, M., {Berger}, E., {Smartt}, S.~J., {et~al.} 2016{\natexlab{b}},
  \apj, 826, 39

\bibitem[{{Nicholl} {et~al.}(2018{\natexlab{b}}){Nicholl}, {Blanchard},
  {Berger}, {Alexander}, {Metzger}, {Bhirombhakdi}, {Chornock}, {Coppejans},
  {Gomez}, {Margalit}, {Margutti}, \& {Terreran}}]{nicholl_15bn_1000d_18}
{Nicholl}, M., {Blanchard}, P.~K., {Berger}, E., {et~al.} 2018{\natexlab{b}},
  \apjl, 866, L24

\bibitem[{{Nicholl} {et~al.}(2017){Nicholl}, {Guillochon}, \&
  {Berger}}]{nicholl_mosfit_mag_17}
{Nicholl}, M., {Guillochon}, J., \& {Berger}, E. 2017, \apj, 850, 55

\bibitem[{{Nicholl} {et~al.}(2013){Nicholl}, {Smartt}, {Jerkstrand}, {Inserra},
  {McCrum}, {Kotak}, {Fraser}, {Wright}, {Chen}, {Smith}, {Young}, {Sim},
  {Valenti}, {Howell}, {Bresolin}, {Kudritzki}, {Tonry}, {Huber}, {Rest},
  {Pastorello}, {Tomasella}, {Cappellaro}, {Benetti}, {Mattila}, {Kankare},
  {Kangas}, {Leloudas}, {Sollerman}, {Taddia}, {Berger}, {Chornock}, {Narayan},
  {Stubbs}, {Foley}, {Lunnan}, {Soderberg}, {Sanders}, {Milisavljevic},
  {Margutti}, {Kirshner}, {Elias-Rosa}, {Morales-Garoffolo}, {Taubenberger},
  {Botticella}, {Gezari}, {Urata}, {Rodney}, {Riess}, {Scolnic}, {Wood-Vasey},
  {Burgett}, {Chambers}, {Flewelling}, {Magnier}, {Kaiser}, {Metcalfe},
  {Morgan}, {Price}, {Sweeney}, \& {Waters}}]{nicholl_slsn_13}
{Nicholl}, M., {Smartt}, S.~J., {Jerkstrand}, A., {et~al.} 2013, \nat, 502, 346

\bibitem[{{O'Connor} \& {Couch}(2018)}]{oconnor_couch_ccsn_3d_18}
{O'Connor}, E.~P. \& {Couch}, S.~M. 2018, \apj, 865, 81

\bibitem[{{Orellana} {et~al.}(2018){Orellana}, {Bersten}, \&
  {Moriya}}]{orellana_pm_18}
{Orellana}, M., {Bersten}, M.~C., \& {Moriya}, T.~J. 2018, \aap, 619, A145

\bibitem[{{Pastorello} {et~al.}(2010){Pastorello}, {Smartt}, {Botticella},
  {Maguire}, {Fraser}, {Smith}, {Kotak}, {Magill}, {Valenti}, {Young},
  {Gezari}, {Bresolin}, {Kudritzki}, {Howell}, {Rest}, {Metcalfe}, {Mattila},
  {Kankare}, {Huang}, {Urata}, {Burgett}, {Chambers}, {Dombeck}, {Flewelling},
  {Grav}, {Heasley}, {Hodapp}, {Kaiser}, {Luppino}, {Lupton}, {Magnier},
  {Monet}, {Morgan}, {Onaka}, {Price}, {Rhoads}, {Siegmund}, {Stubbs},
  {Sweeney}, {Tonry}, {Wainscoat}, {Waterson}, {Waters}, \&
  {Wynn-Williams}}]{pasto_10gx_10}
{Pastorello}, A., {Smartt}, S.~J., {Botticella}, M.~T., {et~al.} 2010, \apjl,
  724, L16

\bibitem[{{Perley} {et~al.}(2016){Perley}, {Quimby}, {Yan}, {Vreeswijk}, {De
  Cia}, {Lunnan}, {Gal-Yam}, {Yaron}, {Filippenko}, {Graham}, {Laher}, \&
  {Nugent}}]{perley_slsn_ic_16}
{Perley}, D.~A., {Quimby}, R.~M., {Yan}, L., {et~al.} 2016, \apj, 830, 13

\bibitem[{{Quimby} {et~al.}(2011){Quimby}, {Kulkarni}, {Kasliwal}, {Gal-Yam},
  {Arcavi}, {Sullivan}, {Nugent}, {Thomas}, {Howell}, {Nakar}, {Bildsten},
  {Theissen}, {Law}, {Dekany}, {Rahmer}, {Hale}, {Smith}, {Ofek}, {Zolkower},
  {Velur}, {Walters}, {Henning}, {Bui}, {McKenna}, {Poznanski}, {Cenko}, \&
  {Levitan}}]{quimby_slsnic_11}
{Quimby}, R.~M., {Kulkarni}, S.~R., {Kasliwal}, M.~M., {et~al.} 2011, \nat,
  474, 487

\bibitem[{{Quimby} {et~al.}(2007){Quimby}, {Wheeler}, {H{\"o}flich}, {Akerlof},
  {Brown}, \& {Rykoff}}]{quimby_06bp_07}
{Quimby}, R.~M., {Wheeler}, J.~C., {H{\"o}flich}, P., {et~al.} 2007, \apj, 666,
  1093

\bibitem[{{Schulze} {et~al.}(2018){Schulze}, {Kr{\"u}hler}, {Leloudas},
  {Gorosabel}, {Mehner}, {Buchner}, {Kim}, {Ibar}, {Amor{\'{\i}}n},
  {Herrero-Illana}, {Anderson}, {Bauer}, {Christensen}, {de Pasquale}, {de
  Ugarte Postigo}, {Gallazzi}, {Hjorth}, {Morrell}, {Malesani}, {Sparre},
  {Stalder}, {Stark}, {Th{\"o}ne}, \& {Wheeler}}]{schulze_slsn_18}
{Schulze}, S., {Kr{\"u}hler}, T., {Leloudas}, G., {et~al.} 2018, \mnras, 473,
  1258

\bibitem[{{Suzuki} \& {Maeda}(2017)}]{suzuki_pm_2d_17}
{Suzuki}, A. \& {Maeda}, K. 2017, \mnras, 466, 2633

\bibitem[{{Swartz} {et~al.}(1995){Swartz}, {Sutherland}, \&
  {Harkness}}]{swartz_gray_95}
{Swartz}, D.~A., {Sutherland}, P.~G., \& {Harkness}, R.~P. 1995, \apj, 446, 766

\bibitem[{{Thompson} {et~al.}(2004){Thompson}, {Chang}, \&
  {Quataert}}]{thompson_pm_04}
{Thompson}, T.~A., {Chang}, P., \& {Quataert}, E. 2004, \apj, 611, 380

\bibitem[{{Tolstov} {et~al.}(2017){Tolstov}, {Zhiglo}, {Nomoto}, {Sorokina},
  {Kozyreva}, \& {Blinnikov}}]{tolstov_slsnic_17}
{Tolstov}, A., {Zhiglo}, A., {Nomoto}, K., {et~al.} 2017, \apjl, 845, L2

\bibitem[{{Usov}(1992)}]{usov_pm_92}
{Usov}, V.~V. 1992, \nat, 357, 472

\bibitem[{{Vartanyan} {et~al.}(2019){Vartanyan}, {Burrows}, {Radice},
  {Skinner}, \& {Dolence}}]{vartanyan_ccsn_3d_19}
{Vartanyan}, D., {Burrows}, A., {Radice}, D., {Skinner}, M.~A., \& {Dolence},
  J. 2019, \mnras, 482, 351

\bibitem[{{Wheeler} {et~al.}(2000){Wheeler}, {Yi}, {H{\"o}flich}, \&
  {Wang}}]{wheeler+00}
{Wheeler}, J.~C., {Yi}, I., {H{\"o}flich}, P., \& {Wang}, L. 2000, \apj, 537,
  810

\bibitem[{{Wongwathanarat} {et~al.}(2015){Wongwathanarat}, {Mueller}, \&
  {Janka}}]{wongwathanarat_15_3d}
{Wongwathanarat}, A., {Mueller}, E., \& {Janka}, H.-T. 2015, \aap, 577, A48

\bibitem[{{Woosley}(2010)}]{woosley_pm_10}
{Woosley}, S.~E. 2010, \apjl, 719, L204

\bibitem[{{Woosley} \& {Heger}(2006)}]{WH06}
{Woosley}, S.~E. \& {Heger}, A. 2006, \apj, 637, 914

\bibitem[{{Yan} {et~al.}(2017{\natexlab{a}}){Yan}, {Lunnan}, {Perley},
  {Gal-Yam}, {Yaron}, {Roy}, {Quimby}, {Sollerman}, {Fremling}, {Leloudas},
  {Cenko}, {Vreeswijk}, {Graham}, {Howell}, {De Cia}, {Ofek}, {Nugent},
  {Kulkarni}, {Hosseinzadeh}, {Masci}, {McCully}, {Rebbapragada}, \&
  {Wo{\'z}niak}}]{yan_slsn_ic_17}
{Yan}, L., {Lunnan}, R., {Perley}, D.~A., {et~al.} 2017{\natexlab{a}}, \apj,
  848, 6

\bibitem[{{Yan} {et~al.}(2017{\natexlab{b}}){Yan}, {Quimby}, {Gal-Yam},
  {Brown}, {Blagorodnova}, {Ofek}, {Lunnan}, {Cooke}, {Cenko}, {Jencson}, \&
  {Kasliwal}}]{yan_Gaia16apd_17}
{Yan}, L., {Quimby}, R., {Gal-Yam}, A., {et~al.} 2017{\natexlab{b}}, \apj, 840,
  57

\bibitem[{{Yan} {et~al.}(2015){Yan}, {Quimby}, {Ofek}, {Gal-Yam}, {Mazzali},
  {Perley}, {Vreeswijk}, {Leloudas}, {De Cia}, {Masci}, {Cenko}, {Cao},
  {Kulkarni}, {Nugent}, {Rebbapragada}, {Wo{\'z}niak}, \&
  {Yaron}}]{yan_slsnic_ha_15}
{Yan}, L., {Quimby}, R., {Ofek}, E., {et~al.} 2015, \apj, 814, 108

\bibitem[{{Yaron} \& {Gal-Yam}(2012)}]{wiserep}
{Yaron}, O. \& {Gal-Yam}, A. 2012, \pasp, 124, 668

\bibitem[{{Yaron} {et~al.}(2017){Yaron}, {Perley}, {Gal-Yam}, {Groh}, {Horesh},
  {Ofek}, {Kulkarni}, {Sollerman}, {Fransson}, {Rubin}, {Szabo}, {Sapir},
  {Taddia}, {Cenko}, {Valenti}, {Arcavi}, {Howell}, {Kasliwal}, {Vreeswijk},
  {Khazov}, {Fox}, {Cao}, {Gnat}, {Kelly}, {Nugent}, {Filippenko}, {Laher},
  {Wozniak}, {Lee}, {Rebbapragada}, {Maguire}, {Sullivan}, \&
  {Soumagnac}}]{yaron_13fs_17}
{Yaron}, O., {Perley}, D.~A., {Gal-Yam}, A., {et~al.} 2017, Nature Physics, 13,
  510

\bibitem[{{Yoon} \& {Langer}(2005)}]{yoon_grb_05}
{Yoon}, S.-C. \& {Langer}, N. 2005, \aap, 443, 643

\bibitem[{{Yoon} {et~al.}(2010){Yoon}, {Woosley}, \& {Langer}}]{yoon_ibc_10}
{Yoon}, S.-C., {Woosley}, S.~E., \& {Langer}, N. 2010, \apj, 725, 940

\end{thebibliography}
\end{document}